\documentclass[11pt]{article}
\usepackage[titletoc,toc]{appendix}
\usepackage{amssymb}
\usepackage[numbers,sort&compress]{natbib}
\usepackage{amsmath,amsfonts,graphicx,epsfig}
\usepackage{bm}
\usepackage{xcolor}
\usepackage{color}
\usepackage{soul}
\def\bea{\begin{eqnarray}}
\def\eea{\end{eqnarray}}
\def\be{\begin{equation}}
\def\ee{\end{equation}}
\def\ba{\begin{array}}
\def\ea{\end{array}}
\def\nn{\nonumber}

\def\p{{\bf p}}
\def\q{{\bf q}}
\def\k{{\bf k}}

\def\x{{\bf x}}

\usepackage{tikz}
\usetikzlibrary{arrows.meta,positioning,shapes.geometric}

\newcommand{\SKSIZE}{9pt}   
\newcommand{\SKLINE}{0.8pt}  
\newcommand{\BOUNDY}{2.6pt} 

\tikzset{
  sk/vertex+/.style = {circle, draw, fill=black, minimum size=\SKSIZE, inner sep=0pt},
  sk/vertex-/.style = {circle, draw, fill=white, minimum size=\SKSIZE, inner sep=0pt},
  sk/source+/.style = {regular polygon, regular polygon sides=5, draw, fill=black, minimum size=\SKSIZE, inner sep=0pt},
  sk/source-/.style = {regular polygon, regular polygon sides=5, draw, fill=white, minimum size=\SKSIZE, inner sep=0pt},
  sk/bound+/.style  = {regular polygon, regular polygon sides=4, draw, fill=black, minimum size=\SKSIZE, inner sep=0pt},
  sk/bound-/.style  = {regular polygon, regular polygon sides=4, draw, fill=white, minimum size=\SKSIZE, inner sep=0pt},
  sk/prop/.style   = {line width=\SKLINE, line cap=round},
  sk/label/.style  = {font=\scriptsize, inner sep=1pt},
  sk/boundary/.style= {line width=\SKLINE}
}

\newcommand{\skvertexp}[2]{\node[sk/vertex+] (#1) at #2 {};}  
\newcommand{\skvertexm}[2]{\node[sk/vertex-] (#1) at #2 {};}  
\newcommand{\skbound}[2]{\node[sk/bound-]  (#1) at #2 {};}   

\colorlet{skgray}{gray!55}            

\newcommand{\sksourceg}[2]{%
  \node[regular polygon, regular polygon sides=5, draw=black, fill=skgray, minimum size=\SKSIZE, inner sep=0pt] (#1) at #2 {};%
}

\newcommand{\G}[2]{\draw[sk/prop] (#1) -- (#2);}
\newcommand{\Garc}[5][]{\draw[sk/prop,#1] (#2) to [out=#4,in=#5] (#3);} 

\newcommand{\SKDOUBLESEP}{1.4pt} 

\tikzset{
  sk/propdouble/.style={
    line width=\SKLINE,
    double,
    double distance=\SKDOUBLESEP,
    line cap=round
  }
}

\newcommand{\Gdouble}[2]{\draw[sk/propdouble] (#1) -- (#2);}

\newcommand{\Gdoublearc}[5][]{%
  \draw[sk/propdouble,#1] (#2) to [out=#4,in=#5] (#3);
}

\newcommand{\SKDOTR}{1.6pt} 

\newcommand{\skvertexd}[2]{%
  \node[sk/vertex-] (#1) at #2 {};%
  \fill (#1.center) circle[radius=\SKDOTR];%
}

\newcommand{\skvertexgd}[2]{%
  \node[circle, draw, fill=skgray, minimum size=\SKSIZE, inner sep=0pt] (#1) at #2 {};%
  \fill (#1.center) circle[radius=\SKDOTR];%
}

\newcommand{\skvertexpdot}[2]{%
  \node[sk/vertex+] (#1) at #2 {};%
  \fill[white] (#1.center) circle[radius=\SKDOTR];%
}

\tikzset{
  sk/propdashed/.style={
    sk/prop,
    dashed
  }
}

\newcommand{\Gdashed}[2]{\draw[sk/propdashed] (#1) -- (#2);}

\newcommand{\Gdashedarc}[5][]{%
  \draw[sk/propdashed,#1] (#2) to[out=#4,in=#5] (#3);
}

\providecommand{\SKXLINE}{0.7pt} 

\makeatletter
\@ifundefined{SKXclipR}{\newlength{\SKXclipR}}{}
\@ifundefined{SKXhalf}{\newlength{\SKXhalf}}{}
\makeatother

\newcommand{\sksourcex}[2]{%
  \node[circle, draw, fill=white, minimum size=\SKSIZE, inner sep=0pt] (#1) at #2 {};%
  \pgfmathsetlength{\SKXclipR}{0.5*\SKSIZE - 0.6*\SKXLINE}%
  \pgfmathsetlength{\SKXhalf}{\SKXclipR/1.41421356}%
  \begin{scope}
    \clip (#1.center) circle[radius=\SKXclipR];%
    \draw[line width=\SKXLINE, draw=black]
      (#1.center) ++(-\SKXhalf,-\SKXhalf) -- ++(2\SKXhalf,2\SKXhalf);%
    \draw[line width=\SKXLINE, draw=black]
      (#1.center) ++(-\SKXhalf,\SKXhalf) -- ++(2\SKXhalf,-2\SKXhalf);%
  \end{scope}%
}

\newcommand{\sksourcexblack}[2]{%
  \node[circle, draw, fill=black, minimum size=\SKSIZE, inner sep=0pt] (#1) at #2 {};%
  \pgfmathsetlength{\SKXclipR}{0.5*\SKSIZE - 0.6*\SKXLINE}%
  \pgfmathsetlength{\SKXhalf}{\SKXclipR/1.41421356}%
  \begin{scope}
    \clip (#1.center) circle[radius=\SKXclipR];%
    \draw[line width=\SKXLINE, draw=white]
      (#1.center) ++(-\SKXhalf,-\SKXhalf) -- ++(2\SKXhalf,2\SKXhalf);%
    \draw[line width=\SKXLINE, draw=white]
      (#1.center) ++(-\SKXhalf,\SKXhalf) -- ++(2\SKXhalf,-2\SKXhalf);%
  \end{scope}%
}

\newcommand{\SKMIXEDSEP}{1.2pt} 

\newcommand{\Gmix}[2]{%
  \begin{scope}[transform canvas={shift={(0,-\SKMIXEDSEP)}}]
    \draw[sk/prop] (#1) -- (#2);
  \end{scope}
  \begin{scope}[transform canvas={shift={(0,\SKMIXEDSEP)}}]
    \draw[sk/prop,dashed] (#1) -- (#2);
  \end{scope}
}

\begin{document}

\setlength\arraycolsep{2pt}

\renewcommand{\theequation}{\arabic{section}.\arabic{equation}}
\setcounter{page}{1}

\begin{titlepage}

\begin{center}

\vskip 1.0 cm

{\LARGE  \bf From the Wavefunction of the Universe to \\[5pt]  In-In-Correlators: A Perturbative Map to All Orders}

\vskip 1.0cm

{\Large
Gonzalo A. Palma
}

\vskip 0.5cm

{\it 
Grupo de Cosmolog\'ia, Departamento de F\'{i}sica, FCFM, \mbox{Universidad de Chile} \\ Blanco Encalada 2008, Santiago, Chile.
}

\vskip 1.5cm

\end{center}

\begin{abstract}

Both the Wavefunction of the Universe and the Schwinger--Keldysh in-in formalism are central tools for analyzing primordial cosmological observables, such as equal-time correlation functions. While their conceptual equivalence is well established, a systematic and explicit map between their diagrammatic expansions has remained elusive. In this article, I construct such a map by analyzing the relation between the two frameworks at the diagrammatic level. I show that diagrams contributing to correlation functions in the Wavefunction of the Universe approach can be uniquely reorganized into Schwinger--Keldysh diagrams. This correspondence holds to all orders in perturbation theory, including arbitrary numbers of interaction vertices and loops.

\end{abstract}

\end{titlepage}

\setcounter{equation}{0}
\section{Introduction}

Understanding the origin of the Universe's large-scale structure is a central goal of modern cosmology. Achieving this objective requires, in particular, a careful control of the systematics involved in the computation of equal-time correlation functions, which describe the statistical properties of primordial fluctuations responsible for the observed structure. In practice, these correlation functions are most commonly computed using the Schwinger--Keldysh in-in formalism~\cite{Schwinger:1960qe, Keldysh:1964ud, Jordan:1986ug, Calzetta:1986ey, Maldacena:2002vr, Weinberg:2005vy, Adshead:2009cb, Chen:2017ryl}. This framework provides a systematic perturbative method for evaluating expectation values in time-dependent backgrounds, such as those relevant during inflation~\cite{Achucarro:2022qrl}. An alternative, and increasingly influential perspective, is provided by the Wavefunction of the Universe approach~\cite{Hartle:1983ai, Halliwell:1984eu, Anninos:2014lwa, Pajer:2020wxk, Goodhew:2024eup}, which encodes the quantum state of cosmological perturbations at late times and organizes their dynamics in terms of wavefunction coefficients. This formalism has proven particularly useful to understand the role of unitarity, locality, and symmetries on the final form of correlators, enabling the use of powerful techniques, such as the cosmological bootstrap program~\cite{Arkani-Hamed:2015bza, Arkani-Hamed:2018kmz, Pajer:2020wnj, Baumann:2021fxj, Baumann:2022jpr, Xianyu:2022jwk, Wang:2022eop} to fix wavefunction coefficients (an approach that is less manifest in the in-in representation).  In what follows, I explore the equivalence between these two frameworks: the Wavefunction of the Universe and the Schwinger--Keldysh in-in formalism. I derive a simple and general connection between them, which can be naturally formulated at the diagrammatic level. 

As is well known, correlation functions computed with the Schwinger--Keldysh formalism are represented in terms of diagrams built from a doubled set of degrees of freedom~(see \cite{Chen:2017ryl} for a recent derivation of the Schwinger--Keldysh rules in the context of cosmology). While this doubling is essential for preserving causality and unitarity, it also leads to a rapid proliferation of diagrams, which can become increasingly difficult to organize as the perturbative order grows. To illustrate this point, consider the following tree-level diagrams contributing to the connected equal-time four-point correlation function of a scalar field $\phi$, evaluated at a final time $t_f$:
\bea 
\label{intro:example-1}
\Big\langle \phi (\k_1) \cdots \phi (\k_4)  \Big\rangle_c 
&\supset& 
\,\,
 \begin{tikzpicture}[baseline=-3.0pt]
  \draw[sk/boundary] (-1.8,\BOUNDY) -- (1.8,\BOUNDY);
  \skbound{B1}{(-1.5,\BOUNDY)}
  \skbound{B2}{( -0.5,\BOUNDY)}
  \skbound{B3}{( 0.5,\BOUNDY)}
    \skbound{B4}{( 1.5,\BOUNDY)}
  \skvertexp{V1}{(-1,-1)}
    \skvertexp{V2}{(1,-1)}
  \G{V1}{B1}
  \G{V1}{B2}
  \G{V2}{B3}
    \G{V2}{B4}
        \G{V1}{V2}
  \node[above=2pt] at (B1) {$\k_1$};
  \node[above=2pt] at (B2) {$\k_2$};
  \node[above=2pt] at (B3) {$\k_3$};
  \node[above=2pt] at (B4) {$\k_4$};
\end{tikzpicture}  
+
 \begin{tikzpicture}[baseline=-3.0pt]
\draw[sk/boundary] (-1.8,\BOUNDY) -- (1.8,\BOUNDY);
  \skbound{B1}{(-1.5,\BOUNDY)}
  \skbound{B2}{( -0.5,\BOUNDY)}
  \skbound{B3}{( 0.5,\BOUNDY)}
    \skbound{B4}{( 1.5,\BOUNDY)}
  \skvertexm{V1}{(-1,-1)}
    \skvertexm{V2}{(1,-1)}
  \G{V1}{B1}
  \G{V1}{B2}
  \G{V2}{B3}
    \G{V2}{B4}
        \G{V1}{V2}
  \node[above=2pt] at (B1) {$\k_1$};
  \node[above=2pt] at (B2) {$\k_2$};
  \node[above=2pt] at (B3) {$\k_3$};
  \node[above=2pt] at (B4) {$\k_4$};
\end{tikzpicture}   
\nn \\
&&
\,\,
+
 \begin{tikzpicture}[baseline=-3.0pt]
  \draw[sk/boundary] (-1.8,\BOUNDY) -- (1.8,\BOUNDY);
  \skbound{B1}{(-1.5,\BOUNDY)}
  \skbound{B2}{( -0.5,\BOUNDY)}
  \skbound{B3}{( 0.5,\BOUNDY)}
    \skbound{B4}{( 1.5,\BOUNDY)}
  \skvertexp{V1}{(-1,-1)}
    \skvertexm{V2}{(1,-1)}
  \G{V1}{B1}
  \G{V1}{B2}
  \G{V2}{B3}
    \G{V2}{B4}
        \G{V1}{V2}
  \node[above=2pt] at (B1) {$\k_1$};
  \node[above=2pt] at (B2) {$\k_2$};
  \node[above=2pt] at (B3) {$\k_3$};
  \node[above=2pt] at (B4) {$\k_4$};
\end{tikzpicture}  
+
 \begin{tikzpicture}[baseline=-3.0pt]
\draw[sk/boundary] (-1.8,\BOUNDY) -- (1.8,\BOUNDY);
  \skbound{B1}{(-1.5,\BOUNDY)}
  \skbound{B2}{( -0.5,\BOUNDY)}
  \skbound{B3}{( 0.5,\BOUNDY)}
    \skbound{B4}{( 1.5,\BOUNDY)}
  \skvertexm{V1}{(-1,-1)}
    \skvertexp{V2}{(1,-1)}
  \G{V1}{B1}
  \G{V1}{B2}
  \G{V2}{B3}
    \G{V2}{B4}
        \G{V1}{V2}
  \node[above=2pt] at (B1) {$\k_1$};
  \node[above=2pt] at (B2) {$\k_2$};
  \node[above=2pt] at (B3) {$\k_3$};
  \node[above=2pt] at (B4) {$\k_4$};
\end{tikzpicture}  
.
\eea
As can be seen, these diagrams are constructed from two classes of three-legged vertices, denoted by black and white solid dots. These two types of vertices arise as a direct consequence of the doubling of degrees of freedom inherent to the Schwinger--Keldysh formalism. The vertices are connected to one another by bulk-to-bulk propagators, and to the boundary, where the external momenta flow into the diagram, by bulk-to-boundary propagators.

The diagrammatic rules determining the precise form of each object appearing in the expressions above will be introduced in more detail later. For the moment, I wish to emphasize the following key aspects. First, the rules associated with white vertices are the complex conjugates of those associated with black vertices. Second, each vertex involves a time integral extending from the infinite past up to the final time $t_f$ at which the correlation function is evaluated. Third, bulk-to-bulk propagators connecting vertices of the same color contain Heaviside step functions of the time variables associated with each vertex. As a result, correlation functions involving diagrams with multiple vertices generally lead to nested time integrals, which are notoriously difficult to handle. By contrast, bulk-to-bulk propagators connecting vertices of different colors, as well as bulk-to-boundary propagators connecting vertices to external legs, do not contain such step functions. Moreover, bulk-to-bulk propagators connecting vertices of different colors can be factorized into products of functions depending independently on each time-integration variable. As a consequence of these properties, the previous set of diagrams can be schematically reorganized in the following way:
\bea
\Big\langle \phi (\k_1) \cdots \phi (\k_4)  \Big\rangle_c 
&\supset& 
\,\,
2 {\rm Re}  \Bigg\{ 
\begin{tikzpicture}[baseline=-3.0pt]
  \draw[sk/boundary] (-1.8,\BOUNDY) -- (1.8,\BOUNDY);
  \skbound{B1}{(-1.5,\BOUNDY)}
  \skbound{B2}{( -0.5,\BOUNDY)}
  \skbound{B3}{( 0.5,\BOUNDY)}
    \skbound{B4}{( 1.5,\BOUNDY)}
  \skvertexp{V1}{(-1,-1)}
    \skvertexp{V2}{(1,-1)}
  \G{V1}{B1}
  \G{V1}{B2}
  \G{V2}{B3}
    \G{V2}{B4}
        \G{V1}{V2}
  \node[above=2pt] at (B1) {$\k_1$};
  \node[above=2pt] at (B2) {$\k_2$};
  \node[above=2pt] at (B3) {$\k_3$};
  \node[above=2pt] at (B4) {$\k_4$};
\end{tikzpicture}  
-
 \begin{tikzpicture}[baseline=-3.0pt]
  \draw[sk/boundary] (-1.8,\BOUNDY) -- (1.8,\BOUNDY);
  \skbound{B1}{(-1.5,\BOUNDY)}
  \skbound{B2}{( -0.5,\BOUNDY)}
   \skbound{S1}{( -0.17, -1.0)}
  \skvertexp{V1}{(-1,-1)}
  \G{V1}{B1}
  \G{V1}{B2}
  \G{V1}{S1}
  \node[above=2pt] at (B1) {$\k_1$};
  \node[above=2pt] at (B2) {$\k_2$};
    \skbound{S2}{( 0.17, -1.0)}
  \skbound{B3}{( 0.5,\BOUNDY)}
    \skbound{B4}{( 1.5,\BOUNDY)}
    \skvertexp{V2}{(1,-1)}
  \G{V2}{S2}
  \G{V2}{B3}
    \G{V2}{B4}
  \node[above=2pt] at (B3) {$\k_3$};
  \node[above=2pt] at (B4) {$\k_4$};
\end{tikzpicture} 
\Bigg\}
\nn 
\\
&&
\,\,
+ 
\int_{\q}  2 {\rm Re}  \Bigg\{ 
 \begin{tikzpicture}[baseline=-3.0pt]
  \draw[sk/boundary] (-1.8,\BOUNDY) -- (0.0,\BOUNDY);
  \skbound{B1}{(-1.5,\BOUNDY)}
  \skbound{B2}{( -0.5,\BOUNDY)}
   \skbound{B3}{( -0.0, -1.0)}
  \skvertexp{V1}{(-1,-1)}
  \G{V1}{B1}
  \G{V1}{B2}
  \G{V1}{B3}
  \node[above=2pt] at (B1) {$\k_1$};
  \node[above=2pt] at (B2) {$\k_2$};
    \node[above=2pt] at (B3) {$\q$};
\end{tikzpicture}  
\Bigg\}
\times
2 {\rm Re}  \Bigg\{ 
\begin{tikzpicture}[baseline=-3.0pt]
  \draw[sk/boundary] (0.0,\BOUNDY) -- (1.8,\BOUNDY);
    \skbound{B2}{( -0.0, -1.0)}
  \skbound{B3}{( 0.5,\BOUNDY)}
    \skbound{B4}{( 1.5,\BOUNDY)}
    \skvertexp{V2}{(1,-1)}
  \G{V2}{B2}
  \G{V2}{B3}
    \G{V2}{B4}
  \node[above=2pt] at (B3) {$\k_3$};
  \node[above=2pt] at (B4) {$\k_4$};
    \node[above=2pt] at (B2) {$\q$};
\end{tikzpicture} 
\Bigg\} 
.
\eea
In this new combination of diagrams, there are bulk-to-boundary propagators that, instead of meeting the boundary, are glued together, with internal momenta flowing through them. This is why, in the previous diagram, there are boundary vertices (\begin{tikzpicture}[baseline=-3.0pt]
  \skbound{B1}{( 0.0, 0.0)} \end{tikzpicture}) that are not attached to the boundary. This gluing arises as a consequence of the factorization of bulk-to-bulk propagators connecting vertices of different colors. Noteworthily, the entire collection of diagrams is now expressed in terms of a single type of vertex.

It turns out that the content inside the brackets of the first line is precisely the Wavefunction of the Universe coefficient $\psi^{(2)}_4(\k_1,\ldots,\k_4)$ at second order in perturbation theory, while the second line contains the product of two wavefunction coefficients, $\psi^{(1)}_3(\k_1,\k_2,\q)$ and $\psi^{(1)}_3(\k_3,\k_4,\q)$, glued together through an integration over the internal momentum $\q$. That is, the four-point function can be schematically written as
\be
\Big\langle \phi (\k_1) \cdots \phi (\k_4)  \Big\rangle_c  \!\! \supset  2 \, {\rm Re} \left\{ \psi^{(2)}_4(\k_1,\ldots,\k_4)\right\}
+ \! \int_{\q} \!  2 {\rm Re} \left\{ \psi^{(1)}_3 (\k_1 , \k_2 , \q) \right\}
\times  2 {\rm Re} \left\{ \psi^{(1)}_3 (\k_3 , \k_4 , \q) \right\}  .
\ee
A crucial step allowing for this relation is the appropriate identification of the bulk-to-bulk propagator appearing in the Wavefunction of the Universe approach. Indeed, the bulk-to-bulk propagator entering the diagrammatic rules used to compute wavefunction coefficients is equal to the Schwinger--Keldysh propagator connecting two black vertices, minus a correction. This modified propagator, which I will denote by a double line, can be written as
\be \label{modified-prop-intro}
\begin{tikzpicture}[baseline=-3.0pt]
  \skvertexp{V1}{(-1, 0)}
    \skvertexp{V2}{(1 , 0)}
        \Gdouble{V1}{V2}
\end{tikzpicture}  
\,\,
= 
\,\,
 \begin{tikzpicture}[baseline=-3.0pt]
  \skvertexp{V1}{(-1, 0)}
    \skvertexp{V2}{(1 , 0)}
        \G{V1}{V2}
\end{tikzpicture}   
\,\,
-
\,\,
 \begin{tikzpicture}[baseline=-3.0pt]
  \skbound{B1}{( -0.15, 0.0)}
  \skvertexp{V1}{(-1.0, 0)}
        \G{V1}{B1}
   \skbound{B2}{( 0.15, 0.0)}
    \skvertexp{V2}{(1.0 , 0)}
        \G{B2}{V2}
\end{tikzpicture}   
\,\,
.
\ee
The second diagram in (\ref{modified-prop-intro}) corresponds to two bulk-to-boundary propagators glued together, forming an effective bulk-to-bulk propagator. This subtraction ensures that the bulk-to-bulk propagator defining wavefunction coefficients vanishes whenever either of its time arguments is evaluated at the boundary time $t_f$. With this identification, the second-order wavefunction coefficient $\psi^{(2)}_4(\k_1,\ldots,\k_4)$ can be represented diagrammatically as
\be
\psi^{(2)}_4(\k_1,\ldots,\k_4)
=
\,\,
\begin{tikzpicture}[baseline=-3.0pt]
  \draw[sk/boundary] (-1.8,\BOUNDY) -- (1.8,\BOUNDY);
  \skbound{B1}{(-1.5,\BOUNDY)}
  \skbound{B2}{( -0.5,\BOUNDY)}
  \skbound{B3}{( 0.5,\BOUNDY)}
    \skbound{B4}{( 1.5,\BOUNDY)}
  \skvertexp{V1}{(-1,-1)}
    \skvertexp{V2}{(1,-1)}
  \G{V1}{B1}
  \G{V1}{B2}
  \G{V2}{B3}
    \G{V2}{B4}
        \Gdouble{V1}{V2}
  \node[above=2pt] at (B1) {$\k_1$};
  \node[above=2pt] at (B2) {$\k_2$};
  \node[above=2pt] at (B3) {$\k_3$};
  \node[above=2pt] at (B4) {$\k_4$};
\end{tikzpicture}   .
\ee

Having introduced the notation for bulk-to-bulk propagators relevant to wavefunction coefficients, let me now consider an example involving loop diagrams. Specifically, consider the one-loop contribution to the equal-time three-point correlation function constructed from three-legged vertices. In the Schwinger--Keldysh formalism, this contribution takes the form
\bea
\label{intro:example-2}
\Big\langle \phi (\k_1) \phi (\k_2) \phi (\k_3)  \Big\rangle_c 
&\supset&
\,\,
 \begin{tikzpicture}[baseline=-3.0pt]
  \draw[sk/boundary] (-1.3,\BOUNDY) -- (1.3,\BOUNDY);
  \skbound{B1}{(-1.0,\BOUNDY)}
  \skbound{B2}{( -0.0,\BOUNDY)}
  \skbound{B3}{( 1.0,\BOUNDY)}
  \skvertexp{V1}{(-0.7,-1)}
    \skvertexp{V2}{(0.7,-1)}
      \skvertexp{V3}{(0 ,-0.65)}
  \G{V1}{B1}
  \G{V3}{B2}
  \G{V2}{B3}
          \Garc{V1}{V2}{-45}{-135}  
           \Garc{V1}{V2}{45}{135}  
  \node[above=2pt] at (B1) {$\k_1$};
  \node[above=2pt] at (B2) {$\k_2$};
  \node[above=2pt] at (B3) {$\k_3$};
\end{tikzpicture}  
+ 
\begin{tikzpicture}[baseline=-3.0pt]
  \draw[sk/boundary] (-1.3,\BOUNDY) -- (1.3,\BOUNDY);
  \skbound{B1}{(-1.0,\BOUNDY)}
  \skbound{B2}{( -0.0,\BOUNDY)}
  \skbound{B3}{( 1.0,\BOUNDY)}
  \skvertexm{V1}{(-0.7,-1)}
    \skvertexm{V2}{(0.7,-1)}
  \G{V1}{B1}
  \G{V3}{B2}
  \G{V2}{B3}
          \Garc{V1}{V2}{-45}{-135}  
           \Garc{V1}{V2}{45}{135}  
             \skvertexm{V3}{(0 ,-0.65)}
  \node[above=2pt] at (B1) {$\k_1$};
  \node[above=2pt] at (B2) {$\k_2$};
  \node[above=2pt] at (B3) {$\k_3$};
\end{tikzpicture} 
\nn 
\\
&&
\,\,
 + 
 \begin{tikzpicture}[baseline=-3.0pt]
  \draw[sk/boundary] (-1.3,\BOUNDY) -- (1.3,\BOUNDY);
  \skbound{B1}{(-1.0,\BOUNDY)}
  \skbound{B2}{( -0.0,\BOUNDY)}
  \skbound{B3}{( 1.0,\BOUNDY)}
  \skvertexm{V1}{(-0.7,-1)}
    \skvertexp{V2}{(0.7,-1)}
  \G{V1}{B1}
  \G{V3}{B2}
  \G{V2}{B3}
          \Garc{V1}{V2}{-45}{-135}  
           \Garc{V1}{V2}{45}{135}  
                 \skvertexp{V3}{(0 ,-0.65)}
  \node[above=2pt] at (B1) {$\k_1$};
  \node[above=2pt] at (B2) {$\k_2$};
  \node[above=2pt] at (B3) {$\k_3$};
\end{tikzpicture}  
+ 
\begin{tikzpicture}[baseline=-3.0pt]
  \draw[sk/boundary] (-1.3,\BOUNDY) -- (1.3,\BOUNDY);
  \skbound{B1}{(-1.0,\BOUNDY)}
  \skbound{B2}{( -0.0,\BOUNDY)}
  \skbound{B3}{( 1.0,\BOUNDY)}
  \skvertexp{V1}{(-0.7,-1)}
    \skvertexm{V2}{(0.7,-1)}
  \G{V1}{B1}
  \G{V3}{B2}
  \G{V2}{B3}
          \Garc{V1}{V2}{-45}{-135}  
           \Garc{V1}{V2}{45}{135}  
             \skvertexm{V3}{(0 ,-0.65)}
  \node[above=2pt] at (B1) {$\k_1$};
  \node[above=2pt] at (B2) {$\k_2$};
  \node[above=2pt] at (B3) {$\k_3$};
\end{tikzpicture} 
  + {\rm perms } ,
\eea
where ``perms.'' represents four additional diagrams obtained by permuting the external momenta of the third and fourth diagrams. Because the first two diagrams contain only vertices of the same color, it is clear that they cannot be factorized into products of lower-order diagrams. By contrast, the remaining six diagrams can all be factorized. The result is a collection of diagrams that can be drawn using a single black vertex:
\bea
\Big\langle \phi (\k_1) \phi (\k_2) \phi (\k_3)  \Big\rangle_c 
&\supset&
\,\,
2 {\rm Re} \Bigg\{
 \begin{tikzpicture}[baseline=-3.0pt]
  \draw[sk/boundary] (-1.3,\BOUNDY) -- (1.3,\BOUNDY);
  \skbound{B1}{(-1.0,\BOUNDY)}
  \skbound{B2}{( -0.0,\BOUNDY)}
  \skbound{B3}{( 1.0,\BOUNDY)}
  \skvertexp{V1}{(-0.7,-1)}
    \skvertexp{V2}{(0.7,-1)}
  \G{V1}{B1}
  \G{V2}{B3}
          \Gdoublearc{V1}{V2}{-45}{-135}  
           \Gdoublearc{V1}{V2}{45}{135}  
                 \skvertexp{V3}{(0 ,-0.65)}
                   \G{V3}{B2}
  \node[above=2pt] at (B1) {$\k_1$};
  \node[above=2pt] at (B2) {$\k_2$};
  \node[above=2pt] at (B3) {$\k_3$};
\end{tikzpicture}  
\Bigg\} 
+
\int_{\q} 2 {\rm Re} \Bigg\{
 \begin{tikzpicture}[baseline=-3.0pt]
  \draw[sk/boundary] (-1.3,\BOUNDY) -- (1.3,\BOUNDY);
  \skbound{B1}{(-1.0,\BOUNDY)}
  \skbound{B2}{( -0.0,\BOUNDY)}
  \skbound{B3}{( 1.0,\BOUNDY)}
  \skvertexp{V1}{(-0.7,-0.8)}
    \skvertexp{V2}{(0.7,-0.8)}
  \G{V1}{B1}
  \G{V2}{B3}
  \skbound{S1}{( -0.15, -1.3)}
  \skbound{S2}{( 0.15, -1.3)}
           \Gdoublearc{V1}{V2}{45}{135}  
            \G{V1}{S1}
            \G{V2}{S2}
                 \skvertexp{V3}{(0 ,-0.45)}
                   \G{V3}{B2}
  \node[above=2pt] at (B1) {$\k_1$};
  \node[above=2pt] at (B2) {$\k_2$};
  \node[above=2pt] at (B3) {$\k_3$};
  \node[above=2pt] at (S1) {$\q$};
    \node[above=2pt] at (S2) {$\q$};
\end{tikzpicture}  
\Bigg\} 
\nn \\
&&  \!\!\!\!\!\!\!\!\!\!\!\!\!\!\!\!\!\!\!\!\!\!\!\!\!\!\!\!\!\!\!\!\!\!\!\!
+\int_{\q_1} \int_{\q_2}  \int_{\q_3} 
2 {\rm Re }
\Bigg\{
\begin{tikzpicture}[baseline=-3.0pt]
  \draw[sk/boundary] (-0.3,\BOUNDY) -- (1.0 ,\BOUNDY);
  \skbound{B3}{( 0.8, -0.5)}
    \skbound{B4}{( 0.8, -1.0)}
             \skvertexp{V1}{(0.0,-0.7)}
  \skbound{B1}{(0.0,\BOUNDY)}
  \node[above=2pt] at (B1) {$\k_1$};
     \G{V1}{B1}
       \G{B4}{V1}
        \G{B3}{V1}
          \node[above=2pt] at (B3) {$\q_1$};
           \node[below=2pt] at (B4) {$\q_2$};
\end{tikzpicture} 
\Bigg\}
\times
2 {\rm Re }
\Bigg\{
\begin{tikzpicture}[baseline=-3.0pt]
  \draw[sk/boundary] (-0.7,\BOUNDY) -- (0.7 ,\BOUNDY);
  \skbound{B3}{( -0.5, -1.0)}
    \skbound{B4}{( +0.5, -1.0)}
             \skvertexp{V1}{(0.0,-0.7)}
  \skbound{B1}{(0.0,\BOUNDY)}
  \node[above=2pt] at (B1) {$\k_3$};
     \G{V1}{B1}
       \G{B4}{V1}
        \G{B3}{V1}
          \node[below=2pt] at (B3) {$\q_2$};
           \node[below=2pt] at (B4) {$\q_3$};
\end{tikzpicture} 
\Bigg\}
\times
2 {\rm Re }
\Bigg\{
\begin{tikzpicture}[baseline=-3.0pt]
  \draw[sk/boundary] (-1,\BOUNDY) -- (0.3 ,\BOUNDY);
  \skbound{B3}{( -0.8, -0.5)}
    \skbound{B4}{( -0.8, -1.0)}
             \skvertexp{V1}{(0.0,-0.7)}
  \skbound{B1}{(0.0,\BOUNDY)}
  \node[above=2pt] at (B1) {$\k_3$};
     \G{V1}{B1}
       \G{B4}{V1}
        \G{B3}{V1}
          \node[above=2pt] at (B3) {$\q_1$};
           \node[below=2pt] at (B4) {$\q_3$};
\end{tikzpicture} 
\Bigg\}
\nn 
\\
&&  \!\!\!\!\!\!\!\!\!\!\!\!\!\!\!\!\!\!\!\!\!\!\!\!\!\!\!\!\!\!\!\!\!\!\!\!
+
\int_{\q_1} \int_{\q_2} 2 \rm{Re} \Bigg\{
\begin{tikzpicture}[baseline=-3.0pt]
  \draw[sk/boundary] (-1.3,\BOUNDY) -- (1 ,\BOUNDY);
  \skbound{B3}{( 0.8, -0.5)}
    \skbound{B4}{( 0.8, -1.0)}
             \skvertexp{V3}{(0 ,-0.5)}
             \skvertexp{V1}{(-0.7,-1.0)}
  \skbound{B1}{(-1.0,\BOUNDY)}
  \skbound{B2}{( -0.0,\BOUNDY)}
  \node[above=2pt] at (B1) {$\k_1$};
  \node[above=2pt] at (B2) {$\k_2$};
     \G{V1}{B1}
  \G{V3}{B2}
   \Gdouble{V1}{V3}
       \G{B4}{V1}
        \G{B3}{V3}
        \node[above=2pt] at (B3) {$\q_1$};
           \node[below=2pt] at (B4) {$\q_2$};
\end{tikzpicture} 
\Bigg\}
\times 
2 {\rm Re }
\Bigg\{
\begin{tikzpicture}[baseline=-3.0pt]
  \draw[sk/boundary] (-1,\BOUNDY) -- (0.3 ,\BOUNDY);
  \skbound{B3}{( -0.8, -0.5)}
    \skbound{B4}{( -0.8, -1.0)}
             \skvertexp{V1}{(0.0,-0.7)}
  \skbound{B1}{(0.0,\BOUNDY)}
  \node[above=2pt] at (B1) {$\k_3$};
     \G{V1}{B1}
       \G{B4}{V1}
        \G{B3}{V1}
          \node[above=2pt] at (B3) {$\q_1$};
           \node[below=2pt] at (B4) {$\q_2$};
\end{tikzpicture} 
\Bigg\} 
+ {\rm perms} , \quad
\eea
where again, ``perms'' denotes additional diagrams obtained from the permutation of the external momenta. In this form, the Schwinger--Keldysh diagrams reorganize themselves into recognizable wavefunction coefficients. For instance, the object inside the first brackets of the first line corresponds to the one-loop corrected three-point wavefunction coefficient $\psi^{(3)}_3(\k_1,\k_2,\k_3)$. The quantity inside the second brackets of the first line corresponds to the tree-level wavefunction $\psi^{(3)}_5(\k_1,\k_2,\k_3, \k_4, \k_5)$ with two of its external momenta glued together.  Finally, the second and third lines contains different combination of wavefunctions encountered in the previous example glued together by appropriately integrating internal momenta. In other words, the previous result can be schematically written as
\bea
\label{example-intro-1loop}
\Big\langle \phi (\k_1) \phi (\k_2) \phi (\k_3)  \Big\rangle_c 
&\supset& 
\,\,
2\,{\rm Re}\!\left\{ \psi^{(3)}_3 (\k_1 , \k_2 , \k_3) \right\}  + \int_{\q} 2\,{\rm Re}\!\left\{ \psi^{(3)}_5 (\k_1 , \k_2 , \k_3, \q , \q) \right\} 
  \nn\\
&&
\!\!\!\!\!\!\!\!\!\!\!\!\!\!\!\!\!\!\!\!\!\!\!\!\!\!\!\!\!\!\!\!\!\!\!\!\!\!\!\!\!\!\!\!\!
+ \int_{\q_1} \int_{\q_2} \int_{\q_3}
2\,{\rm Re}\!\left\{ \psi^{(1)}_3 (\k_1 , \q_1 , \q_2 ) \right\}
\times
2\,{\rm Re}\!\left\{ \psi^{(1)}_3 (\k_2 , \q_2 , \q_3) \right\}
\times
2\,{\rm Re}\!\left\{ \psi^{(1)}_3 (\k_3 , \q_3 , \q_1) \right\}
 \nn\\
&&
\!\!\!\!\!\!\!\!\!\!\!\!\!\!\!\!\!\!\!\!\!\!\!\!\!\!\!\!\!\!\!\!\!\!\!\!\!\!\!\!\!\!\!\!\!
+ \int_{\q_1} \int_{\q_2}
2\,{\rm Re}\!\left\{ \psi^{(2)}_4 (\k_1 , \k_2 , \q_1 , \q_2) \right\}
\times
2\,{\rm Re}\!\left\{ \psi^{(1)}_3 (\k_3 , \q_1 , \q_2) \right\}
+ {\rm perms.}
\eea
A noteworthy feature of this result is that loops at the level of Schwinger--Keldysh diagrams decompose into a combination of loop diagrams at the level of wavefunction coefficients and tree-level wavefunction diagrams glued together. It is well known that loop integrals computed directly in the Wavefunction of the Universe approach are infrared finite. This follows from the fact that bulk-to-bulk propagators defining wavefunction coefficients vanish at the boundary, which is precisely where the infrared limit of the integrals is probed. Consequently, one concludes that infrared divergences in correlation functions, if present, arise exclusively from subdiagrams that are glued together to form loops, as in the present example. The interplay of divergences emerging from loops at the level of wavefunction coefficients versus correlators, and the role of counterterms, have been discussed in~\cite{Creminelli:2024cge, Benincasa:2024ptf, Huenupi:2024ztu}.

I will not burden the reader with additional examples. To arrive at a general statement relating the two perturbative expansions, valid to all orders, I will first review the derivation of the diagrammatic rules used to compute wavefunction coefficients from bulk theories. This review,  presented in Section~\ref{sec:WF-path-integrals}, differs from previous derivations in that it does not rely on a saddle-point approximation. Instead, I work directly with the full path-integral formulation of the wavefunction, to all orders, using standard tools such as generating functionals with sources. To keep the discussion simple, I focus on a bulk theory consisting of a self-interacting scalar field $\phi$, described by an action of the form
$S = \int d^{3}x \int dt \, \mathcal{L}$,
with a Lagrangian $\mathcal{L}$ given by
\be \label{intro:L}
\mathcal L = \mathcal L^{(0)} (\phi , t) + \mathcal L^{\rm int} (\phi , t) \,, \qquad
\mathcal L^{(0)} (\phi , t) =
\frac{1}{2} \dot{\phi}^2
- \frac{1}{2} c_s^2(t) (\nabla \phi)^2
- \frac{1}{2} m^2(t) \phi^2 \, .
\ee
Here, $\mathcal L^{\rm int}(\phi,t)$ denotes an interaction Lagrangian containing higher-order terms in $\phi$, and possibly spatial derivatives acting on the field. This canonically normalized Lagrangian is sufficiently general and already includes single-field inflation as a particular case, provided that $c_s^2(t)$ and $m^2(t)$ are chosen appropriately. The specific form of $\mathcal L^{\rm int}(\phi,t)$ will not play a central role in the discussion. Nevertheless, in order to keep the presentation and intermediate computations as simple as possible, I will specialize to the case of a cubic interaction. In Section~\ref{sec:correlators-from-wfc}, I review how correlation functions are obtained from wavefunction coefficients. I then introduce, in Section~\ref{sec:systematics}, a set of tools that allow for a systematic analysis of these correlators. These tools are subsequently used in Section~\ref{sec:General-map} to derive the general map relating the Wavefunction of the Universe and Schwinger--Keldysh formalisms. The derivation proceeds in the opposite direction to the examples discussed in this introduction: starting from a general collection of diagrams written in the Wavefunction of the Universe formalism, I show how they can be reorganized into a collection of diagrams obeying the Schwinger--Keldysh rules. Finally, in Section~\ref{sec:Examples}, I illustrate how the map works in practice by analyzing several explicit examples.

Throughout this article, I will denote spacetime variables as $x = (\x , t)$. Momenta will appear only in the form of spatial momenta $\p$. In addition, to alleviate the notation, I will use:
\be
\int_{\x} = \int d^3 x , \qquad \int_{\p} = \int \frac{d^3 p}{(2 \pi)^3} ,
\ee
where the first integral corresponds to an integral over spatial volume, and the second corresponds to an integral over momentum space.


\setcounter{equation}{0}
\section{Wavefunction of the Universe path integral}
\label{sec:WF-path-integrals}

In this section, I review the derivation of the diagrammatic rules that allow for the computation of $n$-point wavefunction coefficients $\psi_n$, both in configuration space and in momentum space. To begin, recall that these coefficients are defined at the boundary time $t_f$, at which we are interested in computing correlation functions, and they parametrize the wavefunction $\Psi[\varphi,t_f]$ as
\be \label{coefficients-def}
\Psi[\varphi,t_f]
\propto
\exp\Bigg\{
\sum_{n=2}^{\infty} \frac{1}{n!}
\int_{\x_1} \cdots \int_{\x_n}
\psi_n(\x_1,\cdots,\x_n; t_f)\,
\varphi(\x_1)\cdots\varphi(\x_n)
\Bigg\} .
\ee
The wavefunction $\Psi[\varphi,t_f]$ contains all the relevant information about the state of the system at the time $t_f$. In particular, it determines the probability density functional $\rho[\varphi,t_f] = |\Psi[\varphi,t_f]|^2$, which gives the probability of observing the bulk quantum field $\hat\phi(\x)$ in a given spatial configuration $\varphi(\x)$ at the time $t_f$.

\subsection{Path integral form for the wavefunction}

To determine the form of the coefficients $\psi_n(\x_1,\cdots,\x_n;t_f)$ in configuration space, we need to understand how the system evolves from the infinite past up to the time $t_f$. This, in turn, requires specifying the initial state of the system. Since we are interested in applications of the Wavefunction of the Universe to the computation of primordial correlation functions, I will assume that the initial state in the infinite past corresponds to the vacuum. With this assumption, the wavefunction can be written as the projection
$\Psi[\varphi,t_f] = \langle \varphi | \hat U(t_f,-\infty) | \Omega \rangle$,
where $|\Omega\rangle$ denotes the vacuum state, $\hat U(t_f,-\infty)$ is the unitary evolution operator evolving the system from the infinite past up to the final time $t_f$, and $|\varphi\rangle$ is a basis state satisfying
$\hat\phi(\x)|\varphi\rangle = \varphi(\x)|\varphi\rangle$
in the Schr\"odinger picture.

By expressing $\hat U(t_f,-\infty)$ as a succession of infinitesimal unitary evolution operators, it is straightforward to obtain
\be \label{Psi-path-integral}
\Psi[\varphi,t_f]
=
\mathcal N
\int_{\phi(t_f)=\varphi}
\!\!\!\!\!\!\!\!\!\!\!\!
\mathcal D\phi \;
\exp\!\left[
i \int_{-\infty}^{t_f} dt \int_{\x}
\mathcal L_\epsilon(\phi,t)
\right] .
\ee
Here, the symbol $\mathcal D\phi$ denotes a functional integration over all possible field configurations $\phi(t,\x)$ defined from the infinite past up to the boundary time $t_f$. At the boundary, the field is constrained to match the spatial configuration $\varphi(\x)$. The Lagrangian $\mathcal L_\epsilon(\phi,t)$ appearing in the exponent is the same Lagrangian introduced in Eq.~(\ref{intro:L}), supplemented with an $\epsilon$-prescription that selects the vacuum state in the infinite past. More explicitly, the quadratic Lagrangian appearing in Eq.~(\ref{Psi-path-integral}), incorporating the $\epsilon$-prescription, is given by
\be
\mathcal L^{(0)}_\epsilon(\phi,t)
=
\frac{1}{2}\dot\phi^2
-
\frac{1}{2}c_s^2(t)(\nabla\phi)^2
-
\frac{1}{2}(1-i\epsilon)m^2(t)\phi^2 ,
\ee
where $\epsilon$ is a positive infinitesimal parameter.

A result that will be useful later, and that can be proven directly from Eq.~(\ref{Psi-path-integral}), is
\be \label{Psi-path-integral-der}
\frac{\delta}{\delta \varphi(\x_1)} \cdots \frac{\delta}{\delta \varphi(\x_n)}
\Psi[\varphi,t_f]
=
\mathcal N
\int_{\phi(t_f)=\varphi}
\!\!\!\!\!\!\!\!\!\!\!\!
\mathcal D\phi \;
e^{ i \int_{-\infty}^{t_f} dt \int_{\x} \mathcal L_\epsilon(\phi,t) }
\,
i\Pi_\phi(\x_1,t_f)\cdots i\Pi_\phi(\x_n,t_f) ,
\ee
where $\Pi_\phi(\x,t)$ denotes the canonical momentum conjugate to $\phi$, as inferred from Eq.~(\ref{intro:L}). In the particular case in which the interaction Lagrangian does not contain time derivatives of $\phi$, one simply has $\Pi_\phi(\x,t)=\dot\phi(\x,t)$, which I will assume throughout for simplicity.

\subsection{Generating functional}

The challenge now is to derive a perturbative scheme to compute the right-hand side of Eq.~(\ref{Psi-path-integral-der}). To this end, let us introduce a generating functional $Z[\varphi,J,t_f]$, which depends both on the final field configuration $\varphi(\x)$ at time $t_f$ and on a bulk external source $J(x)$, with $x=(\x,t)$. Omitting the explicit dependence on $t_f$, this functional is defined as
\be \label{Psi-path-integral-J}
Z[\varphi,J]
\equiv
\mathcal N
\int_{\phi(t_f)=\varphi}
\!\!\!\!\!\!\!\!\!\!\!\!
\mathcal D\phi \;
\exp\!\left[
i \int_{-\infty}^{t_f} dt \int_{\x}
\Big( \mathcal L_\epsilon(\phi,t) + \phi(x) J(x) \Big)
\right] .
\ee
We can now split the theory into its free and interacting parts, as in Eq.~(\ref{intro:L}), to rewrite the generating functional as
\be \label{Psi-path-integral-J-split}
Z[\varphi,J]
=
\exp\!\left\{
i \int_{-\infty}^{t_f} dt \int_{\x}
\mathcal L^{\rm int}\!\left( - i \frac{\delta}{\delta J(x)} , t \right)
\right\}
Z_0[\varphi,J] ,
\ee
where the free generating functional $Z_0[\varphi,J]$ is given by
\be \label{Psi-path-integral-J-0}
Z_0[\varphi,J]
=
\mathcal N
\int_{\phi(t_f)=\varphi}
\!\!\!\!\!\!\!\!\!\!\!\!
\mathcal D\phi \;
\exp\!\left[
i \int_{-\infty}^{t_f} dt \int_{\x}
\Big( \mathcal L^{(0)}_\epsilon(\phi,t) + \phi(x) J(x) \Big)
\right] .
\ee
Since $\mathcal L^{(0)}_\epsilon(\phi,t)$ is quadratic in the field, the functional integral in Eq.~(\ref{Psi-path-integral-J-0}) can be evaluated explicitly. To do so, it is convenient to perform the following field redefinition:
\be \label{xi-phi}
\phi(x)
\;\to\;
\xi(x)
=
\phi(x)
-
i \int_{-\infty}^{t_f} dt' \int_{\x'}
G(x,x')\, J(x') ,
\ee
where $G(x,x')=G(x',x)$ is a function symmetric under the interchange of spacetime arguments, which will shortly be identified as a bulk-to-bulk propagator. We impose two conditions on this function. First, it must be a Green's function for the free equation of motion:
\be \label{G-cond-1}
\Big[
\frac{d^2}{dt^2}
-
c_s^2(t)\nabla^2
+
m^2(t)(1-i\epsilon)
\Big]
G(x,x')
=
- i\,\delta^{(3)}(\x-\x')\,\delta(t-t') .
\ee
Second, it must vanish whenever either of its time arguments is evaluated at the infinite past or at the boundary time $t_f$:
\be \label{G-cond-2}
\lim_{t,t'\to -\infty} G(x,x') = 0 ,
\qquad
\lim_{t,t'\to t_f} G(x,x') = 0 .
\ee
As shown in Appendix~A of Ref.~\cite{Goodhew:2020hob}, the Green's function satisfying these two properties is given by
\be \label{G-bulk-bulk-wf}
G(x , x')
=
\int_{\k}
\Big[
\phi_k(t )\phi_k^*(t')\,\theta(t - t')
+
\phi_k(t')\phi_k^*(t)\,\theta(t' - t)
-
\frac{\phi_k(t_f)}{\phi_k^*(t_f)}
\,\phi_k^*(t')\phi_k^*(t)
\Big]
e^{-i\k\cdot(\x -\x')} , \quad
\ee
where $\phi_k(t)$ is the mode function satisfying the equation of motion
\be
\Big[
\frac{d^2}{dt^2}
+
c_s^2(t) k^2
+
m^2(t)(1-i\epsilon)
\Big]
\phi_k(t)
=
0 ,
\ee
and normalized according to the Wronskian condition
\be \label{wronskian-cond}
\phi_k(t)\dot\phi_k^*(t) - \phi_k^*(t)\dot\phi_k(t) = i .
\ee
As usual, the Green's function $G(x,x')$ can be visualized in terms of a propagator connecting the spacetime points $x$ and $x'$ by a line. In the present discussion, we will denote this bulk-to-bulk propagator by a double line, as follows:
\be
x\,\, 
\begin{tikzpicture}[baseline=-3.0pt]
\coordinate (V1) at (-1, 0);
\coordinate (V2) at (1, 0);
     \Gdouble{V1}{V2}
\end{tikzpicture}  
\,\, x'
\quad
 \longrightarrow 
 \quad
  G (x , x') .
\ee

By applying the field reparameterization~(\ref{xi-phi}) to the functional $Z_0[\varphi,J]$, with $G(x,x')$ satisfying conditions~(\ref{G-cond-1}) and~(\ref{G-cond-2}), one readily finds
\bea \label{Z0-J-Gaussian}
Z_0[\varphi,J]
&=&
\Psi^{(0)}[\varphi,t_f]
\times
\exp\!\left[
\int_{-\infty}^{t_f} dt \int_{\x}
\int_{-\infty}^{t_f} dt'  \int_{\x'}
\Big( \varphi(\x)\,\frac{d}{dt}\delta(t -t_f) \Big)
G(x ,x')\, J(x')
\right]
\nn\\
&&
\times
\exp\!\left[
-\frac{1}{2}
\int_{-\infty}^{t_f} dt  \int_{\x}
\int_{-\infty}^{t_f} dt' \int_{\x'}
J(x )\, G(x ,x')\, J(x')
\right] .
\eea
The first line in Eq.~(\ref{Z0-J-Gaussian}) contains a contribution that depends on both $\varphi(\x)$ and $J(x)$. This term arises from boundary contributions generated by partial integrations of the action in Eq.~(\ref{Psi-path-integral-J-0}). We can reexpress (\ref{Z0-J-Gaussian}) diagrammatically by introducing rules that specify how to represent bulk sources $J(x)$ and the external field $\varphi(\x)$ using graphical symbols. We will adopt the following two assignments involving crossed dots:
\bea
\begin{tikzpicture}[baseline=-3.0pt]
\coordinate (s1) at (0,0);
  \Gdouble{s1}{1.5, 0.0} 
    \sksourcexblack{s1}{(0,0)}
\end{tikzpicture} 
\quad 
 & \longrightarrow &
\quad   \int_{-\infty}^{t_f}  \!\! \!\!  dt \!  \int_{\x} \!\! i \, J (x) \Big[ \cdots \Big] , \\
 \label{field-insertions}
\begin{tikzpicture}[baseline=-3.0pt]
\coordinate (s1) at (0,0);
  \Gdouble{s1}{1.5, 0.0}
  \sksourcex{s1}{(0,0)}
\end{tikzpicture} 
\quad 
& \longrightarrow &
\quad   \int_{-\infty}^{t_f}  \!\! \!\!  dt \!  \int_{\x} \!\!  \, \varphi (\x) \frac{d}{dt} \Big[ \cdots \Big] . 
\eea
In the previous expressions, the notation $\big[ \cdots \big]$ stands for functions of the integration variables $t$ and $\x$, arising from propagators meeting sources and fields. With the help of these rules, it is possible to rewrite Eq.~(\ref{Z0-J-Gaussian}) in the following diagrammatic form:
\be
Z_0[\varphi , J ]
=
\Psi^{(0)}[\varphi , t_f]\,
\exp\Bigg\{
\begin{tikzpicture}[baseline=-3.0pt]
  \sksourcex{s1}{(0.0,0.0)}
  \sksourcexblack{s2}{(1.5,0.0)}
  \Gdouble{s1}{s2}
  \sksourcex{s1}{(0.0,0.0)}
  \sksourcexblack{s2}{(1.5,0.0)}
\end{tikzpicture}
+
\begin{tikzpicture}[baseline=-3.0pt]
  \sksourcexblack{s1}{(0.0,0.0)}
  \sksourcexblack{s2}{(1.5,0.0)}
  \Gdouble{s1}{s2}
  \sksourcexblack{s1}{(0.0,0.0)}
  \sksourcexblack{s2}{(1.5,0.0)}
\end{tikzpicture}
\Bigg\} .
\ee
Here, the first diagram (where a field $\varphi$ is connected to a source $J$) represents the argument of the first exponential in Eq.~(\ref{Z0-J-Gaussian}). The argument of the second exponential is instead represented by the diagram in which two sources are connected. When translating these diagrams into analytic expressions, one must multiply the result by a factor $1/S_D$, where $S_D$ denotes the symmetry factor of the diagram. For the first diagram the symmetry factor is equal to $1$, while for the second diagram it is $2!$. Although I will not show it explicitly here, the free wavefunction $\Psi^{(0)}[\varphi,t_f]$ can also be expressed diagrammatically. Including this contribution, the final diagrammatic representation of $Z_0[\varphi,J]$ reads
\be \label{Z0-diagrams}
Z_0[\varphi , J ]
\propto
\exp\Bigg\{
\begin{tikzpicture}[baseline=-3.0pt]
  \sksourcex{s1}{(0.0,0.0)}
  \sksourcex{s2}{(1.5,0.0)}
  \Gdouble{s1}{s2}
  \sksourcex{s1}{(0.0,0.0)}
  \sksourcex{s2}{(1.5,0.0)}
\end{tikzpicture}
+
\begin{tikzpicture}[baseline=-3.0pt]
  \sksourcex{s1}{(0.0,0.0)}
  \sksourcexblack{s2}{(1.5,0.0)}
  \Gdouble{s1}{s2}
  \sksourcex{s1}{(0.0,0.0)}
  \sksourcexblack{s2}{(1.5,0.0)}
\end{tikzpicture}
+
\begin{tikzpicture}[baseline=-3.0pt]
  \sksourcexblack{s1}{(0.0,0.0)}
  \sksourcexblack{s2}{(1.5,0.0)}
  \Gdouble{s1}{s2}
  \sksourcexblack{s1}{(0.0,0.0)}
  \sksourcexblack{s2}{(1.5,0.0)}
\end{tikzpicture}
\Bigg\} .
\ee
Again, one must remember to multiply the corresponding analytic expression associated with the first diagram by a factor of $1/2!$.

Now, inserting Eq.~(\ref{Z0-diagrams}) back into Eq.~(\ref{Psi-path-integral-J-split}), we can expand the full generating functional $Z[\varphi,J]$ perturbatively in powers of $J$ using standard diagrammatic rules. To illustrate how this expansion works, and to define the rules that allow one to write down diagrams, let us consider the particular case in which the interaction Lagrangian $\mathcal L^{\rm int}(\phi,t)$ consists of a simple cubic interaction,
\be
\mathcal L_{\rm int}(\phi,t)
=
- \frac{1}{3!}\,\alpha(t)\,\phi^3 .
\label{L-interaction-alpha}
\ee
More general interactions can be straightforwardly treated by extending this example. This cubic interaction implies the existence of three-legged vertices at which propagators meet. The diagrammatic rule specifying how such a vertex translates into an analytic expression is
\be
\begin{tikzpicture}[baseline=-3.0pt]
\coordinate (v) at (0,0);
  \Gdouble{v}{-1.0,0.0}
  \Gdouble{v}{0.4,0.7}
  \Gdouble{v}{0.4,-0.7}
  \skvertexp{v}{(0,0)}
\end{tikzpicture} t
\quad
\longrightarrow
\quad
- i \int_{-\infty}^{t_f} \! dt \int_{\x} \alpha(t) \big[ \cdots \big] ,
\ee
where $\big[ \cdots \big]$ denotes functions of the integration variables $t$ and $\x$, arising from propagators joining at the vertex. I will refer to these vertices as bulk vertices.

With these rules in place, one can now expand $Z[\varphi,J]$ in terms of diagrams. The result is conveniently expressed as
\be
Z[\varphi,J] = \exp W[\varphi,J] ,
\ee
where $W[\varphi,J]$ is the generating functional of connected diagrams. That is, $W[\varphi,J]$ consists of the sum of all connected diagrams constructed using the three-legged bulk vertex introduced above, following the diagrammatic rules described previously. As usual, when writing down diagrams, the corresponding analytic expression must be multiplied by a factor $1/S_D$, where $S_D$ denotes the symmetry factor of the diagram. At the lowest order in sources, vertices, and field insertions, the generating functional $W[\varphi,J]$ takes the form
 \vspace{-0.5cm}
 \bea \label{W-diagrams}
 W[\varphi , J ]  &  = & 
 \begin{tikzpicture}[baseline=-3.0pt]
  \sksourcexblack{s1}{(0.0,  0.0)} 
  \sksourcexblack{s2}{(1.5, 0.0)}
  \Gdouble{s1}{s2} 
    \sksourcexblack{s1}{(0.0,  0.0)} 
  \sksourcexblack{s2}{(1.5, 0.0)}
\end{tikzpicture} 
 + 
 \begin{tikzpicture}[baseline=-3.0pt]
   \sksourcex{s1}{(0.0,  0.0)} 
  \sksourcexblack{s2}{(1.5, 0.0)}
  \Gdouble{s1}{s2} 
     \sksourcex{s1}{(0.0,  0.0)} 
  \sksourcexblack{s2}{(1.5, 0.0)}
\end{tikzpicture} 
 + 
 \begin{tikzpicture}[baseline=-3.0pt]
   \sksourcex{s1}{(0.0,  0.0)} 
  \sksourcex{s2}{(1.5, 0.0)}
  \Gdouble{s1}{s2} 
     \sksourcex{s1}{(0.0,  0.0)} 
 \sksourcex{s2}{(1.5, 0.0)}
\end{tikzpicture} 
+
\begin{tikzpicture}[baseline=-3.0pt]
  \skvertexp{v}{(0,0)}
  \sksourcexblack{J}{(-1.0,0)}
  \Gdouble{J}{v}
 \draw[sk/propdouble] (v.25) to[out=45,in=-45,looseness=20] (v.-25);
  \skvertexp{v}{(0,0)}
  \sksourcexblack{J}{(-1.0,0)}
\end{tikzpicture} 
+
\begin{tikzpicture}[baseline=-3.0pt]
  \skvertexp{v}{(0,0)}
  \sksourcex{J}{(-1.0,0)}
  \Gdouble{J}{v}
 \draw[sk/propdouble] (v.25) to[out=45,in=-45,looseness=20] (v.-25);
  \skvertexp{v}{(0,0)}
   \sksourcex{J}{(-1.0,0)}
\end{tikzpicture} 
\nn  \\[-10pt]
&& + 
\begin{tikzpicture}[baseline=-3.0pt]
\coordinate (v) at (0,0);
  \Gdouble{v}{-1.0, 0.0}
  \Gdouble{v}{0.4, 0.7}
  \Gdouble{v}{0.4,-0.7}
   \skvertexp{v}{(0,0)}
   \sksourcexblack{s1}{(-1.0, 0.0)}
    \sksourcexblack{s2}{(0.4, 0.7)}
     \sksourcexblack{s3}{(0.4,-0.7)}
\end{tikzpicture}
+
\begin{tikzpicture}[baseline=-3.0pt]
\coordinate (v) at (0,0);
  \Gdouble{v}{-1.0, 0.0}
  \Gdouble{v}{0.4, 0.7}
  \Gdouble{v}{0.4,-0.7}
   \skvertexp{v}{(0,0)}
   \sksourcex{s1}{(-1.0, 0.0)}
    \sksourcexblack{s2}{(0.4, 0.7)}
     \sksourcexblack{s3}{(0.4,-0.7)}
\end{tikzpicture}
 +
\begin{tikzpicture}[baseline=-3.0pt]
\coordinate (v) at (0,0);
  \Gdouble{v}{-1.0, 0.0}
  \Gdouble{v}{0.4, 0.7}
  \Gdouble{v}{0.4,-0.7}
   \skvertexp{v}{(0,0)}
   \sksourcex{s1}{(-1.0, 0.0)}
    \sksourcex{s2}{(0.4, 0.7)}
     \sksourcexblack{s3}{(0.4,-0.7)}
\end{tikzpicture}
+
\begin{tikzpicture}[baseline=-3.0pt]
\coordinate (v) at (0,0);
  \Gdouble{v}{-1.0, 0.0}
  \Gdouble{v}{0.4, 0.7}
  \Gdouble{v}{0.4,-0.7}
   \skvertexp{v}{(0,0)}
   \sksourcex{s1}{(-1.0, 0.0)}
    \sksourcex{s2}{(0.4, 0.7)}
     \sksourcex{s3}{(0.4,-0.7)} 
\end{tikzpicture}
+
\begin{tikzpicture}[baseline=-3.0pt]
\coordinate (v1) at (-0.5,0);
\coordinate (v2) at (0.5,0);
   \Gdouble{v1}{v2}
  \Gdouble{v1}{-0.9, 0.7}
  \Gdouble{v1}{-0.9, -0.7}
   \sksourcexblack{s1}{(-0.9, 0.7)}
    \sksourcexblack{s2}{(-0.9, -0.7)}
      \skvertexp{v2}{(0.5,0)}
   \skvertexp{v1}{(-0.5,0)}
    \draw[sk/propdouble] (v2.25) to[out=45,in=-45,looseness=20] (v2.-25);
\end{tikzpicture}
\nn \\
&& + 
\begin{tikzpicture}[baseline=-3.0pt]
\coordinate (v1) at (-0.5,0);
\coordinate (v2) at (0.5,0);
   \Gdouble{v1}{v2}
  \Gdouble{v1}{-0.9, 0.7}
  \Gdouble{v1}{-0.9, -0.7}
   \sksourcexblack{s1}{(-0.9, 0.7)}
    \sksourcex{s2}{(-0.9, -0.7)}
      \skvertexp{v2}{(0.5,0)}
   \skvertexp{v1}{(-0.5,0)}
    \draw[sk/propdouble] (v2.25) to[out=45,in=-45,looseness=20] (v2.-25);
\end{tikzpicture}
+
\begin{tikzpicture}[baseline=-3.0pt]
\coordinate (v1) at (-0.5,0);
\coordinate (v2) at (0.5,0);
   \Gdouble{v1}{v2}
  \Gdouble{v1}{-0.9, 0.7}
  \Gdouble{v1}{-0.9, -0.7}
   \sksourcex{s1}{(-0.9, 0.7)}
    \sksourcex{s2}{(-0.9, -0.7)}
      \skvertexp{v2}{(0.5,0)}
   \skvertexp{v1}{(-0.5,0)}
    \draw[sk/propdouble] (v2.25) to[out=45,in=-45,looseness=20] (v2.-25);
\end{tikzpicture}
+
\begin{tikzpicture}[baseline=-0.3pt]
  \skvertexp{vL}{(0,0)}     
  \skvertexp{vR}{(1.0,0)}    
  \sksourceg{JL}{(-0.8,0)}   
  \sksourceg{JR}{(1.8,0)}   
  \Gdouble{JL}{vL}
  \Gdouble{vR}{JR}
  \Gdoublearc{vL}{vR}{45}{135}    
  \Gdoublearc{vL}{vR}{-45}{-135}   
  \skvertexp{vL}{(0,0)}     
  \skvertexp{vR}{(1.0,0)}  
   \sksourcexblack{JL}{(-0.8,0)}   
  \sksourcexblack{JR}{(1.8,0)} 
\end{tikzpicture} 
+ 
\begin{tikzpicture}[baseline=-0.3pt]
  \skvertexp{vL}{(0,0)}     
  \skvertexp{vR}{(1.0,0)}    
  \sksourceg{JL}{(-0.8,0)}   
   \sksourcex{JR}{(1.8,0)}    
  \Gdouble{JL}{vL}
  \Gdouble{vR}{JR}
  \Gdoublearc{vL}{vR}{45}{135}    
  \Gdoublearc{vL}{vR}{-45}{-135}   
  \skvertexp{vL}{(0,0)}     
  \skvertexp{vR}{(1.0,0)}  
   \sksourcexblack{JL}{(-0.8,0)}   
  \sksourcex{JR}{(1.8,0)} 
\end{tikzpicture} 
\nn
\\
&& + 
\begin{tikzpicture}[baseline=-0.3pt]
  \skvertexp{vL}{(0,0)}     
  \skvertexp{vR}{(1.0,0)}    
  \sksourcex{JL}{(-0.8,0)}   
  \sksourcex{JR}{(1.8,0)}   
  \Gdouble{JL}{vL}
  \Gdouble{vR}{JR}
  \Gdoublearc{vL}{vR}{45}{135}    
  \Gdoublearc{vL}{vR}{-45}{-135}   
  \skvertexp{vL}{(0,0)}     
  \skvertexp{vR}{(1.0,0)}  
   \sksourcex{JL}{(-0.8,0)}   
  \sksourcex{JR}{(1.8,0)} 
\end{tikzpicture} 
+
\begin{tikzpicture}[baseline=-3.0pt]
\coordinate (v1) at (-0.5,0);
\coordinate (v2) at (0.5,0);
  \Gdouble{v2}{0.9, 0.7}
  \Gdouble{v2}{0.9,-0.7}
   \Gdouble{v1}{v2}
  \Gdouble{v1}{-0.9, 0.7}
  \Gdouble{v1}{-0.9, -0.7}
    \sksourcexblack{s1}{(-0.9, 0.7)}
     \sksourcexblack{s2}{(-0.9, -0.7)}
     \sksourcexblack{s3}{(0.9, 0.7)}
      \sksourcexblack{s4}{(0.9,-0.7)}
      \skvertexp{v2}{(0.5,0)}
   \skvertexp{v1}{(-0.5,0)}
\end{tikzpicture}
+
\begin{tikzpicture}[baseline=-3.0pt]
\coordinate (v1) at (-0.5,0);
\coordinate (v2) at (0.5,0);
  \Gdouble{v2}{0.9, 0.7}
  \Gdouble{v2}{0.9,-0.7}
   \Gdouble{v1}{v2}
  \Gdouble{v1}{-0.9, 0.7}
  \Gdouble{v1}{-0.9, -0.7}
    \sksourcexblack{s1}{(-0.9, 0.7)}
     \sksourcexblack{s2}{(-0.9, -0.7)}
     \sksourcex{s3}{(0.9, 0.7)}
      \sksourcexblack{s4}{(0.9,-0.7)} 
      \skvertexp{v2}{(0.5,0)}
   \skvertexp{v1}{(-0.5,0)}
\end{tikzpicture}
+
\begin{tikzpicture}[baseline=-3.0pt]
\coordinate (v1) at (-0.5,0);
\coordinate (v2) at (0.5,0);
  \Gdouble{v2}{0.9, 0.7}
  \Gdouble{v2}{0.9,-0.7}
   \Gdouble{v1}{v2}
  \Gdouble{v1}{-0.9, 0.7}
  \Gdouble{v1}{-0.9, -0.7}
    \sksourcexblack{s1}{(-0.9, 0.7)}
     \sksourcexblack{s2}{(-0.9, -0.7)}
    \sksourcex{s3}{(0.9, 0.7)}
     \sksourcex{s4}{(0.9,-0.7)} 
      \skvertexp{v2}{(0.5,0)}
   \skvertexp{v1}{(-0.5,0)}
\end{tikzpicture}
\nn \\
&& + 
\begin{tikzpicture}[baseline=-3.0pt]
\coordinate (v1) at (-0.5,0);
\coordinate (v2) at (0.5,0);
  \Gdouble{v2}{0.9, 0.7}
  \Gdouble{v2}{0.9,-0.7}
   \Gdouble{v1}{v2}
  \Gdouble{v1}{-0.9, 0.7}
  \Gdouble{v1}{-0.9, -0.7}
    \sksourcexblack{s1}{(-0.9, 0.7)}
    \sksourcex{s2}{(-0.9, -0.7)}
     \sksourcexblack{s3}{(0.9, 0.7)}
     \sksourcex{s4}{(0.9,-0.7)}
      \skvertexp{v2}{(0.5,0)}
   \skvertexp{v1}{(-0.5,0)}
\end{tikzpicture}
+
\begin{tikzpicture}[baseline=-3.0pt]
\coordinate (v1) at (-0.5,0);
\coordinate (v2) at (0.5,0);
  \Gdouble{v2}{0.9, 0.7}
  \Gdouble{v2}{0.9,-0.7}
   \Gdouble{v1}{v2}
  \Gdouble{v1}{-0.9, 0.7}
  \Gdouble{v1}{-0.9, -0.7}
    \sksourcexblack{s1}{(-0.9, 0.7)}
    \sksourcex{s2}{(-0.9, -0.7)}
    \sksourcex{s3}{(0.9, 0.7)}
     \sksourcex{s4}{(0.9,-0.7)} 
      \skvertexp{v2}{(0.5,0)}
   \skvertexp{v1}{(-0.5,0)}
\end{tikzpicture}
 +
\begin{tikzpicture}[baseline=-3.0pt]
\coordinate (v1) at (-0.5,0);
\coordinate (v2) at (0.5,0);
  \Gdouble{v2}{0.9, 0.7}
  \Gdouble{v2}{0.9,-0.7}
   \Gdouble{v1}{v2}
  \Gdouble{v1}{-0.9, 0.7}
  \Gdouble{v1}{-0.9, -0.7}
   \sksourcex{s1}{(-0.9, 0.7)}
    \sksourcex{s2}{(-0.9, -0.7)}
    \sksourcex{s3}{(0.9, 0.7)}
     \sksourcex{s4}{(0.9,-0.7)} 
      \skvertexp{v2}{(0.5,0)}
   \skvertexp{v1}{(-0.5,0)}
\end{tikzpicture}
+ \cdots ,
 \eea
where the ellipses denote higher-order contributions involving additional bulk vertices.

\subsection{Wavefunction coefficients in configuration space}
\label{sec:wf-coeff}

Recall that Eq.~(\ref{Psi-path-integral-der}) tells us how to compute functional derivatives of the wavefunction $\Psi[\varphi,t_f]$ with respect to $\varphi$. Since we now have a perturbative expression for $\Psi_J[\varphi,t_f]$, we can rewrite Eq.~(\ref{Psi-path-integral-der}) in the following form:
\be \label{Psi-path-integral-der-2}
\frac{\delta}{\delta \varphi(\x_1)} \cdots \frac{\delta}{\delta \varphi(\x_n)} \Psi[\varphi,t_f]
=
\left( \frac{d}{dt_f} \frac{\delta}{\delta J(x_1)} \right)
\cdots
\left( \frac{d}{dt_f} \frac{\delta}{\delta J(x_n)} \right)
Z[\varphi,J]\Big|_{J=0} .
\ee
According to Eq.~(\ref{coefficients-def}), the wavefunction coefficients can be expressed in terms of derivatives of $\ln\Psi[\varphi,t_f]$ as
\be \label{psi-der-Psi}
\psi_n(\x_1,\cdots,\x_n;t_f)
=
\frac{\delta}{\delta \varphi(\x_1)} \cdots \frac{\delta}{\delta \varphi(\x_n)}
\ln\Psi[\varphi,t_f]\Big|_{\varphi=0} .
\ee
Using Eq.~(\ref{Psi-path-integral-der-2}) in Eq.~(\ref{psi-der-Psi}), we obtain the following relation between wavefunction coefficients and $J$-derivatives of the generating functional $W[\varphi,J]$:
\be \label{psi-W}
\psi_n(\x_1,\cdots,\x_n;t_f)
=
\left( \frac{d}{dt_f} \frac{\delta}{\delta J(x_1)} \right)
\cdots
\left( \frac{d}{dt_f} \frac{\delta}{\delta J(x_n)} \right)
W[\varphi,J]\Big|_{J=\varphi=0} .
\ee
This relation provides the desired diagrammatic rules for computing wavefunction coefficients. Since Eq.~(\ref{psi-W}) requires evaluating $W[\varphi,J]$ at $\varphi=0$, the diagrams contributing to wavefunction coefficients are simply those appearing in Eq.~(\ref{W-diagrams}) with only sources $J$ attached to their external legs. Moreover, the derivatives with respect to $J$ in Eq.~(\ref{psi-W}) are accompanied by derivatives with respect to the boundary time $t_f$. This operation acts on the propagators and gives rise to a new diagrammatic rule, defining bulk-to-boundary propagators that connect bulk vertices to the boundary surface at which wavefunction coefficients are evaluated. The corresponding assignment is:
\be \label{external-K}
\x \,\,
\begin{tikzpicture}[baseline=-3.0pt]
  \skbound{s1}{(0.0,  0.0)} 
   \skvertexp{v1}{(1.5,  0.0)} 
  \draw[sk/propdouble] (s1) -- (v1); 
  \skbound{s1}{(0.0,  0.0)} 
   \skvertexp{v1}{(1.5,  0.0)} 
\end{tikzpicture} 
\quad 
 \longrightarrow 
\quad  K (\x , x')  \equiv  i \frac{d}{dt}  G (x  , x') \bigg|_{t = t_f} .
\ee
It is straightforward to show that, thanks to the Wronskian condition~(\ref{wronskian-cond}), the bulk-to-boundary propagator takes the form
\bea \label{K-xx}
K(\x,x')
&=&
\int_{\k}
\frac{\phi_k^*(t')}{\phi_k^*(t_f)}
\, e^{-i\k\cdot(\x-\x')} .
\eea
One may also define an additional rule in which the boundary is connected to itself by a single propagator. This object necessarily coincides with the wavefunction coefficient of the free theory:

\be \label{external-K-free}
\x \,\,
\begin{tikzpicture}[baseline=-3.0pt]
  \skbound{s1}{(0.0,  0.0)} 
   \skbound{v1}{(1.5,  0.0)} 
  \draw[sk/propdouble] (s1) -- (v1); 
   \skbound{s1}{(0.0,  0.0)} 
   \skbound{v1}{(1.5,  0.0)} 
\end{tikzpicture} 
\,\, \x'
\quad 
 \longrightarrow 
\quad  \psi_2^{\rm free} (\x , \x' ; t_f)  \equiv -  \frac{d}{dt}  \frac{d}{dt'}  G (x  , x') \bigg|_{t , t' = t_f} .
\ee

With these additional rules in place, we can now compute any desired wavefunction coefficient diagrammatically, to arbitrary order in perturbation theory. A few illustrative examples are:
\bea \label{n-point-wc-1}
 \psi_2  (\x_1 , \x_2  ; t_f) 
 &=&  
 \,\,
  \begin{tikzpicture}[baseline=-3.0pt]
  \draw[sk/boundary] (-1.3,\BOUNDY) -- (1.3,\BOUNDY);
  \skbound{B1}{(-1.0,\BOUNDY)}
  \skbound{B3}{( 1.0,\BOUNDY)}
       \Gdoublearc{B1}{B3}{-60}{-120}  
  \node[above=2pt] at (B1) {$\x_1$};
  \node[above=2pt] at (B3) {$\x_2$};
   \skbound{B1}{(-1.0,\BOUNDY)}
  \skbound{B3}{( 1.0,\BOUNDY)}
\end{tikzpicture} 
+
 \begin{tikzpicture}[baseline=-3.0pt]
  \draw[sk/boundary] (-1.3,\BOUNDY) -- (1.3,\BOUNDY);
  \skbound{B1}{(-1.0,\BOUNDY)}
  \skbound{B3}{( 1.0,\BOUNDY)}
  \skvertexp{V1}{(-0.7,-1)}
    \skvertexp{V2}{(0.7,-1)}
  \Gdouble{V1}{B1}
  \Gdouble{V2}{B3}
          \Gdoublearc{V1}{V2}{-45}{-135}  
           \Gdoublearc{V1}{V2}{45}{135}  
  \node[above=2pt] at (B1) {$\x_1$};
  \node[above=2pt] at (B3) {$\x_2$};
    \skvertexp{V1}{(-0.7,-1)}
    \skvertexp{V2}{(0.7,-1)}
    \skbound{B1}{(-1.0,\BOUNDY)}
  \skbound{B3}{( 1.0,\BOUNDY)}
\end{tikzpicture}  
+ \cdots ,
  \\
  \psi_3  (\x_1 , \x_2, \x_3  ; t_f)  &=& 
  \,\,
 \begin{tikzpicture}[baseline=-3.0pt]
  \draw[sk/boundary] (-1.3,\BOUNDY) -- (1.3,\BOUNDY);
  \skbound{B1}{(-1.0,\BOUNDY)}
  \skbound{B2}{( 0.0,\BOUNDY)}
  \skbound{B3}{( 1.0,\BOUNDY)}
  \skvertexp{V}{(0,-1)}
  \Gdouble{V}{B1}
  \Gdouble{V}{B2}
  \Gdouble{V}{B3}
  \node[above=2pt] at (B1) {$\x_1$};
  \node[above=2pt] at (B2) {$\x_2$};
  \node[above=2pt] at (B3) {$\x_3$};
    \skvertexp{V}{(0,-1)}
    \skbound{B1}{(-1.0,\BOUNDY)}
  \skbound{B2}{( 0.0,\BOUNDY)}
  \skbound{B3}{( 1.0,\BOUNDY)}
\end{tikzpicture} 
+
 \begin{tikzpicture}[baseline=-3.0pt]
  \draw[sk/boundary] (-1.3,\BOUNDY) -- (1.3,\BOUNDY);
  \skbound{B1}{(-1.0,\BOUNDY)}
  \skbound{B2}{( -0.0,\BOUNDY)}
  \skbound{B3}{( 1.0,\BOUNDY)}
  \skvertexp{V1}{(-0.7,-1)}
    \skvertexp{V2}{(0.7,-1)}
  \Gdouble{V1}{B1}
  \Gdouble{V2}{B3}
          \Gdoublearc{V1}{V2}{-45}{-135}  
           \Gdoublearc{V1}{V2}{45}{135}  
                 \skvertexp{V3}{(0 ,-0.65)}
                   \Gdouble{V3}{B2}
  \node[above=2pt] at (B1) {$\x_1$};
  \node[above=2pt] at (B2) {$\x_2$};
  \node[above=2pt] at (B3) {$\x_3$};
    \skvertexp{V1}{(-0.7,-1)}
    \skvertexp{V2}{(0.7,-1)}
\skvertexp{V3}{(0 ,-0.65)}
    \skbound{B1}{(-1.0,\BOUNDY)}
  \skbound{B2}{( -0.0,\BOUNDY)}
  \skbound{B3}{( 1.0,\BOUNDY)}
\end{tikzpicture}  
+ \cdots ,
 \\
   \psi_4 (\x_1 , \x_2, \x_3 , \x_4  ; t_f)   &=& 
   \,\,
 \begin{tikzpicture}[baseline=-3.0pt]
  \draw[sk/boundary] (-1.8,\BOUNDY) -- (1.8,\BOUNDY);
  \skbound{B1}{(-1.5,\BOUNDY)}
  \skbound{B2}{( -0.5,\BOUNDY)}
  \skbound{B3}{( 0.5,\BOUNDY)}
    \skbound{B4}{( 1.5,\BOUNDY)}
  \skvertexp{V1}{(-1,-1)}
    \skvertexp{V2}{(1,-1)}
  \Gdouble{V1}{B1}
  \Gdouble{V1}{B2}
  \Gdouble{V2}{B3}
    \Gdouble{V2}{B4}
        \Gdouble{V1}{V2}
  \node[above=2pt] at (B1) {$\x_1$};
  \node[above=2pt] at (B2) {$\x_2$};
  \node[above=2pt] at (B3) {$\x_3$};
  \node[above=2pt] at (B4) {$\x_4$};
    \skvertexp{V1}{(-1,-1)}
    \skvertexp{V2}{(1,-1)}
    \skbound{B1}{(-1.5,\BOUNDY)}
  \skbound{B2}{( -0.5,\BOUNDY)}
  \skbound{B3}{( 0.5,\BOUNDY)}
    \skbound{B4}{( 1.5,\BOUNDY)}
\end{tikzpicture} 
+  {\rm perms} + \cdots . 
 \label{n-point-wc-3}
\eea
In passing, it is worth noting that these expressions could also have been obtained by directly differentiating $W[\varphi,J]$ with respect to $\varphi$ rather than $J$. That is:
\be \label{psi-W-2}
\psi_n(\x_1,\cdots,\x_n;t_f)
=
\frac{\delta}{\delta \varphi(\x_1)} \cdots \frac{\delta}{\delta \varphi(\x_n)}
W[\varphi,J]\Big|_{J=\varphi=0} .
\ee
Then, thanks to the diagrammatic rule~(\ref{field-insertions}), it is straightforward to verify that the external legs obtained from Eq.~(\ref{psi-W-2}) coincide precisely with those defined in Eq.~(\ref{external-K}).

\subsection{Wavefunction coefficients in momentum space}

Let me now present the diagrammatic rules for computing wavefunction coefficients in momentum space. To begin with, three-legged bulk vertices are assigned according to
\be
\label{bulk-vertex-black}
\begin{tikzpicture}[baseline=-3.0pt]
\coordinate (v) at (0,0);
  \Gdouble{v}{-1.0, 0.0}
  \Gdouble{v}{0.4, 0.7}
  \Gdouble{v}{0.4,-0.7}
   \skvertexp{v}{(0,0)}
\end{tikzpicture} \; t
\quad
\longrightarrow
\quad
- i (2\pi)^3 \delta^{(3)}(\k_1+\k_2+\k_3)
\int_{-\infty}^{t_f} \! dt \, \alpha(t) \, \big[ \cdots \big] ,
\ee
where $\k_1$, $\k_2$, and $\k_3$ denote the momenta flowing into the vertex. These vertices can be joined by bulk-to-bulk propagators labeled by the momentum flowing through them:
\be
\label{G-bb}
t \;\;
\begin{tikzpicture}[baseline=-3.0pt]
\coordinate (V1) at (-1, 0);
\coordinate (V2) at (1, 0);
  \Gdouble{V1}{V2}
  \skvertexp{v1}{(-1,0.0)}
  \skvertexp{v2}{(1,0.0)}
\end{tikzpicture}
\;\; t'
\quad
\longrightarrow
\quad
G(k,t ,t') ,
\ee
where $G(k,t ,t')$ is the Fourier representation of the Green's function introduced in Eq.~(\ref{G-bulk-bulk-wf}), explicitly given by
\be \label{b-t-b-k}
G(k,t ,t')
=
\phi_k(t')\phi_k^*(t )\,\theta(t' -t)
+
\phi_k(t)\phi_k^*(t')\,\theta(t -t')
-
\frac{\phi_k(t_f)}{\phi_k^*(t_f)}
\,\phi_k^*(t )\phi_k^*(t') .
\ee
Next, bulk vertices can be connected to the boundary by bulk-to-boundary propagators, with the assignment
\be \label{external-K-fourier}
\k \;\;
\begin{tikzpicture}[baseline=-3.0pt]
  \skbound{s1}{(0.0,0.0)}
  \skvertexp{v1}{(2.0,0.0)}
  \draw[sk/propdouble] (s1) -- (v1);
  \skbound{s1}{(0.0,0.0)}
  \skvertexp{v1}{(2.0,0.0)}
\end{tikzpicture}
\;\; t
\quad
\longrightarrow
\quad
K(k,t) ,
\ee
where $K(k,t)$ is the Fourier transform of Eq.~(\ref{K-xx}), given by
\be
K(k,t) = \frac{\phi_k^*(t)}{\phi_k^*(t_f)} .
\ee
After assembling a given diagram, one must integrate over all internal momenta $\q$ (that is, momenta flowing between pairs of vertices) using the measure $\int_{\q}$. In addition, each diagram must be multiplied by its corresponding symmetry factor. The resulting expression yields the wavefunction coefficient $\psi_n(\k_1,\ldots,\k_n;t_f)$, which, due to momentum conservation at each vertex, is proportional to an overall Dirac delta function. For this reason, it is convenient to introduce reduced amplitudes $\psi_n'(\k_1,\ldots,\k_n;t_f)$ defined by
\be\label{reduced-amp}
\psi_n(\k_1,\ldots,\k_n;t_f)
=
(2\pi)^3 \delta^{(3)}(\k_1+\cdots+\k_n)\,
\psi_n'(\k_1,\ldots,\k_n;t_f) .
\ee

\subsection{Free theory wavefunction}
\label{sec:free-wf}

Before examining the computation of correlation functions, it is useful to have an explicit expression for the free two-point wavefunction coefficient $\psi^{\rm free}_2(\x,\x')$ introduced in Eq.~(\ref{external-K-free}). According to (\ref{reduced-amp}), one can write $\psi^{\rm free}_2(\x,\x')$ in terms of  $\psi'_2(k,t_f) \equiv \psi'_2(\k , - \k ; t_f)$ as:
\bea
\psi^{\rm free}_2(\x,\x')
=
\int_{\k}
\psi'_2(k,t_f)\,
e^{-i\k\cdot(\x-\x')} .
\eea
A direct computation of Eq.~(\ref{external-K-free}) then shows that $\psi'_2(k,t_f)$ is given by
\be
\psi'_2(k,t_f)
=
i\,\frac{\dot\phi_k^*(t_f)}{\phi_k^*(t_f)} .
\ee
However, as we shall see shortly, the quantity relevant for the computation of correlation functions is the real part of this coefficient. Taking the real part of the expression above and using the Wronskian condition~(\ref{wronskian-cond}), one finds
\be
2\,{\rm Re}\!\left[\psi'_2(k,t_f)\right]
=
-\,\frac{1}{|\phi_k(t_f)|^2} .
\ee
I will return to this result momentarily.


\setcounter{equation}{0}
\section{Correlators from wavefunction coefficients}
\label{sec:correlators-from-wfc}

In this section, I review how to obtain equal-time $n$-point correlation functions from the wavefunction $\Psi[\varphi,t_f]$. The wavefunction defines the probability distribution functional $\rho[\varphi,t_f] \equiv \big| \Psi[\varphi,t_f] \big|^2$,  which, in terms of wavefunction coefficients, can be written as
\be
\rho[\varphi,t_f]
\propto
\exp\Bigg\{
\sum_{n=2}^{\infty} \frac{1}{n!}
\int_{\x_1} \cdots \int_{\x_n}
\Big[ 2\,{\rm Re}\,\psi_n(\x_1,\ldots,\x_n;t_f) \Big]
\varphi(\x_1)\cdots\varphi(\x_n)
\Bigg\} .
\ee
Equal-time correlation functions are then obtained by performing the functional integral
\be \label{rho-cor}
\Big\langle \varphi(\x_1)\cdots\varphi(\x_n) \Big\rangle
=
\int \mathcal D\varphi \;
\rho[\varphi,t_f]\,
\varphi(\x_1)\cdots\varphi(\x_n) .
\ee
Note that, in this case, the functional integral is performed over all spatial configurations $\varphi(\x)$ at the boundary time $t_f$. This is to be contrasted with the path integral in Eq.~(\ref{Psi-path-integral}), where integration over the full bulk spacetime plays a central role. To alleviate the notation, in what follows I will omit the explicit dependence of wavefunction coefficients on the boundary time $t_f$.

\subsection{Generating functional}

To evaluate Eq.~(\ref{rho-cor}), it is convenient to introduce a new generating functional:
\be \label{Z-gen-rho}
Z[J]
=
\int \mathcal D\varphi \;
\rho[\varphi ]\,
e^{\int_{\x} J(\x)\,\varphi(\x)} .
\ee
Note that, in this case, the source $J(\x)$ depends only on spatial coordinates. In terms of this generating functional, equal-time connected $n$-point correlation functions can be written as
\be \label{connected-corr-W}
\big\langle \varphi(\x_1)\cdots\varphi(\x_n) \big\rangle_c
=
\frac{\delta}{\delta J(\x_1)} \cdots \frac{\delta}{\delta J(\x_n)}
\, W[J]\Big|_{J=0} ,
\ee
where $W[J]=\ln Z[J]$ is the generating functional of connected diagrams.

To obtain an explicit diagrammatic representation, let us decompose the two-point wavefunction coefficient as
\be \label{free-split-psi}
\psi_2(\x,\x')
=
\psi^{\rm free}_2(\x,\x')
+
\psi^{\rm int}_2(\x,\x') ,
\ee
where $\psi^{\rm free}_2(\x,\x')$ is the free-theory two-point coefficient previously introduced in Eq.~(\ref{external-K-free}), while $\psi^{\rm int}_2(\x,\x')$ is constructed from bulk vertices, in the same manner as the second diagram in Eq.~(\ref{n-point-wc-1}). This decomposition allows us to define the zeroth-order generating functional
\be \label{Z-0-wf}
Z_0[J]
\propto
\int \mathcal D\varphi \;
\exp\Bigg\{
\frac{1}{2}
\int_{\x}\int_{\x'}
\Big[ 2\,{\rm Re}\,\psi^{\rm free}_2(\x,\x') \Big]
\varphi(\x)\varphi(\x')
+
\int_{\x} J(\x)\varphi(\x)
\Bigg\} .
\ee
To evaluate this Gaussian integral, it is useful to perform the field reparameterization
\be \label{field-para-psi}
\varphi(\x)
\;\to\;
\xi(\x)
=
\varphi(\x)
+
\int_{\x'} \Delta(\x,\x')\, J(\x') ,
\ee
where $\Delta(\x,\x')$ is a symmetric Green'ss function satisfying
\be \label{cond-Delta-psi}
\int_{\x''}
\Big[ 2\,{\rm Re}\,\psi^{\rm free}_2(\x,\x'') \Big]
\Delta(\x'',\x')
=
-\,\delta(\x-\x') .
\ee
Substituting Eq.~(\ref{field-para-psi}) into Eq.~(\ref{Z-0-wf}) and using the condition~(\ref{cond-Delta-psi}), one finds
\bea
Z_0[J]
&=&
Z_0[0]\,
\exp\Bigg\{
\frac{1}{2}
\int_{\x}\int_{\x'}
J(\x)\,\Delta(\x,\x')\,J(\x')
\Bigg\} .
\eea
To determine the explicit form of $\Delta(\x,\x')$, it is convenient to introduce its Fourier transform $\Delta(k)$. Using the results of Section~\ref{sec:free-wf}, one finds
\bea \label{Delta-prop-coeff}
\Delta(\x,\x')
&=&
\int_{\k}
e^{-i\k\cdot(\x-\x')}
\,\Delta(k) ,
\qquad
\Delta(k)
=
|\phi_k(t_f)|^2 .
\eea

Having obtained an explicit expression for $Z_0[J]$, we can now derive a diagrammatic representation for the full generating functional $Z[J]$ defined in Eq.~(\ref{Z-gen-rho}). This can be written as
\be
Z[J]
\propto
\exp\Bigg\{
\sum_{n=2}^{\infty} \frac{1}{n!}
\int_{\x_1}\cdots\int_{\x_n}
\Big[ 2\,{\rm Re}\,\psi_n(\x_1,\cdots,\x_n) \Big]
\frac{\delta}{\delta J(\x_1)}\cdots\frac{\delta}{\delta J(\x_n)}
\Bigg\}
Z_0[J] .
\ee
Since the free contribution $\psi^{\rm free}_2(\x,\x')$ has already been incorporated into $Z_0[J]$, the $n=2$ term in the sum corresponds only to the interacting piece $\psi^{\rm int}_2(\x,\x')$ introduced in Eq.~(\ref{free-split-psi}). It then follows directly that $W[J]=\ln Z[J]$ consists of the sum of all connected diagrams constructed from boundary vertices with $n$ legs, determined by the coefficients $\psi_n$, with sources $J$ attached to them. Consequently, a connected $n$-point correlation function computed via Eq.~(\ref{connected-corr-W}) is given by the sum of all possible $n$-legged diagrams built from wavefunction coefficients acting as vertices.

\subsection{Correlation functions in momentum space} 
\label{sec:wfc-diagrammatic-rules-correlators}

I now present the diagrammatic rules for computing equal-time connected $n$-point correlation functions. In Fourier space, an $n$-legged boundary vertex is assigned according to
\be
\begin{tikzpicture}[baseline=-3.0pt]
  \skvertexgd{v}{(0,0)}
  \draw[sk/propdashed] (v) -- (-1.0, 0.0);
  \draw[sk/propdashed] (v) -- (-0.4, 0.7);
  \draw[sk/propdashed] (v) -- ( 0.4, 0.7);
  \draw[sk/propdashed] (v) -- (-0.4,-0.7); 
  \draw[sk/propdashed] (v) -- ( 1.0, 0.0);
   \node[above=2pt] at (0.3,-0.8) {$\psi_{n}$};
\end{tikzpicture}
\quad 
\longrightarrow  
\quad 
(2\pi)^3 \delta^{(3)}(\k_1+\cdots+\k_n)
\Big[ 2\,{\rm Re}\,\psi'_n(\k_1,\cdots,\k_n) \Big] ,
\ee
where $\k_1,\ldots,\k_n$ denote the momenta flowing into the $n$-legged vertex. Vertices can be connected to one another by propagators obeying the assignment
\be
\begin{tikzpicture}[baseline=-3.0pt]
  \draw[sk/propdashed] (0,0) -- (2,0.0);
  \skvertexgd{v1}{(0.0,0.0)} 
  \skvertexgd{v2}{(2,0.0)} 
\end{tikzpicture}
\quad
\longrightarrow
\quad
\Delta(k) ,
\ee
where $\Delta(k)$ is the propagator introduced in Eq.~(\ref{Delta-prop-coeff}). These propagators also connect boundary vertices to external legs representing the fields entering the correlation function. The corresponding rule is
\be \label{external-K-fourier-2}
\k \;\;
\begin{tikzpicture}[baseline=-3.0pt]
  \draw[sk/propdashed] (0,0) -- (2,0.0); 
  \skbound{s1}{(0.0,0.0)} 
  \skvertexgd{v1}{(2,0.0)} 
\end{tikzpicture}
\quad
\longrightarrow
\quad
\Delta(k) .
\ee
After translating a diagram into the corresponding analytical expression, one must integrate over all internal momenta $\q$ (that is, momenta flowing between pairs of vertices) using the measure $\int_{\q}$. As usual, each diagram must also be multiplied by the appropriate symmetry factor determined by its topology.


\setcounter{equation}{0}
\section{Systematics with wavefunction coefficients}
\label{sec:systematics}

Before deriving the general map between the two formalisms in Section~\ref{sec:General-map}, it is useful to introduce a number of concepts related to the structure of the bulk diagrams that define wavefunction coefficients.

\subsection{Expansion in terms of bulk vertices}
\label{sec:partitioning}

An arbitrary connected $n$-point correlation function can be organized as an expansion in the number of bulk interaction vertices,
\bea
\label{n-point-general-V}
\Big\langle \varphi(\k_1)\cdots\varphi(\k_n) \Big\rangle_c
&=&
\sum_{V=1}^{\infty}
\Big\langle \varphi(\k_1)\cdots\varphi(\k_n) \Big\rangle_c^{(V)} ,
\eea
where $V$ denotes the number of bulk vertices (as defined by the diagrammatic rules introduced in Section~\ref{sec:wf-coeff}) entering the construction of the corresponding wavefunction coefficients. It is therefore natural to expand the wavefunction coefficients themselves according to the number of bulk vertices contributing to them,
\be
\psi_n
=
\psi_n^{(1)}+\psi_n^{(2)}+\psi_n^{(3)}+\cdots .
\ee
Not all terms in this expansion are non-vanishing. In particular, since we are considering cubic interactions, one finds that $\psi_n^{(V)}=0$ whenever $n$ is even and $V$ is odd, or whenever $n$ is odd and $V$ is even.

As a simple illustration, consider the connected three-point correlation function. When expanded in powers of bulk vertices, only odd values of $V$ contribute. The lowest non-vanishing contribution arises at $V=1$ and takes the form
\be
\label{example-result-three-point-lowest}
\Big\langle \varphi(\k_1) \varphi(\k_2)  \varphi(\k_3) \Big\rangle_c^{(1)}
= 
\,\,
 \begin{tikzpicture}[baseline=-3.0pt]
  \draw[sk/boundary] (-1.3,\BOUNDY) -- (1.3,\BOUNDY);
  \skbound{B1}{(-1.0,\BOUNDY)}
  \skbound{B2}{( 0.0,\BOUNDY)}
  \skbound{B3}{( 1.0,\BOUNDY)}
  \skvertexgd{V1}{(0,-1)}
  \Gdashed{V1}{B1}
  \Gdashed{V1}{B2}
  \Gdashed{V1}{B3}
  \node[above=2pt] at (B1) {$\k_1$};
  \node[above=2pt] at (B2) {$\k_2$};
  \node[above=2pt] at (B3) {$\k_3$};
      \node[above=2pt] at (0.6,-1.5) {$\psi_{3}^{(1)}$};
\end{tikzpicture} .
\ee
On the other hand, the number of boundary diagrams contributing to the three-point function at third order increases significantly. In this case one finds
\bea 
\label{example-result-three-point-full}
\Big\langle \varphi(\k_1) \varphi(\k_2)  \varphi(\k_3) \Big\rangle_c^{(3)}
&=&
\,\,
 \begin{tikzpicture}[baseline=-3.0pt]
  \draw[sk/boundary] (-1.3,\BOUNDY) -- (1.3,\BOUNDY);
  \skbound{B1}{(-1.0,\BOUNDY)}
  \skbound{B2}{( 0.0,\BOUNDY)}
  \skbound{B3}{( 1.0,\BOUNDY)}
  \skvertexgd{V1}{(0,-1)}
  \Gdashed{V1}{B1}
  \Gdashed{V1}{B2}
  \Gdashed{V1}{B3}
  \node[above=2pt] at (B1) {$\k_1$};
  \node[above=2pt] at (B2) {$\k_2$};
  \node[above=2pt] at (B3) {$\k_3$};
      \node[above=2pt] at (0.6,-1.5) {$\psi_{3}^{(3)}$};
\end{tikzpicture} 
+
 \begin{tikzpicture}[baseline=-3.0pt]
  \draw[sk/boundary] (-1.3,\BOUNDY) -- (1.3,\BOUNDY);
  \skbound{B1}{(-1.0,\BOUNDY)}
  \skbound{B2}{( -0.0,\BOUNDY)}
  \skbound{B3}{( 1.0,\BOUNDY)}
  \skvertexgd{V1}{(0,-1)}
  \Gdashed{V1}{B1}
  \Gdashed{V1}{B2}
  \Gdashed{V1}{B3}
  \node[above=2pt] at (B1) {$\k_1$};
  \node[above=2pt] at (B2) {$\k_2$};
  \node[above=2pt] at (B3) {$\k_3$};
    \node[above=2pt] at (-0.8,-1.7) {$\psi_{5}^{(3)}$};
     \draw[sk/propdashed] (V1) to[out=-55,in=-125,looseness=20] (V1);
\end{tikzpicture}
+
\begin{tikzpicture}[baseline=-3.0pt]
  \draw[sk/boundary] (-1.3,\BOUNDY) -- (1.3,\BOUNDY);
  \skbound{B1}{(-1.0,\BOUNDY)}
  \skbound{B2}{( -0.0,\BOUNDY)}
  \skbound{B3}{( 1.0,\BOUNDY)}
  \skvertexgd{V1}{(-0.5,-1)}
    \skvertexgd{V2}{(0.5,-1)}
  \Gdashed{V1}{B1}
  \Gdashed{V1}{B2}
  \Gdashed{V2}{B3}
    \Gdashedarc{V1}{V2}{-25}{-155}  
          \Gdashedarc{V1}{V2}{25}{155}  
  \node[above=2pt] at (B1) {$\k_1$};
  \node[above=2pt] at (B2) {$\k_2$};
  \node[above=2pt] at (B3) {$\k_3$};
   \node[above=2pt] at (-1.1,-1.5) {$\psi_{4}^{(2)}$};
   \node[above=2pt] at (1.1,-1.5) {$\psi_{3}^{(1)}$};
\end{tikzpicture}  
\nn \\[-10pt]
&&
\,\,
+
 \begin{tikzpicture}[baseline=-3.0pt]
  \draw[sk/boundary] (-1.3,\BOUNDY) -- (1.3,\BOUNDY);
  \skbound{B1}{(-1.0,\BOUNDY)}
  \skbound{B2}{( -0.0,\BOUNDY)}
  \skbound{B3}{( 1.0,\BOUNDY)}
  \skvertexgd{V1}{(-0.7,-1)}
    \skvertexgd{V2}{(0.7,-1)}
  \Gdashed{V1}{B1}
  \Gdashed{V3}{B2}
          \Gdashedarc{V1}{V2}{-35}{-145}  
           \Gdashedarc{V1}{V2}{35}{145}  
              \skvertexgd{V3}{(0 ,-0.7)}
                  \Gdashed{V2}{B3}
  \node[above=2pt] at (B1) {$\k_1$};
  \node[above=2pt] at (B2) {$\k_2$};
  \node[above=2pt] at (B3) {$\k_3$};
     \node[above=2pt] at (-1.3,-1.5) {$\psi_{3}^{(1)}$};
   \node[above=2pt] at (1.3,-1.5) {$\psi_{3}^{(1)}$};
     \node[above=2pt] at (0.45,-0.85) {$\psi_{3}^{(1)}$};
\end{tikzpicture}  
+
\begin{tikzpicture}[baseline=-3.0pt]
  \draw[sk/boundary] (-1.3,\BOUNDY) -- (1.3,\BOUNDY);
  \skbound{B1}{(-1.0,\BOUNDY)}
  \skbound{B2}{( -0.0,\BOUNDY)}
  \skbound{B3}{( 1.0,\BOUNDY)}
  \skvertexgd{V1}{(-0.5,-1)}
    \skvertexgd{V2}{(0.5,-1)}
  \Gdashed{V1}{B1}
  \Gdashed{V1}{B2}
  \Gdashed{V2}{B3}
  \Gdashed{V1}{V2}
  \node[above=2pt] at (B1) {$\k_1$};
  \node[above=2pt] at (B2) {$\k_2$};
  \node[above=2pt] at (B3) {$\k_3$};
   \node[above=2pt] at (-1.1,-1.5) {$\psi_{3}^{(1)}$};
   \node[above=2pt] at (1.1,-1.5) {$\psi_{2}^{(2)}$};
\end{tikzpicture}  
+
 \begin{tikzpicture}[baseline=-3.0pt]
  \draw[sk/boundary] (-1.8,\BOUNDY) -- (0.8,\BOUNDY);
  \skbound{B1}{(-1.5,\BOUNDY)}
  \skbound{B2}{( -0.5 ,\BOUNDY)}
  \skbound{B3}{( 0.5,\BOUNDY)}
  \skvertexgd{V1}{(-1.5,-1)}
    \skvertexgd{V2}{(-0.7,-1)}
      \skvertexgd{V3}{(0 ,-1)}
  \Gdashed{V1}{B1}
  \Gdashed{V3}{B2}
    \Gdashed{V2}{V3}
      \Gdashed{V3}{B3}
     \Gdashedarc{V1}{V2}{-45}{-135}  
          \Gdashedarc{V1}{V2}{45}{135}  
  \node[above=2pt] at (B1) {$\k_1$};
  \node[above=2pt] at (B2) {$\k_2$};
  \node[above=2pt] at (B3) {$\k_3$};
  \node[above=2pt] at (-1.5,-2.0) {$\psi_{3}^{(1)}$};
   \node[above=2pt] at (-0.7,-2.0) {$\psi_{3}^{(1)}$};
    \node[above=2pt] at (0,-2.0) {$\psi_{3}^{(1)}$};
\end{tikzpicture} 
\nn
\\[-10pt]
&&
\,\,
+
 \begin{tikzpicture}[baseline=-3.0pt]
  \draw[sk/boundary] (-1.3,\BOUNDY) -- (1.3,\BOUNDY);
  \skbound{B1}{(-1.0,\BOUNDY)}
  \skbound{B2}{( -0.0,\BOUNDY)}
  \skbound{B3}{( 1.0,\BOUNDY)}
  \skvertexgd{V1}{(-0,-0.8)}
  \skvertexgd{V2}{(-0.0,-1.4)}
  \Gdashed{V1}{B1}
  \Gdashed{V1}{B2}
  \Gdashed{V1}{B3}
    \Gdashed{V1}{V2}
     \draw[sk/propdashed] (V2.25) to[out=45,in=-45,looseness=20] (V2.-25);
  \node[above=2pt] at (B1) {$\k_1$};
  \node[above=2pt] at (B2) {$\k_2$};
  \node[above=2pt] at (B3) {$\k_3$};
  \node[above=2pt] at (-0.6,-1.3) {$\psi_{4}^{(2)}$};
   \node[above=2pt] at (-0.5,-2.0) {$\psi_{3}^{(1)}$};
\end{tikzpicture}  
+
 \begin{tikzpicture}[baseline=-3.0pt]
  \draw[sk/boundary] (-1.8,\BOUNDY) -- (0.8,\BOUNDY);
  \skbound{B1}{(-1.5,\BOUNDY)}
  \skbound{B2}{( -0.5,\BOUNDY)}
  \skbound{B3}{( 0.5,\BOUNDY)}
  \skvertexgd{V1}{(-1,-0.8)}
  \skvertexgd{V2}{(-1.0,-1.4)}
   \skvertexgd{V3}{(0 ,-0.8)}
  \Gdashed{V1}{B1}
  \Gdashed{V3}{B2}
    \Gdashed{V3}{B3}
    \Gdashed{V1}{V3}
    \Gdashed{V1}{V2}
     \draw[sk/propdashed] (V2.25) to[out=45,in=-45,looseness=20] (V2.-25);
  \node[above=2pt] at (B1) {$\k_1$};
  \node[above=2pt] at (B2) {$\k_2$};
  \node[above=2pt] at (B3) {$\k_3$};
  \node[above=2pt] at (-1.5,-1.2) {$\psi_{3}^{(1)}$};
   \node[above=2pt] at (-1.5,-2.0) {$\psi_{3}^{(1)}$};
     \node[above=2pt] at (0.5,-1.5) {$\psi_{3}^{(1)}$};
\end{tikzpicture} 
+ {\rm perms} . \quad
\eea
where ``perms'' denotes additional diagrams obtained by interchanging the external momenta, whenever such permutations lead to inequivalent contributions.

As another example, consider the four-point correlation function at the lowest order in the number of bulk vertices. This corresponds to $V=2$, and is given by
\be \label{four-point-example-gray-2}
\Big\langle \varphi(\k_1)\cdots\varphi(\k_4) \Big\rangle_c^{(2)}
=
\,\,
 \begin{tikzpicture}[baseline=-3.0pt]
  \draw[sk/boundary] (-1.8,\BOUNDY) -- (1.8,\BOUNDY);
  \skbound{B1}{(-1.5,\BOUNDY)}
  \skbound{B2}{( -0.5,\BOUNDY)}
  \skbound{B3}{( 0.5,\BOUNDY)}
    \skbound{B4}{( 1.5,\BOUNDY)}
  \skvertexgd{V1}{(0,-1)}
  \Gdashed{V1}{B1}
  \Gdashed{V1}{B2}
  \Gdashed{V1}{B3}
    \Gdashed{V1}{B4}
  \node[above=2pt] at (B1) {$\k_1$};
  \node[above=2pt] at (B2) {$\k_2$};
  \node[above=2pt] at (B3) {$\k_3$};
  \node[above=2pt] at (B4) {$\k_4$};
    \node[above=2pt] at (0.0,-2.0) {$\psi_{4}^{(2)}$};
\end{tikzpicture} 
 +
  \begin{tikzpicture}[baseline=-3.0pt]
  \draw[sk/boundary] (-1.8,\BOUNDY) -- (1.8,\BOUNDY);
  \skbound{B1}{(-1.5,\BOUNDY)}
  \skbound{B2}{( -0.5,\BOUNDY)}
  \skbound{B3}{( 0.5,\BOUNDY)}
    \skbound{B4}{( 1.5,\BOUNDY)}
  \skvertexgd{V1}{(-1,-1)}
    \skvertexgd{V2}{(1,-1)}
  \Gdashed{V1}{B1}
  \Gdashed{V1}{B2}
  \Gdashed{V2}{B3}
    \Gdashed{V2}{B4}
        \Gdashed{V1}{V2}
  \node[above=2pt] at (B1) {$\k_1$};
  \node[above=2pt] at (B2) {$\k_2$};
  \node[above=2pt] at (B3) {$\k_3$};
  \node[above=2pt] at (B4) {$\k_4$};
   \node[above=2pt] at (-1.0,-2.0) {$\psi_{3}^{(1)}$};
   \node[above=2pt] at (1.0,-2.0) {$\psi_{3}^{(1)}$};
\end{tikzpicture}   
+ {\rm perms} 
.
\ee
As in the case of the three-point function, the number of diagrams contributing to the four-point function at the next-to-leading order increases significantly. In this case, one finds
\bea \label{four-point-example-gray-3}
\Big\langle \varphi(\k_1)\cdots\varphi(\k_4) \Big\rangle_c^{(4)}
&=&
\begin{tikzpicture}[baseline=-3.0pt]
  \draw[sk/boundary] (-1.8,\BOUNDY) -- (1.8,\BOUNDY);
  \skbound{B1}{(-1.5,\BOUNDY)}
  \skbound{B2}{( -0.5,\BOUNDY)}
  \skbound{B3}{( 0.5,\BOUNDY)}
    \skbound{B4}{( 1.5,\BOUNDY)}
  \skvertexgd{V1}{(0,-1)}
  \Gdashed{V1}{B1}
  \Gdashed{V1}{B2}
  \Gdashed{V1}{B3}
    \Gdashed{V1}{B4}
  \node[above=2pt] at (B1) {$\k_1$};
  \node[above=2pt] at (B2) {$\k_2$};
  \node[above=2pt] at (B3) {$\k_3$};
  \node[above=2pt] at (B4) {$\k_4$};
    \node[above=2pt] at (0.0,-2.0) {$\psi_{4}^{(4)}$};
\end{tikzpicture} 
+
 \begin{tikzpicture}[baseline=-3.0pt]
  \draw[sk/boundary] (-1.8,\BOUNDY) -- (1.8,\BOUNDY);
  \skbound{B1}{(-1.5,\BOUNDY)}
  \skbound{B2}{( -0.5,\BOUNDY)}
  \skbound{B3}{( 0.5,\BOUNDY)}
    \skbound{B4}{( 1.5,\BOUNDY)}
  \skvertexgd{V1}{(0,-1)}
  \Gdashed{V1}{B1}
  \Gdashed{V1}{B2}
  \Gdashed{V1}{B3}
    \Gdashed{V1}{B4}
  \node[above=2pt] at (B1) {$\k_1$};
  \node[above=2pt] at (B2) {$\k_2$};
  \node[above=2pt] at (B3) {$\k_3$};
  \node[above=2pt] at (B4) {$\k_4$};
    \node[above=2pt] at (-0.8,-1.7) {$\psi_{6}^{(4)}$};
     \draw[sk/propdashed] (V1) to[out=-55,in=-125,looseness=20] (V1);
\end{tikzpicture} 
 +
  \begin{tikzpicture}[baseline=-3.0pt]
  \draw[sk/boundary] (-1.8,\BOUNDY) -- (1.8,\BOUNDY);
  \skbound{B1}{(-1.5,\BOUNDY)}
  \skbound{B2}{( -0.5,\BOUNDY)}
  \skbound{B3}{( 0.5,\BOUNDY)}
    \skbound{B4}{( 1.5,\BOUNDY)}
  \skvertexgd{V1}{(-1,-1)}
    \skvertexgd{V2}{(1,-1)}
  \Gdashed{V1}{B1}
  \Gdashed{V1}{B2}
  \Gdashed{V2}{B3}
    \Gdashed{V2}{B4}
        \Gdashed{V1}{V2}
  \node[above=2pt] at (B1) {$\k_1$};
  \node[above=2pt] at (B2) {$\k_2$};
  \node[above=2pt] at (B3) {$\k_3$};
  \node[above=2pt] at (B4) {$\k_4$};
   \node[above=2pt] at (-1.0,-2.0) {$\psi_{3}^{(1)}$};
   \node[above=2pt] at (1.0,-2.0) {$\psi_{3}^{(3)}$};
\end{tikzpicture}   
\nn
\\[-12pt]
&&
\!\!\!\!\!\!\!\!\!\!\!\!\!\!\!\!\!\!\!\!\!\!\!\!\!\!\!\!\!\!\!\!\!
+
\begin{tikzpicture}[baseline=-3.0pt]
  \draw[sk/boundary] (-1.8,\BOUNDY) -- (1.8,\BOUNDY);
  \skbound{B1}{(-1.5,\BOUNDY)}
  \skbound{B2}{( -0.5,\BOUNDY)}
  \skbound{B3}{( 0.5,\BOUNDY)}
    \skbound{B4}{( 1.5,\BOUNDY)}
  \skvertexgd{V1}{(-0.5,-1)}
    \skvertexgd{V2}{(1,-1)}
  \Gdashed{V1}{B1}
  \Gdashed{V1}{B2}
  \Gdashed{V1}{B3}
    \Gdashed{V2}{B4}
        \Gdashed{V1}{V2}
  \node[above=2pt] at (B1) {$\k_1$};
  \node[above=2pt] at (B2) {$\k_2$};
  \node[above=2pt] at (B3) {$\k_3$};
  \node[above=2pt] at (B4) {$\k_4$};
     \node[above=2pt] at (-0.5,-2.0) {$\psi_{4}^{(2)}$};
   \node[above=2pt] at (1.0,-2.0) {$\psi_{2}^{(2)}$};
\end{tikzpicture} 
+
  \begin{tikzpicture}[baseline=-3.0pt]
  \draw[sk/boundary] (-1.8,\BOUNDY) -- (1.8,\BOUNDY);
  \skbound{B1}{(-1.5,\BOUNDY)}
  \skbound{B2}{( -0.5,\BOUNDY)}
  \skbound{B3}{( 0.5,\BOUNDY)}
    \skbound{B4}{( 1.5,\BOUNDY)}
  \skvertexgd{V1}{(-1,-1)}
    \skvertexgd{V2}{(1,-1)}
  \Gdashed{V1}{B1}
  \Gdashed{V1}{B2}
  \Gdashed{V2}{B3}
    \Gdashed{V2}{B4}
    \Gdashedarc{V1}{V2}{-25}{-155}  
          \Gdashedarc{V1}{V2}{25}{155}  
  \node[above=2pt] at (B1) {$\k_1$};
  \node[above=2pt] at (B2) {$\k_2$};
  \node[above=2pt] at (B3) {$\k_3$};
  \node[above=2pt] at (B4) {$\k_4$};
   \node[above=2pt] at (-1.0,-2.0) {$\psi_{4}^{(2)}$};
   \node[above=2pt] at (1.0,-2.0) {$\psi_{4}^{(2)}$};
\end{tikzpicture}   
+
  \begin{tikzpicture}[baseline=-3.0pt]
  \draw[sk/boundary] (-1.8,\BOUNDY) -- (1.8,\BOUNDY);
  \skbound{B1}{(-1.5,\BOUNDY)}
  \skbound{B2}{( -0.5,\BOUNDY)}
  \skbound{B3}{( 0.5,\BOUNDY)}
    \skbound{B4}{( 1.5,\BOUNDY)}
  \skvertexgd{V1}{(-1,-1)}
    \skvertexgd{V2}{(1,-1)}
  \Gdashed{V1}{B1}
  \Gdashed{V1}{B2}
  \Gdashed{V2}{B3}
    \Gdashed{V2}{B4}
        \Gdashed{V1}{V2}
  \node[above=2pt] at (B1) {$\k_1$};
  \node[above=2pt] at (B2) {$\k_2$};
  \node[above=2pt] at (B3) {$\k_3$};
  \node[above=2pt] at (B4) {$\k_4$};
   \node[above=2pt] at (-1.0,-2.0) {$\psi_{3}^{(1)}$};
   \node[above=2pt] at (1.7,-1.5) {$\psi_{5}^{(3)}$};
        \draw[sk/propdashed] (V2) to[out=-55,in=-125,looseness=20] (V2); 
\end{tikzpicture}   
\nn
\\[-12pt]
&& 
\!\!\!\!\!\!\!\!\!\!\!\!\!\!\!\!\!\!\!\!\!\!\!\!\!\!\!\!\!\!\!\!\!
+
\begin{tikzpicture}[baseline=-3.0pt]
  \draw[sk/boundary] (-1.8,\BOUNDY) -- (1.8,\BOUNDY);
  \skbound{B1}{(-1.5,\BOUNDY)}
  \skbound{B2}{( -0.5,\BOUNDY)}
  \skbound{B3}{( 0.5,\BOUNDY)}
    \skbound{B4}{( 1.5,\BOUNDY)}
  \skvertexgd{V1}{(-0.5,-1)}
    \skvertexgd{V2}{(1,-1)}
  \Gdashed{V1}{B1}
  \Gdashed{V1}{B2}
  \Gdashed{V1}{B3}
    \Gdashed{V2}{B4}
          \Gdashedarc{V1}{V2}{-25}{-155}  
          \Gdashedarc{V1}{V2}{25}{155}  
  \node[above=2pt] at (B1) {$\k_1$};
  \node[above=2pt] at (B2) {$\k_2$};
  \node[above=2pt] at (B3) {$\k_3$};
  \node[above=2pt] at (B4) {$\k_4$};
     \node[above=2pt] at (-0.5,-2.0) {$\psi_{5}^{(3)}$};
   \node[above=2pt] at (1.0,-2.0) {$\psi_{3}^{(1)}$};
\end{tikzpicture} 
+ 
\begin{tikzpicture}[baseline=-3.0pt]
  \draw[sk/boundary] (-1.8,\BOUNDY) -- (1.8,\BOUNDY);
  \skbound{B1}{(-1.5,\BOUNDY)}
  \skbound{B2}{( -0.5,\BOUNDY)}
  \skbound{B3}{( 0.5,\BOUNDY)}
    \skbound{B4}{( 1.5,\BOUNDY)}
  \skvertexgd{V1}{(-1,-1)}
  \skvertexgd{V4}{(1 ,-1)}
  \Gdashedarc{V1}{V4}{-25}{-155}  
           \Gdashedarc{V1}{V4}{25}{155}  
    \skvertexgd{V2}{(-0.35,-0.75)}
      \skvertexgd{V3}{(0.35 ,-0.75)}  
  \Gdashed{V1}{B1}
  \Gdashed{V2}{B2}
  \Gdashed{V3}{B3}
    \Gdashed{V4}{B4}
  \node[above=2pt] at (B1) {$\k_1$};
  \node[above=2pt] at (B2) {$\k_2$};
  \node[above=2pt] at (B3) {$\k_3$};
  \node[above=2pt] at (B4) {$\k_4$};
   \node[above=2pt] at (-1.0,-2.0) {$\psi_{3}^{(1)}$};
   \node[above=2pt] at (1.0,-2.0) {$\psi_{3}^{(1)}$};
   \node[above=2pt] at (0.85,-0.9) {$\psi_{3}^{(1)}$};
   \node[above=2pt] at (0.05,-0.75) {$\psi_{3}^{(1)}$};
\end{tikzpicture} 
+
  \begin{tikzpicture}[baseline=-3.0pt]
  \draw[sk/boundary] (-1.8,\BOUNDY) -- (1.8,\BOUNDY);
  \skbound{B1}{(-1.5,\BOUNDY)}
  \skbound{B2}{( -0.5,\BOUNDY)}
  \skbound{B3}{( 0.5,\BOUNDY)}
    \skbound{B4}{( 1.5,\BOUNDY)}
 \skvertexgd{V1}{(-1,-1)}
  \skvertexgd{V3}{(1,-1)}
  \Gdashed{V1}{B1}
    \Gdashed{V3}{B4}
    \Gdashedarc{V1}{V3}{-25}{-155}  
          \Gdashedarc{V1}{V3}{25}{155}  
            \skvertexgd{V2}{(0,-0.73)}
               \Gdashed{V2}{B2}
  \Gdashed{V2}{B3}
  \node[above=2pt] at (B1) {$\k_1$};
  \node[above=2pt] at (B2) {$\k_2$};
  \node[above=2pt] at (B3) {$\k_3$};
  \node[above=2pt] at (B4) {$\k_4$};
   \node[above=2pt] at (-1.6,-1.5) {$\psi_{3}^{(1)}$};
   \node[above=2pt] at (1.6,-1.5) {$\psi_{3}^{(1)}$};
    \node[above=2pt] at (0.7,-0.9) {$\psi_{4}^{(2)}$};
\end{tikzpicture}   
\nn 
\\[-5pt]
&&
\!\!\!\!\!\!\!\!\!\!\!\!\!\!\!\!\!\!\!\!\!\!\!\!\!\!\!\!\!\!\!\!\!
+
 \begin{tikzpicture}[baseline=-3.0pt]
  \draw[sk/boundary] (-1.8,\BOUNDY) -- (1.8,\BOUNDY);
  \skbound{B1}{(-1.5,\BOUNDY)}
  \skbound{B2}{( -0.5,\BOUNDY)}
  \skbound{B3}{( 0.5,\BOUNDY)}
    \skbound{B4}{( 1.5,\BOUNDY)}
  \skvertexgd{V1}{(-1,-1)}
    \skvertexgd{V2}{(0,-1)}
     \skvertexgd{V3}{(1,-1)}
  \Gdashed{V1}{B1}
  \Gdashed{V1}{B2}
  \Gdashed{V2}{B3}
    \Gdashed{V3}{B4}
        \Gdashed{V1}{V2}
         \Gdashedarc{V2}{V3}{-25}{-155}  
          \Gdashedarc{V2}{V3}{25}{155}  
  \node[above=2pt] at (B1) {$\k_1$};
  \node[above=2pt] at (B2) {$\k_2$};
  \node[above=2pt] at (B3) {$\k_3$};
  \node[above=2pt] at (B4) {$\k_4$};
   \node[above=2pt] at (-1.0,-2.0) {$\psi_{3}^{(1)}$};
    \node[above=2pt] at (0.0,-2.0) {$\psi_{4}^{(2)}$};
   \node[above=2pt] at (1.0,-2.0) {$\psi_{3}^{(1)}$};
\end{tikzpicture}  
+
 \begin{tikzpicture}[baseline=-3.0pt]
  \draw[sk/boundary] (-1.8,\BOUNDY) -- (1.8,\BOUNDY);
  \skbound{B1}{(-1.5,\BOUNDY)}
  \skbound{B2}{( -0.5,\BOUNDY)}
  \skbound{B3}{( 0.5,\BOUNDY)}
    \skbound{B4}{( 1.5,\BOUNDY)}
  \skvertexgd{V1}{(-1,-1)}
    \skvertexgd{V2}{(0,-1)}
     \skvertexgd{V3}{(1,-1)}
  \Gdashed{V1}{B1}
  \Gdashed{V1}{B2}
  \Gdashed{V3}{B3}
    \Gdashed{V3}{B4}
        \Gdashed{V1}{V2}
         \Gdashedarc{V2}{V3}{-25}{-155}  
          \Gdashedarc{V2}{V3}{25}{155}  
  \node[above=2pt] at (B1) {$\k_1$};
  \node[above=2pt] at (B2) {$\k_2$};
  \node[above=2pt] at (B3) {$\k_3$};
  \node[above=2pt] at (B4) {$\k_4$};
   \node[above=2pt] at (-1.0,-2.0) {$\psi_{3}^{(1)}$};
    \node[above=2pt] at (0.0,-2.0) {$\psi_{3}^{(1)}$};
   \node[above=2pt] at (1.0,-2.0) {$\psi_{4}^{(2)}$};
\end{tikzpicture}  
+
 \begin{tikzpicture}[baseline=-3.0pt]
  \draw[sk/boundary] (-1.8,\BOUNDY) -- (1.8,\BOUNDY);
  \skbound{B1}{(-1.5,\BOUNDY)}
  \skbound{B2}{( -0.5,\BOUNDY)}
  \skbound{B3}{( 0.5,\BOUNDY)}
    \skbound{B4}{( 1.5,\BOUNDY)}
  \skvertexgd{V1}{(-1,-0.8)}
   \skvertexgd{V2}{(-1,-1.4)}
    \skvertexgd{V4}{(1 ,-0.8)}
  \Gdashed{V1}{B1}
  \Gdashed{V1}{B2}
  \Gdashed{V4}{B3}
    \Gdashed{V4}{B4}
    \Gdashed{V1}{V2}
    \Gdashed{V1}{V4}
     \draw[sk/propdashed] (V2.25) to[out=45,in=-45,looseness=20] (V2.-25);
  \node[above=2pt] at (B1) {$\k_1$};
  \node[above=2pt] at (B2) {$\k_2$};
  \node[above=2pt] at (B3) {$\k_3$};
  \node[above=2pt] at (B4) {$\k_4$};
  \node[above=2pt] at (-1.5,-1.3) {$\psi_{4}^{(2)}$};
   \node[above=2pt] at (-1.5,-2.0) {$\psi_{3}^{(1)}$};
     \node[above=2pt] at (1.5,-1.5) {$\psi_{3}^{(1)}$};
\end{tikzpicture} 
\nn 
\\[-5pt]
&&
\!\!\!\!\!\!\!\!\!\!\!\!\!\!\!\!\!\!\!\!\!\!\!\!\!\!\!\!\!\!\!\!\!
+
 \begin{tikzpicture}[baseline=-3.0pt]
  \draw[sk/boundary] (-1.8,\BOUNDY) -- (1.8,\BOUNDY);
  \skbound{B1}{(-1.5,\BOUNDY)}
  \skbound{B2}{( -0.5,\BOUNDY)}
  \skbound{B3}{( 0.5,\BOUNDY)}
    \skbound{B4}{( 1.5,\BOUNDY)}
   \skvertexgd{V1}{(-1,-1)}
    \skvertexgd{V2}{(0.2,-1)}
      \skvertexgd{V4}{(1 ,-1)}
  \Gdashed{V1}{B1}
  \Gdashed{V4}{B3}
    \Gdashed{V4}{B4}
    \Gdashed{V2}{V4}
     \Gdashedarc{V1}{V2}{-35}{-145}  
          \Gdashedarc{V1}{V2}{35}{145}  
          \skvertexgd{V3}{(-0.4 ,-0.75)}
  \Gdashed{V3}{B2}
  \node[above=2pt] at (B1) {$\k_1$};
  \node[above=2pt] at (B2) {$\k_2$};
  \node[above=2pt] at (B3) {$\k_3$};
  \node[above=2pt] at (B4) {$\k_4$};
   \node[above=2pt] at (-1.2,-2.0) {$\psi_{3}^{(1)}$};
   \node[above=2pt] at (0.25,-2.0) {$\psi_{3}^{(1)}$};
    \node[above=2pt] at (0.0,-0.85) {$\psi_{3}^{(1)}$};
     \node[above=2pt] at (1.2,-2.0) {$\psi_{3}^{(1)}$};
\end{tikzpicture}
+
 \begin{tikzpicture}[baseline=-3.0pt]
  \draw[sk/boundary] (-1.8,\BOUNDY) -- (1.8,\BOUNDY);
  \skbound{B1}{(-1.5,\BOUNDY)}
  \skbound{B2}{( -0.5,\BOUNDY)}
  \skbound{B3}{( 0.5,\BOUNDY)}
    \skbound{B4}{( 1.5,\BOUNDY)}
   \skvertexgd{V1}{(-1,-1)}
    \skvertexgd{V2}{(-0.4,-1)}
     \skvertexgd{V3}{(0.4 ,-1)}
      \skvertexgd{V4}{(1 ,-1)}
  \Gdashed{V1}{B1}
  \Gdashed{V1}{B2}
  \Gdashed{V4}{B3}
    \Gdashed{V4}{B4}
    \Gdashed{V1}{V2}
    \Gdashed{V3}{V4}
     \Gdashedarc{V2}{V3}{-45}{-135}  
          \Gdashedarc{V2}{V3}{45}{135}  
  \node[above=2pt] at (B1) {$\k_1$};
  \node[above=2pt] at (B2) {$\k_2$};
  \node[above=2pt] at (B3) {$\k_3$};
  \node[above=2pt] at (B4) {$\k_4$};
   \node[above=2pt] at (-1.0,-2.0) {$\psi_{3}^{(1)}$};
   \node[above=2pt] at (-0.33,-2.0) {$\psi_{3}^{(1)}$};
    \node[above=2pt] at (0.33,-2.0) {$\psi_{3}^{(1)}$};
     \node[above=2pt] at (1.0,-2.0) {$\psi_{3}^{(1)}$};
\end{tikzpicture}  
+
 \begin{tikzpicture}[baseline=-3.0pt]
  \draw[sk/boundary] (-1.8,\BOUNDY) -- (1.8,\BOUNDY);
  \skbound{B1}{(-1.5,\BOUNDY)}
  \skbound{B2}{( -0.5 ,\BOUNDY)}
  \skbound{B3}{( 0.5,\BOUNDY)}
    \skbound{B4}{( 1.5,\BOUNDY)}
  \skvertexgd{V1}{(-1.5,-1)}
    \skvertexgd{V2}{(-0.7,-1)}
      \skvertexgd{V3}{(0 ,-1)}
       \skvertexgd{V4}{(1 ,-1)}
  \Gdashed{V1}{B1}
  \Gdashed{V3}{B2}
  \Gdashed{V4}{B3}
    \Gdashed{V4}{B4}
    \Gdashed{V2}{V3}
    \Gdashed{V3}{V4}
     \Gdashedarc{V1}{V2}{-45}{-135}  
          \Gdashedarc{V1}{V2}{45}{135}  
  \node[above=2pt] at (B1) {$\k_1$};
  \node[above=2pt] at (B2) {$\k_2$};
  \node[above=2pt] at (B3) {$\k_3$};
  \node[above=2pt] at (B4) {$\k_4$};
  \node[above=2pt] at (-1.5,-2.0) {$\psi_{3}^{(1)}$};
   \node[above=2pt] at (-0.7,-2.0) {$\psi_{3}^{(1)}$};
    \node[above=2pt] at (0,-2.0) {$\psi_{3}^{(1)}$};
     \node[above=2pt] at (1.0,-2.0) {$\psi_{3}^{(1)}$};
\end{tikzpicture} 
\nn \\[-5pt]
&&
\!\!\!\!\!\!\!\!\!\!\!\!\!\!\!\!\!\!\!\!\!\!\!\!\!\!\!\!\!\!\!\!\! 
+
 \begin{tikzpicture}[baseline=-3.0pt]
  \draw[sk/boundary] (-1.8,\BOUNDY) -- (1.8,\BOUNDY);
  \skbound{B1}{(-1.5,\BOUNDY)}
  \skbound{B2}{( -0.5,\BOUNDY)}
  \skbound{B3}{( 0.5,\BOUNDY)}
    \skbound{B4}{( 1.5,\BOUNDY)}
  \skvertexgd{V1}{(-1,-0.8)}
   \skvertexgd{V2}{(0.0,-1.4)}
    \skvertexgd{V3}{(0 ,-0.8)}
    \skvertexgd{V4}{(1 ,-0.8)}
  \Gdashed{V1}{B1}
  \Gdashed{V1}{B2}
  \Gdashed{V4}{B3}
    \Gdashed{V4}{B4}
    \Gdashed{V1}{V3}
    \Gdashed{V3}{V2}
    \Gdashed{V3}{V4}
     \draw[sk/propdashed] (V2.25) to[out=45,in=-45,looseness=20] (V2.-25);
  \node[above=2pt] at (B1) {$\k_1$};
  \node[above=2pt] at (B2) {$\k_2$};
  \node[above=2pt] at (B3) {$\k_3$};
  \node[above=2pt] at (B4) {$\k_4$};
  \node[above=2pt] at (-1.5,-1.5) {$\psi_{3}^{(1)}$};
   \node[above=2pt] at (-0.5,-2.0) {$\psi_{3}^{(1)}$};
     \node[above=2pt] at (1.5,-1.5) {$\psi_{3}^{(1)}$};
     \node[above=2pt] at (0,-0.75) {$\psi_{3}^{(1)}$};
\end{tikzpicture} 
+
 \begin{tikzpicture}[baseline=-3.0pt]
  \draw[sk/boundary] (-1.8,\BOUNDY) -- (1.8,\BOUNDY);
  \skbound{B1}{(-1.5,\BOUNDY)}
  \skbound{B2}{( -0.5,\BOUNDY)}
  \skbound{B3}{( 0.5,\BOUNDY)}
    \skbound{B4}{( 1.5,\BOUNDY)}
  \skvertexgd{V1}{(-1,-0.8)}
  \skvertexgd{V2}{(-1.0,-1.4)}
   \skvertexgd{V3}{(0 ,-0.8)}
   \skvertexgd{V4}{(1 ,-0.8)}
  \Gdashed{V1}{B1}
  \Gdashed{V3}{B2}
  \Gdashed{V4}{B3}
    \Gdashed{V4}{B4}
    \Gdashed{V1}{V3}
    \Gdashed{V1}{V2}
    \Gdashed{V3}{V4}
     \draw[sk/propdashed] (V2.25) to[out=45,in=-45,looseness=20] (V2.-25);
  \node[above=2pt] at (B1) {$\k_1$};
  \node[above=2pt] at (B2) {$\k_2$};
  \node[above=2pt] at (B3) {$\k_3$};
  \node[above=2pt] at (B4) {$\k_4$};
  \node[above=2pt] at (-1.5,-1.2) {$\psi_{3}^{(1)}$};
   \node[above=2pt] at (-1.5,-2.0) {$\psi_{3}^{(1)}$};
     \node[above=2pt] at (1.5,-1.5) {$\psi_{3}^{(1)}$};
     \node[above=2pt] at (0.2,-1.7) {$\psi_{3}^{(1)}$};
\end{tikzpicture} 
+ {\rm perms}
. \qquad
\eea
where, again, ``perms'' stands for additional diagrams obtained by interchanging external momenta, whenever such an operation is required. The diagrams contributing to (\ref{example-result-three-point-lowest}) and (\ref{four-point-example-gray-2}) are tree-level diagrams, including the bulk diagrams entering the computation of $\psi_3^{(1)}$ and $\psi_4^{(2)}$. In contrast, every term appearing in (\ref{example-result-three-point-full}) and (\ref{four-point-example-gray-3}) involves a loop integral. Indeed, tree-level diagrams in (\ref{example-result-three-point-full}) and (\ref{four-point-example-gray-3}) necessarily contain at least one wavefunction coefficient that includes a loop integral, whereas loop diagrams are built entirely from tree-level wavefunction coefficients.

\subsection{Topology of wavefunction diagrams}
\label{sec:topology-wf}

A wavefunction coefficient $\psi_n^{(V)}(\k_1 , \ldots , \k_n)$, which is fully symmetric under permutations of its external momenta, can be further decomposed according to the topology of the bulk diagrams contributing to its computation. I will consider the following expansion in terms of sub-coefficients that are sensitive to diagram topology:
\be \label{wfc-topology}
\psi_n^{(V)} (\k_1 , \ldots , \k_n )
=
\sum_t \Big[ \psi_{n,t}^{(V)} (\k_1 , \ldots , \k_n ) + {\rm perms} \Big] .
\ee
Here, $t$ labels the topology of the bulk diagram entering the computation of $\psi_n^{(V)}$, a choice that is purely conventional. The term ``perms'' denotes additional contributions obtained by permuting the external momenta of $\psi_{n,t}^{(V)} (\k_1 , \ldots , \k_n )$, whenever such permutations produce inequivalent functions. This prescription is necessary because individual contributions $\psi_{n,t}^{(V)}$ need not be symmetric under the interchange of its external momenta, even though their sum must be.

Let us illustrate this decomposition with a few examples. At second order in the number of bulk vertices, there are two distinct diagram topologies contributing to the two-point wavefunction coefficient. These contributions may be labeled as follows:
\bea
 \psi_{2,1}^{(2)}  (\k_1 , \k_2)   &=& \,\,
 \begin{tikzpicture}[baseline=-3.0pt]
  \draw[sk/boundary] (-1.3,\BOUNDY) -- (1.3,\BOUNDY);
  \skbound{B1}{(-1.0,\BOUNDY)}
  \skbound{B3}{( 1.0,\BOUNDY)}
  \skvertexp{V1}{(-0.7,-1)}
    \skvertexp{V2}{(0.7,-1)}
  \Gdouble{V1}{B1}
  \Gdouble{V2}{B3}
          \Gdoublearc{V1}{V2}{-45}{-135}  
           \Gdoublearc{V1}{V2}{45}{135}  
  \node[above=2pt] at (B1) {$\k_1$};
  \node[above=2pt] at (B3) {$\k_2$};
    \skvertexp{V1}{(-0.7,-1)}
    \skvertexp{V2}{(0.7,-1)}
    \skbound{B1}{(-1.0,\BOUNDY)}
  \skbound{B3}{( 1.0,\BOUNDY)}
\end{tikzpicture}, 
\\
 \psi_{2,2}^{(2)}  (\k_1 , \k_2)  &=& \,\,
 \begin{tikzpicture}[baseline=-3.0pt]
  \draw[sk/boundary] (-1.3,\BOUNDY) -- (1.3,\BOUNDY);
  \skbound{B1}{(-1.0,\BOUNDY)}
  \skbound{B3}{( 1.0,\BOUNDY)}
  \skvertexp{V1}{(0,-0.5)}
    \skvertexp{V2}{(0,-1)}
  \Gdouble{V1}{B1}
  \Gdouble{V1}{B3}
  \Gdouble{V1}{V2}
  \node[above=2pt] at (B1) {$\k_1$};
  \node[above=2pt] at (B3) {$\k_2$};
    \skvertexp{V1}{(0,-0.5)}
    \skbound{B1}{(-1.0,\BOUNDY)}
  \skbound{B3}{( 1.0,\BOUNDY)}
   \draw[sk/propdouble] (V2.25) to[out=45,in=-45,looseness=20] (V2.-25);
       \skvertexp{V2}{(0,-1)}
\end{tikzpicture} . \qquad
\label{psi-2-2}
\eea
Both of these diagrams are symmetric under the interchange of external momenta. Therefore, according to Eq.~(\ref{wfc-topology}), the corresponding wavefunction coefficient can be written as
\be
\psi_{2}^{(2)}(\k_1 , \k_2)
=
\psi_{2,1}^{(2)}(\k_1 , \k_2)
+
\psi_{2,2}^{(2)}(\k_1 , \k_2) .
\ee
To continue, at the one-vertex level there is a single diagram contributing to the three-point wavefunction coefficient:
\be
\label{psi-3-1}
 \psi_{3,1}^{(1)}  (\k_1 , \k_2, \k_3)   = \,\,
 \begin{tikzpicture}[baseline=-3.0pt]
  \draw[sk/boundary] (-1.3,\BOUNDY) -- (1.3,\BOUNDY);
  \skbound{B1}{(-1.0,\BOUNDY)}
  \skbound{B2}{( 0.0,\BOUNDY)}
  \skbound{B3}{( 1.0,\BOUNDY)}
  \skvertexp{V}{(0,-1)}
  \Gdouble{V}{B1}
  \Gdouble{V}{B2}
  \Gdouble{V}{B3}
  \node[above=2pt] at (B1) {$\k_1$};
  \node[above=2pt] at (B2) {$\k_2$};
  \node[above=2pt] at (B3) {$\k_3$};
    \skvertexp{V}{(0,-1)}
    \skbound{B1}{(-1.0,\BOUNDY)}
  \skbound{B2}{( 0.0,\BOUNDY)}
  \skbound{B3}{( 1.0,\BOUNDY)}
\end{tikzpicture} .
\ee
In this case, since there is only a single contributing diagram and it is fully symmetric under the interchange of external momenta, there is no need to introduce an explicit topological label. Nevertheless, I find it necessary to stick to the present convention. Accordingly, we have the trivial identity
\be
\psi_{3}^{(1)}(\k_1 , \k_2 , \k_3)
=
\psi_{3,1}^{(1)}(\k_1 , \k_2 , \k_3) .
\ee
On the other hand, at second order in the number of bulk vertices, there are three distinct diagrams contributing to the three-point wavefunction coefficient. These are:
\bea
 \psi_{3,1}^{(3)} (\k_1 , \k_2 , \k_3 )   &=& \,\,
 \begin{tikzpicture}[baseline=-3.0pt]
  \draw[sk/boundary] (-1.3,\BOUNDY) -- (1.3,\BOUNDY);
  \skbound{B1}{(-1.0,\BOUNDY)}
  \skbound{B2}{( -0.0,\BOUNDY)}
  \skbound{B3}{( 1.0,\BOUNDY)}
  \skvertexp{V1}{(-0.7,-1)}
    \skvertexp{V2}{(0.7,-1)}
  \Gdouble{V1}{B1}
  \Gdouble{V2}{B3}
          \Gdoublearc{V1}{V2}{-45}{-135}  
           \Gdoublearc{V1}{V2}{45}{135}  
                 \skvertexp{V3}{(0 ,-0.65)}
                   \Gdouble{V3}{B2}
  \node[above=2pt] at (B1) {$\k_1$};
  \node[above=2pt] at (B2) {$\k_2$};
  \node[above=2pt] at (B3) {$\k_3$};
    \skvertexp{V1}{(-0.7,-1)}
    \skvertexp{V2}{(0.7,-1)}
    \skvertexp{V3}{(0 ,-0.65)}
    \skbound{B1}{(-1.0,\BOUNDY)}
  \skbound{B2}{( -0.0,\BOUNDY)}
  \skbound{B3}{( 1.0,\BOUNDY)}
\end{tikzpicture}  ,
\\
\psi_{3,2}^{(3)}  (\k_1, \k_2, \k_3) 
&=& \,\,
 \begin{tikzpicture}[baseline=-3.0pt]
  \draw[sk/boundary] (-1.8,\BOUNDY) -- (1.8,\BOUNDY);
  \skbound{B1}{(-1.5,\BOUNDY)}
  \skbound{B2}{( 0.5 ,\BOUNDY)}
  \skbound{B3}{( 1.5,\BOUNDY)}
  \skvertexp{V1}{(-1,-1)}
    \skvertexp{V2}{(0,-1)}
      \skvertexp{V3}{(1 ,-1)}
  \Gdouble{V1}{B1}
  \Gdouble{V3}{B2}
    \Gdouble{V3}{B3}
    \Gdouble{V2}{V3}
     \Gdoublearc{V1}{V2}{-45}{-135}  
          \Gdoublearc{V1}{V2}{45}{135}  
      \skbound{B1}{(-1.5,\BOUNDY)}
  \skbound{B2}{(0.5,\BOUNDY)}
  \skbound{B3}{( 1.5,\BOUNDY)}
  \skvertexp{V1}{(-1,-1)}
    \skvertexp{V2}{(0,-1)}
      \skvertexp{V3}{(1 ,-1)}
  \node[above=2pt] at (B1) {$\k_1$};
  \node[above=2pt] at (B2) {$\k_2$};
  \node[above=2pt] at (B3) {$\k_3$};
\end{tikzpicture} 
,
\\
\psi_{3,3}^{(3)}  (\k_1, \k_2, \k_3) 
&=& \,\,
 \begin{tikzpicture}[baseline=-3.0pt]
  \draw[sk/boundary] (-1.8,\BOUNDY) -- (0.8,\BOUNDY);
  \skbound{B1}{(-1.5,\BOUNDY)}
  \skbound{B2}{( -0.5,\BOUNDY)}
  \skbound{B3}{( 0.5,\BOUNDY)}
  \skvertexp{V1}{(-1,-0.7)}
    \skvertexp{V2}{(-1.0,-1.3)}
      \skvertexp{V3}{(0 ,-0.7)}
  \Gdouble{V1}{B1}
  \Gdouble{V3}{B2}
    \Gdouble{V3}{B3}
    \Gdouble{V1}{V3}
    \Gdouble{V1}{V2}
     \draw[sk/propdouble] (V2.25) to[out=45,in=-45,looseness=20] (V2.-25);
      \skbound{B1}{(-1.5,\BOUNDY)}
  \skbound{B2}{( -0.5,\BOUNDY)}
  \skbound{B3}{( 0.5,\BOUNDY)}
  \skvertexp{V1}{(-1,-0.7)}
     \skvertexp{V2}{(-1.0,-1.3)}
      \skvertexp{V3}{(0 ,-0.7)}
  \node[above=2pt] at (B1) {$\k_1$};
  \node[above=2pt] at (B2) {$\k_2$};
  \node[above=2pt] at (B3) {$\k_3$};
\end{tikzpicture}   
.
\eea
It should be clear that $\psi_{3,2}^{(3)}(\k_1,\k_2,\k_3)$ and
$\psi_{3,3}^{(3)}(\k_1,\k_2,\k_3)$ are symmetric under the interchange
$\k_2 \leftrightarrow \k_3$, but are not symmetric under the interchange of
$\k_1$ with either of the remaining momenta. As a consequence,
Eq.~(\ref{wfc-topology}) implies that
\bea
\psi_3^{(3)}(\k_1,\k_2,\k_3)
&=&
\psi_{3,1}^{(3)}(\k_1,\k_2,\k_3)
+ \psi_{3,2}^{(3)}(\k_1,\k_2,\k_3)
+ \psi_{3,2}^{(3)}(\k_2,\k_1,\k_3)
+ \psi_{3,2}^{(3)}(\k_3,\k_2,\k_1)
\nn\\
&&
+ \psi_{3,3}^{(3)}(\k_1,\k_2,\k_3)
+ \psi_{3,3}^{(3)}(\k_2,\k_1,\k_3)
+ \psi_{3,3}^{(3)}(\k_3,\k_2,\k_1) \, .
\eea
In Appendix~\ref{Sec:App} I provide additional examples of diagrams contributing
to wavefunction coefficients at various orders. These examples will be useful
for the analysis presented in Section~\ref{sec:Examples}.

\subsection{Introducing graphs}

While this step may appear unnecessary, it is useful to enforce the expansion (\ref{wfc-topology}) within each wavefunction coefficient $\psi_n^{(V)}$ appearing in the computation of correlation functions. Doing so leads to a proliferation of boundary vertices, each labeled not only by the number of bulk vertices $V$, but also by the topology $t$ of the corresponding bulk diagram. In practice, this amounts to defining new diagrammatic rules for computing correlation functions, analogous to those
introduced in Section~\ref{sec:wfc-diagrammatic-rules-correlators}, but now with multiple distinct boundary vertices labeled by $\psi_{n,t}^{(V)}$:
\be
\begin{tikzpicture}[baseline=-3.0pt]
  \skvertexgd{v}{(0,0)}
  \draw[sk/propdashed] (v) -- (-1.0, 0.0);
  \draw[sk/propdashed] (v) -- (-0.4, 0.7);
  \draw[sk/propdashed] (v) -- ( 0.4, 0.7);
  \draw[sk/propdashed] (v) -- (-0.4,-0.7); 
  \draw[sk/propdashed] (v) -- ( 1.0, 0.0);
   \node[above=2pt] at (0.3,-1.0) {$\psi_{n,t}^{(V)}$};
\end{tikzpicture}
\quad 
\longrightarrow  
\quad 
(2\pi)^3 \delta^{(3)}(\k_1+\cdots+\k_n)
\Big[ 2\,{\rm Re}\,\psi^{(V)'}_{n,t}(\k_1,\cdots,\k_n) \Big] .
\ee
Note that, once this decomposition is implemented, the resulting boundary vertices are no longer symmetric under permutations of the incoming momenta.

\begin{figure}[t!]
\centering
\includegraphics[width=\linewidth]{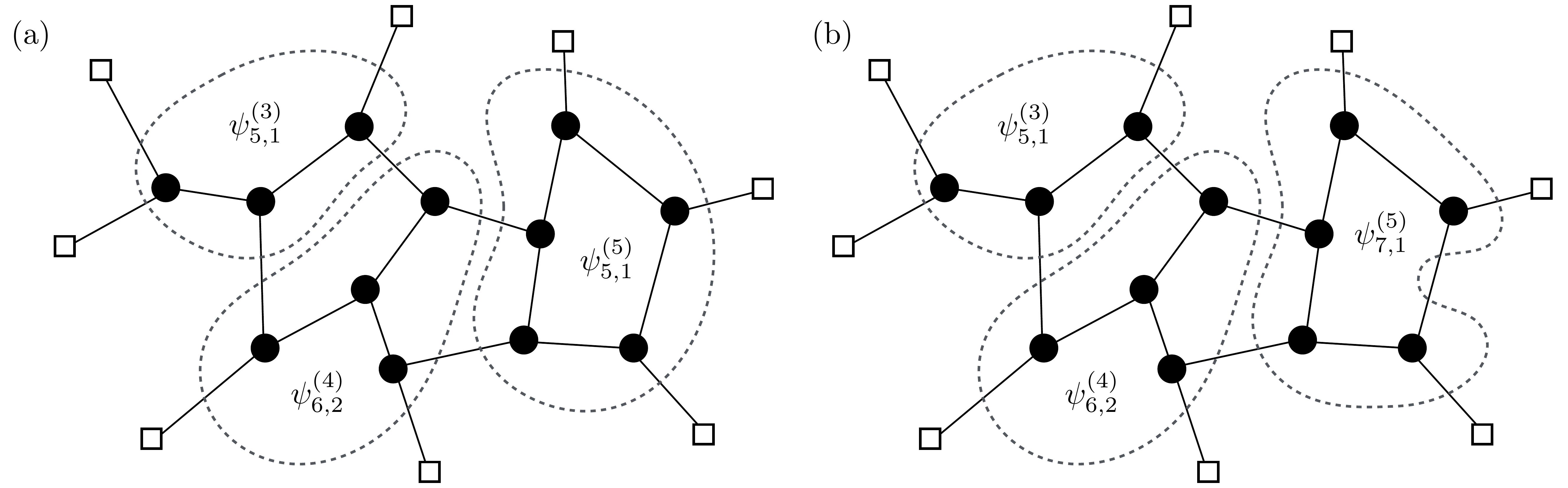}
\caption{\footnotesize (a) A graph with $V=12$ vertices and $n=8$ external legs partitioned into $P=3$ groups. Each group is labeled by the number of enclosed vertices and the number of legs intersecting the corresponding partition boundary. (b) The same graph, but now allowing the partition to cross internal legs connecting vertices within the same group. This changes the labeling of the corresponding partition from $n_p=5$ to $n_p=7$.}
\label{figure_01}
\end{figure}

With this refined decomposition in place, let me introduce a convenient way of visualizing an arbitrary diagram contributing to $\big\langle \varphi(\k_1)\cdots\varphi(\k_n) \big\rangle_c^{(V)}$, built from a specific combination of boundary vertices $\psi_{n,t}^{(V)}$ with fixed topology. Consider a fully connected graph of a given topology, constructed from $V$ three-valent bulk vertices and $n$ external legs. Such a graph contains $(3V+n)/2$ internal and external legs. We may partition this graph into $P$ groups of connected vertices, which we label by $p=1,\ldots,P$. Each partition $p$ defines a connected subgraph characterized by the number of enclosed bulk
vertices $V_p$, the number of legs $n_p$ intersecting the boundary of the partition, and the topology $t_p$ of the enclosed subgraph. An example is shown in Fig.~\ref{figure_01}(a).
Each partition $p$ may be interpreted as capturing the internal structure of a diagram contributing to the wavefunction coefficient $\psi_{n_p,t_p}^{(V_p)}$. In the example shown in Fig.~\ref{figure_01}(a), the partition reveals contributions to $\psi_{5,1}^{(3)}$, $\psi_{6,2}^{(4)}$, and $\psi_{5,1}^{(5)}$ (see Appendix~\ref{Sec:App}). 

The procedure described so far does not exhaust all possible combinations of
wavefunction coefficients contributing to
$\big\langle \varphi(\k_1)\cdots\varphi(\k_n) \big\rangle_c^{(V)}$, as it misses
the possibility of generating loops by contracting pairs of external legs
belonging to the same wavefunction coefficient. This additional class of
contributions can be incorporated by allowing partition boundaries to cross an
internal edge of the graph twice. An example of this situation is shown in
Fig.~\ref{figure_01}(b). When such double crossings are allowed, the labeling of
the partition changes, since the number of intersections between the boundary
and the legs increases. In the example of Fig.~\ref{figure_01}(b), this procedure
leads to a different combination of subdiagrams contributing to
$\psi_{5,1}^{(3)}$, $\psi_{6,2}^{(4)}$, and $\psi_{7,1}^{(5)}$. We should allow this type of crossing only as long as the resulting partition continues to
enclose a fully connected set of vertices. This condition is satisfied only if
the crossed edge belongs to a closed path entirely contained within the
partition. By contrast, if a partition encloses a tree-level subgraph, such a
crossing would effectively split the partition into two disconnected
sub-partitions, thereby increasing the total number of partitions. Configurations
of this type are already accounted for by considering larger values of $P$.


\setcounter{equation}{0}
\section{A general map}
\label{sec:General-map}

I now have all the tools required to derive a general map connecting the wavefunction and Schwinger--Keldysh diagrammatic rules. I will show that the Wavefunction of the Universe diagrams contributing to correlation functions can be reorganized, in a unique way, into Schwinger--Keldysh diagrams.

\subsection{Conjugate wavefunction coefficients}

To begin, note that the coefficient vertices $2\,\mathrm{Re}[\psi_n]$ and the $\Delta$-propagators are real quantities. Nevertheless, it is useful to decompose boundary vertices into separate contributions associated with $\psi_n$ and its complex conjugate $\psi_n^{*}$, in the following schematic way:
\be
\begin{tikzpicture}[baseline=-3.0pt]
  \skvertexgd{v}{(0,0)}
  \draw[sk/propdashed] (v) -- (-1.0, 0.0);
  \draw[sk/propdashed] (v) -- (-0.4, 0.7);
  \draw[sk/propdashed] (v) -- ( 0.4, 0.7);
  \draw[sk/propdashed] (v) -- (-0.4,-0.7); 
  \draw[sk/propdashed] (v) -- ( 1.0, 0.0);
   \node[above=2pt] at (0.3,-0.8) {$\psi_{n}$};
\end{tikzpicture}
=
\begin{tikzpicture}[baseline=-3.0pt]
  \skvertexpdot{v}{(0,0)}
  \draw[sk/propdashed] (v) -- (-1.0, 0.0);
  \draw[sk/propdashed] (v) -- (-0.4, 0.7);
  \draw[sk/propdashed] (v) -- ( 0.4, 0.7);
  \draw[sk/propdashed] (v) -- (-0.4,-0.7); 
  \draw[sk/propdashed] (v) -- ( 1.0, 0.0);
   \node[above=2pt] at (0.3,-0.8) {$\psi_{n}$};
\end{tikzpicture} 
+ 
\begin{tikzpicture}[baseline=-3.0pt]
  \skvertexd{v}{(0,0)}
  \draw[sk/propdashed] (v) -- (-1.0, 0.0);
  \draw[sk/propdashed] (v) -- (-0.4, 0.7);
  \draw[sk/propdashed] (v) -- ( 0.4, 0.7);
  \draw[sk/propdashed] (v) -- (-0.4,-0.7); 
  \draw[sk/propdashed] (v) -- ( 1.0, 0.0);
   \node[above=2pt] at (0.3,-0.8) {$\psi_{n}^*$};
\end{tikzpicture}
.
\ee
Here, the black $n$-legged vertex is generated by the coefficient $\psi_n$, whereas the white vertex is generated by its complex conjugate $\psi_n^{*}$, according to the following assignments:
\bea
\begin{tikzpicture}[baseline=-3.0pt]
  \skvertexpdot{v}{(0,0)}
  \draw[sk/propdashed] (v) -- (-1.0, 0.0);
  \draw[sk/propdashed] (v) -- (-0.4, 0.7);
  \draw[sk/propdashed] (v) -- ( 0.4, 0.7);
  \draw[sk/propdashed] (v) -- (-0.4,-0.7); 
  \draw[sk/propdashed] (v) -- ( 1.0, 0.0);
   \node[above=2pt] at (0.3,-0.8) {$\psi_{n}$};
\end{tikzpicture}
\quad 
& \longrightarrow  &
\quad 
(2 \pi)^{3} \delta^{(3)} (\k_1 + \cdots + \k_n)\, \psi'_n(\k_1,\cdots,\k_n)  ,\\
\begin{tikzpicture}[baseline=-3.0pt]
  \skvertexd{v}{(0,0)}
  \draw[sk/propdashed] (v) -- (-1.0, 0.0);
  \draw[sk/propdashed] (v) -- (-0.4, 0.7);
  \draw[sk/propdashed] (v) -- ( 0.4, 0.7);
  \draw[sk/propdashed] (v) -- (-0.4,-0.7); 
  \draw[sk/propdashed] (v) -- ( 1.0, 0.0);
   \node[above=2pt] at (0.3,-0.8) {$\psi_{n}^*$};
\end{tikzpicture}
\quad 
 &\longrightarrow  &
\quad 
(2 \pi)^{3} \delta^{(3)} (\k_1 + \cdots + \k_n)\, \big[\psi'_n(\k_1,\cdots,\k_n)\big]^{*}  .
\eea
This splitting allows any diagram contributing to a correlation function to be expanded into a sum of diagrams containing black and white vertices, representing wavefunction coefficients and their complex conjugates. As an illustration, the $V=2$ contribution to the four-point correlator in Eq.~(\ref{four-point-example-gray-2}) can be rewritten as
\bea \label{four-point-example-bw}
\Big \langle \varphi (\k_1 ) \cdots  \varphi (\k_4 )  \Big \rangle_c^{(2)} &=& 
 \begin{tikzpicture}[baseline=-3.0pt]
  \draw[sk/boundary] (-1.8,\BOUNDY) -- (1.8,\BOUNDY);
  \skbound{B1}{(-1.5,\BOUNDY)}
  \skbound{B2}{( -0.5,\BOUNDY)}
  \skbound{B3}{( 0.5,\BOUNDY)}
    \skbound{B4}{( 1.5,\BOUNDY)}
  \skvertexpdot{V1}{(0,-1)}
  \Gdashed{V1}{B1}
  \Gdashed{V1}{B2}
  \Gdashed{V1}{B3}
    \Gdashed{V1}{B4}
  \node[above=2pt] at (B1) {$\k_1$};
  \node[above=2pt] at (B2) {$\k_2$};
  \node[above=2pt] at (B3) {$\k_3$};
  \node[above=2pt] at (B4) {$\k_4$};
      \node[above=2pt] at (0.0,-2.0) {$\psi_{4}^{(2)}$};
\end{tikzpicture} 
 +
  \begin{tikzpicture}[baseline=-3.0pt]
  \draw[sk/boundary] (-1.8,\BOUNDY) -- (1.8,\BOUNDY);
  \skbound{B1}{(-1.5,\BOUNDY)}
  \skbound{B2}{( -0.5,\BOUNDY)}
  \skbound{B3}{( 0.5,\BOUNDY)}
    \skbound{B4}{( 1.5,\BOUNDY)}
  \skvertexd{V1}{(0,-1)}
  \Gdashed{V1}{B1}
  \Gdashed{V1}{B2}
  \Gdashed{V1}{B3}
    \Gdashed{V1}{B4}
  \node[above=2pt] at (B1) {$\k_1$};
  \node[above=2pt] at (B2) {$\k_2$};
  \node[above=2pt] at (B3) {$\k_3$};
  \node[above=2pt] at (B4) {$\k_4$};
      \node[above=2pt] at (0.0,-2.0) {$\psi_{4}^{(2)*}$};
\end{tikzpicture}  +
  \begin{tikzpicture}[baseline=-3.0pt]
  \draw[sk/boundary] (-1.8,\BOUNDY) -- (1.8,\BOUNDY);
  \skbound{B1}{(-1.5,\BOUNDY)}
  \skbound{B2}{( -0.5,\BOUNDY)}
  \skbound{B3}{( 0.5,\BOUNDY)}
    \skbound{B4}{( 1.5,\BOUNDY)}
  \skvertexpdot{V1}{(-1,-1)}
    \skvertexpdot{V2}{(1,-1)}
  \Gdashed{V1}{B1}
  \Gdashed{V1}{B2}
  \Gdashed{V2}{B3}
    \Gdashed{V2}{B4}
        \Gdashed{V1}{V2}
  \node[above=2pt] at (B1) {$\k_1$};
  \node[above=2pt] at (B2) {$\k_2$};
  \node[above=2pt] at (B3) {$\k_3$};
  \node[above=2pt] at (B4) {$\k_4$};
     \node[above=2pt] at (-1.0,-2.0) {$\psi_{3}^{(1)}$};
   \node[above=2pt] at (1.0,-2.0) {$\psi_{3}^{(1)}$};
\end{tikzpicture}   
\nn \\
&& \!\!\!\!\!\!\!\!\!\!\!\!\!\!\!\!\!\!\!\!\!\!\!\!
+
  \begin{tikzpicture}[baseline=-3.0pt]
  \draw[sk/boundary] (-1.8,\BOUNDY) -- (1.8,\BOUNDY);
  \skbound{B1}{(-1.5,\BOUNDY)}
  \skbound{B2}{( -0.5,\BOUNDY)}
  \skbound{B3}{( 0.5,\BOUNDY)}
    \skbound{B4}{( 1.5,\BOUNDY)}
  \skvertexd{V1}{(-1,-1)}
    \skvertexd{V2}{(1,-1)}
  \Gdashed{V1}{B1}
  \Gdashed{V1}{B2}
  \Gdashed{V2}{B3}
    \Gdashed{V2}{B4}
        \Gdashed{V1}{V2}
  \node[above=2pt] at (B1) {$\k_1$};
  \node[above=2pt] at (B2) {$\k_2$};
  \node[above=2pt] at (B3) {$\k_3$};
  \node[above=2pt] at (B4) {$\k_4$};
     \node[above=2pt] at (-1.0,-2.0) {$\psi_{3}^{(1)*}$};
   \node[above=2pt] at (1.0,-2.0) {$\psi_{3}^{(1)*}$};
\end{tikzpicture}   
 +
  \begin{tikzpicture}[baseline=-3.0pt]
  \draw[sk/boundary] (-1.8,\BOUNDY) -- (1.8,\BOUNDY);
  \skbound{B1}{(-1.5,\BOUNDY)}
  \skbound{B2}{( -0.5,\BOUNDY)}
  \skbound{B3}{( 0.5,\BOUNDY)}
    \skbound{B4}{( 1.5,\BOUNDY)}
  \skvertexpdot{V1}{(-1,-1)}
    \skvertexd{V2}{(1,-1)}
  \Gdashed{V1}{B1}
  \Gdashed{V1}{B2}
  \Gdashed{V2}{B3}
    \Gdashed{V2}{B4}
        \Gdashed{V1}{V2}
  \node[above=2pt] at (B1) {$\k_1$};
  \node[above=2pt] at (B2) {$\k_2$};
  \node[above=2pt] at (B3) {$\k_3$};
  \node[above=2pt] at (B4) {$\k_4$};
     \node[above=2pt] at (-1.0,-2.0) {$\psi_{3}^{(1)}$};
   \node[above=2pt] at (1.0,-2.0) {$\psi_{3}^{(1)*}$};
\end{tikzpicture}   
+
  \begin{tikzpicture}[baseline=-3.0pt]
  \draw[sk/boundary] (-1.8,\BOUNDY) -- (1.8,\BOUNDY);
  \skbound{B1}{(-1.5,\BOUNDY)}
  \skbound{B2}{( -0.5,\BOUNDY)}
  \skbound{B3}{( 0.5,\BOUNDY)}
    \skbound{B4}{( 1.5,\BOUNDY)}
  \skvertexd{V1}{(-1,-1)}
    \skvertexpdot{V2}{(1,-1)}
  \Gdashed{V1}{B1}
  \Gdashed{V1}{B2}
  \Gdashed{V2}{B3}
    \Gdashed{V2}{B4}
        \Gdashed{V1}{V2}
  \node[above=2pt] at (B1) {$\k_1$};
  \node[above=2pt] at (B2) {$\k_2$};
  \node[above=2pt] at (B3) {$\k_3$};
  \node[above=2pt] at (B4) {$\k_4$};
     \node[above=2pt] at (-1.0,-2.0) {$\psi_{3}^{(1)*}$};
   \node[above=2pt] at (1.0,-2.0) {$\psi_{3}^{(1)}$};
\end{tikzpicture}   
.
\eea

While white boundary vertices simply represent the complex conjugates of black vertices, it will be useful to treat them as distinct objects endowed with their own diagrammatic rules. In other words, we may define rules to compute $\psi_n^{*}$ independently of $\psi_n$. To this end, let us introduce white bulk vertices representing the cubic bulk interaction through the assignment
\be 
\label{bulk-vertex-white}
\begin{tikzpicture}[baseline=-3.0pt]
\coordinate (v) at (0,0);
  \Gdouble{v}{-1.0, 0.0}
  \Gdouble{v}{0.4, 0.7}
  \Gdouble{v}{0.4,-0.7}
   \skvertexm{v}{(0,0)}
\end{tikzpicture} t
\quad
 \longrightarrow  
\quad 
+ i (2 \pi)^{3} \delta^{(3)} (\k_1 + \k_2 + \k_3)
\int_{-\infty}^{t_f} \! dt \, \alpha(t)\, \big[ \cdots \big] .
\ee
This rule is simply the complex-conjugate counterpart of Eq.~(\ref{bulk-vertex-black}). These bulk vertices are joined by bulk-to-bulk propagators obeying
\be \label{G-ww}
t \,\, 
\begin{tikzpicture}[baseline=-3.0pt]
\coordinate (V1) at (-1, 0);
\coordinate (V2) at (1, 0);
     \Gdouble{V1}{V2}
     \skvertexm{v1}{(-1,  0.0)} 
     \skvertexm{v2}{(1,  0.0)} 
\end{tikzpicture}  
\,\, t'
\quad
 \longrightarrow 
 \quad
  G^{*}(k,t,t') ,
\ee
and, finally, we define bulk-to-boundary propagators through the assignment
\be \label{external-K-fourier-white}
\k \,\,
\begin{tikzpicture}[baseline=-3.0pt]
  \skbound{s1}{(0.0,  0.0)} 
   \skvertexm{v1}{(2.0,  0.0)} 
  \draw[sk/propdouble] (s1) -- (v1); 
  \skbound{s1}{(0.0,  0.0)} 
   \skvertexm{v1}{(2.0 ,  0.0)} 
\end{tikzpicture} 
\,
t
\quad 
 \longrightarrow 
\quad  K^{*}(k,t) .
\ee

\subsection{Composite propagators}
\label{sec:new-prop}

Now that we have rules to compute $\psi_n^{*}$, notice that in correlator diagrams bulk vertices can be connected in several distinct ways. In particular, pairs of black bulk vertices contributing to $\psi_n$ are joined by $G$-propagators, whereas pairs of white bulk vertices contributing to $\psi_n^{*}$ are joined by $G^{*}$-propagators. However, bulk vertices belonging to different boundary vertices can also be connected through $\Delta$-propagators. As an illustration, consider the third diagram in Eq.~(\ref{four-point-example-bw}). If we expand the boundary vertices $\psi_{3}^{(1)}$ and $\psi_{3}^{(1)}$ into bulk vertices and bulk-to-boundary propagators, we obtain
\bea \label{exchange-new-prop-1}
  \begin{tikzpicture}[baseline=-3.0pt]
  \draw[sk/boundary] (-1.8,\BOUNDY) -- (1.8,\BOUNDY);
  \skbound{B1}{(-1.5,\BOUNDY)}
  \skbound{B2}{( -0.5,\BOUNDY)}
  \skbound{B3}{( 0.5,\BOUNDY)}
    \skbound{B4}{( 1.5,\BOUNDY)}
  \skvertexpdot{V1}{(-1,-1)}
    \skvertexpdot{V2}{(1,-1)}
  \Gdashed{V1}{B1}
  \Gdashed{V1}{B2}
  \Gdashed{V2}{B3}
    \Gdashed{V2}{B4}
        \Gdashed{V1}{V2}
  \node[above=2pt] at (B1) {$\k_1$};
  \node[above=2pt] at (B2) {$\k_2$};
  \node[above=2pt] at (B3) {$\k_3$};
  \node[above=2pt] at (B4) {$\k_4$};
     \node[above=2pt] at (-1.0,-2.0) {$\psi_{3}^{(1)}$};
   \node[above=2pt] at (1.0,-2.0) {$\psi_{3}^{(1)}$};
\end{tikzpicture}   
=
  \begin{tikzpicture}[baseline=-3.0pt]
  \draw[sk/boundary] (-2,\BOUNDY) -- (2,\BOUNDY);
  \skbound{B1}{(-1.7,\BOUNDY)}
  \skbound{B2}{( -0.7,\BOUNDY)}
  \skbound{B3}{( 0.7,\BOUNDY)}
    \skbound{B4}{( 1.7,\BOUNDY)}
  \skvertexp{V1}{(-1.2,-1.5)}
    \skvertexp{V2}{(1.2,-1.5)}
        \skbound{L1}{(-0.5,-1.5)}
          \skbound{L2}{(-1.7,-0.7)}
           \skbound{L3}{(-0.7,-0.7)}
            \skbound{R1}{(0.5,-1.5)}
             \skbound{R2}{(1.7,-0.7)}
           \skbound{R3}{(0.7,-0.7)}
           \Gdouble{V1}{L1}
            \Gdashed{L1}{R1}
             \Gdouble{R1}{V2}
               \Gdouble{V1}{L2}
  \Gdouble{V1}{L3}
  \Gdouble{V2}{R2}
    \Gdouble{V2}{R3}
     \Gdashed{L2}{B1}
      \Gdashed{L3}{B2}
      \Gdashed{R3}{B3}
      \Gdashed{R2}{B4}
  \node[above=2pt] at (B1) {$\k_1$};
  \node[above=2pt] at (B2) {$\k_2$};
  \node[above=2pt] at (B3) {$\k_3$};
  \node[above=2pt] at (B4) {$\k_4$};
  \skvertexp{V1}{(-1.2,-1.5)}
    \skvertexp{V2}{(1.2,-1.5)}
        \skbound{L1}{(-0.5,-1.5)}
          \skbound{L2}{(-1.7,-0.7)}
           \skbound{L3}{(-0.7,-0.7)}
            \skbound{R1}{(0.5,-1.5)}
             \skbound{R2}{(1.7,-0.7)}
           \skbound{R3}{(0.7,-0.7)}
\end{tikzpicture}  
.
\eea
Similarly, expanding the boundary vertices $\psi_{3}^{(1)}$ and $\psi_{3}^{(1)*}$ appearing in the fifth diagram of Eq.~(\ref{four-point-example-bw}), one finds:
\bea  \label{exchange-new-prop-2}
  \begin{tikzpicture}[baseline=-3.0pt]
  \draw[sk/boundary] (-1.8,\BOUNDY) -- (1.8,\BOUNDY);
  \skbound{B1}{(-1.5,\BOUNDY)}
  \skbound{B2}{( -0.5,\BOUNDY)}
  \skbound{B3}{( 0.5,\BOUNDY)}
    \skbound{B4}{( 1.5,\BOUNDY)}
  \skvertexpdot{V1}{(-1,-1)}
    \skvertexd{V2}{(1,-1)}
  \Gdashed{V1}{B1}
  \Gdashed{V1}{B2}
  \Gdashed{V2}{B3}
    \Gdashed{V2}{B4}
        \Gdashed{V1}{V2}
  \node[above=2pt] at (B1) {$\k_1$};
  \node[above=2pt] at (B2) {$\k_2$};
  \node[above=2pt] at (B3) {$\k_3$};
  \node[above=2pt] at (B4) {$\k_4$};
     \node[above=2pt] at (-1.0,-2.0) {$\psi_{3}^{(1)}$};
   \node[above=2pt] at (1.0,-2.0) {$\psi_{3}^{(1)*}$};
\end{tikzpicture}   
=
  \begin{tikzpicture}[baseline=-3.0pt]
  \draw[sk/boundary] (-2,\BOUNDY) -- (2,\BOUNDY);
  \skbound{B1}{(-1.7,\BOUNDY)}
  \skbound{B2}{( -0.7,\BOUNDY)}
  \skbound{B3}{( 0.7,\BOUNDY)}
    \skbound{B4}{( 1.7,\BOUNDY)}
  \skvertexp{V1}{(-1.2,-1.5)}
    \skvertexm{V2}{(1.2,-1.5)}
        \skbound{L1}{(-0.5,-1.5)}
          \skbound{L2}{(-1.7,-0.7)}
           \skbound{L3}{(-0.7,-0.7)}
            \skbound{R1}{(0.5,-1.5)}
             \skbound{R2}{(1.7,-0.7)}
           \skbound{R3}{(0.7,-0.7)}
           \Gdouble{V1}{L1}
            \Gdashed{L1}{R1}
             \Gdouble{R1}{V2}
               \Gdouble{V1}{L2}
  \Gdouble{V1}{L3}
  \Gdouble{V2}{R2}
    \Gdouble{V2}{R3}
     \Gdashed{L2}{B1}
      \Gdashed{L3}{B2}
      \Gdashed{R3}{B3}
      \Gdashed{R2}{B4}
  \node[above=2pt] at (B1) {$\k_1$};
  \node[above=2pt] at (B2) {$\k_2$};
  \node[above=2pt] at (B3) {$\k_3$};
  \node[above=2pt] at (B4) {$\k_4$};
    \skvertexp{V1}{(-1.2,-1.5)}
    \skvertexm{V2}{(1.2,-1.5)}
        \skbound{L1}{(-0.5,-1.5)}
          \skbound{L2}{(-1.7,-0.7)}
           \skbound{L3}{(-0.7,-0.7)}
            \skbound{R1}{(0.5,-1.5)}
             \skbound{R2}{(1.7,-0.7)}
           \skbound{R3}{(0.7,-0.7)}
\end{tikzpicture}  
.
\eea
These mixings of $K$-propagators and $\Delta$-propagators motivate the introduction of new composite propagators, defined diagrammatically as follows:
\bea
  \begin{tikzpicture}[baseline=-3.0pt]
  \skvertexp{V1}{(-1.0,0)}
    \skvertexp{V2}{(1.0,0)}
             \Gmix{V1}{V2}
\end{tikzpicture} 
\quad
&\equiv&
\quad
  \begin{tikzpicture}[baseline=-3.0pt]
  \skvertexp{V1}{(-1.2,0)}
    \skvertexp{V2}{(1.2,0)}
        \skbound{L1}{(-0.5,0)}
            \skbound{R1}{(0.5,0)}
           \Gdouble{V1}{L1}
            \Gdashed{L1}{R1}
             \Gdouble{R1}{V2}
             \skvertexp{V1}{(-1.2,0)}
    \skvertexp{V2}{(1.2,0)}
        \skbound{L1}{(-0.5,0)}
            \skbound{R1}{(0.5,0)}
\end{tikzpicture}  
,
\\ 
  \begin{tikzpicture}[baseline=-3.0pt]
  \skvertexm{V1}{(-1,0)}
    \skvertexm{V2}{(1,0)}
             \Gmix{V1}{V2}
\end{tikzpicture} 
\quad
&\equiv&
\quad
  \begin{tikzpicture}[baseline=-3.0pt]
  \skvertexm{V1}{(-1.2,0)}
    \skvertexm{V2}{(1.2,0)}
        \skbound{L1}{(-0.5,0)}
            \skbound{R1}{(0.5,0)}
           \Gdouble{V1}{L1}
            \Gdashed{L1}{R1}
             \Gdouble{R1}{V2}
              \skvertexm{V1}{(-1.2,0)}
    \skvertexm{V2}{(1.2,0)}
        \skbound{L1}{(-0.5,0)}
            \skbound{R1}{(0.5,0)}
\end{tikzpicture}
,  
\\
  \begin{tikzpicture}[baseline=-3.0pt]
  \skvertexp{V1}{(-1,0)}
    \skvertexm{V2}{(1,0)}
             \G{V1}{V2}
\end{tikzpicture} 
\quad
&\equiv&
\quad
  \begin{tikzpicture}[baseline=-3.0pt]
  \skvertexp{V1}{(-1.2,0)}
    \skvertexm{V2}{(1.2,0)}
        \skbound{L1}{(-0.5,0)}
            \skbound{R1}{(0.5,0)}
           \Gdouble{V1}{L1}
            \Gdashed{L1}{R1}
             \Gdouble{R1}{V2}
             \skvertexp{V1}{(-1.2,0)}
    \skvertexm{V2}{(1.2,0)}
        \skbound{L1}{(-0.5,0)}
            \skbound{R1}{(0.5,0)}
\end{tikzpicture}  
,
\\
  \begin{tikzpicture}[baseline=-3.0pt]
  \skvertexm{V1}{(-1,0)}
    \skvertexp{V2}{(1,0)}
             \G{V1}{V2}
\end{tikzpicture} 
\quad
&\equiv&
\quad
  \begin{tikzpicture}[baseline=-3.0pt]
  \skvertexm{V1}{(-1.2,0)}
    \skvertexp{V2}{(1.2,0)}
        \skbound{L1}{(-0.5,0)}
            \skbound{R1}{(0.5,0)}
           \Gdouble{V1}{L1}
            \Gdashed{L1}{R1}
             \Gdouble{R1}{V2}
             \skvertexm{V1}{(-1.2,0)}
    \skvertexp{V2}{(1.2,0)}
        \skbound{L1}{(-0.5,0)}
            \skbound{R1}{(0.5,0)}
\end{tikzpicture}  
, \\
\begin{tikzpicture}[baseline=-3.0pt]
  \skbound{s1}{(-1,  0.0)} 
   \skvertexp{v1}{(1,  0.0)} 
  \draw[sk/prop] (s1) -- (v1); 
\end{tikzpicture} 
\quad 
&\equiv&
\quad 
\begin{tikzpicture}[baseline=-3.0pt]
  \skbound{s1}{(-1.2,  0.0)} 
   \skvertexp{v1}{(1.24,  0.0)} 
      \skbound{v}{(0.0,  0.0)} 
  \draw[sk/propdashed] (s1) -- (v); 
   \draw[sk/propdouble] (v1) -- (v); 
  \skbound{s1}{(-1.2,  0.0)} 
   \skvertexp{v1}{(1.24,  0.0)} 
    \skbound{v}{(0.0,  0.0)} 
\end{tikzpicture} 
,
 \\
\begin{tikzpicture}[baseline=-3.0pt]
  \skbound{s1}{(-1,  0.0)} 
   \skvertexm{v1}{(1,  0.0)} 
  \draw[sk/prop] (s1) -- (v1); 
\end{tikzpicture} 
\quad 
&\equiv&
\quad 
\begin{tikzpicture}[baseline=-3.0pt]
  \skbound{s1}{(-1.2,  0.0)} 
   \skvertexm{v1}{(1.24,  0.0)} 
      \skbound{v}{(0.0,  0.0)} 
  \draw[sk/propdashed] (s1) -- (v); 
   \draw[sk/propdouble] (v1) -- (v); 
  \skbound{s1}{(-1.2,  0.0)} 
   \skvertexm{v1}{(1.24,  0.0)} 
   \skbound{v}{(0.0,  0.0)} 
\end{tikzpicture} .
\eea
Note that I have deliberately kept propagators joining bulk vertices of the same color (but belonging to different coefficients) as double lines (solid--dashed double lines), while propagators joining vertices of different colors are represented by a single line. The same convention applies to bulk-to-boundary propagators. With these definitions, the diagram in Eq.~(\ref{exchange-new-prop-1}) can be rewritten as
\be
 \begin{tikzpicture}[baseline=-3.0pt]
  \draw[sk/boundary] (-1.8,\BOUNDY) -- (1.8,\BOUNDY);
  \skbound{B1}{(-1.5,\BOUNDY)}
  \skbound{B2}{( -0.5,\BOUNDY)}
  \skbound{B3}{( 0.5,\BOUNDY)}
    \skbound{B4}{( 1.5,\BOUNDY)}
  \skvertexpdot{V1}{(-1,-1)}
    \skvertexpdot{V2}{(1,-1)}
  \Gdashed{V1}{B1}
  \Gdashed{V1}{B2}
  \Gdashed{V2}{B3}
    \Gdashed{V2}{B4}
        \Gdashed{V1}{V2}
  \node[above=2pt] at (B1) {$\k_1$};
  \node[above=2pt] at (B2) {$\k_2$};
  \node[above=2pt] at (B3) {$\k_3$};
  \node[above=2pt] at (B4) {$\k_4$};
     \node[above=2pt] at (-1.0,-2.0) {$\psi_{3}^{(1)}$};
   \node[above=2pt] at (1.0,-2.0) {$\psi_{3}^{(1)}$};
\end{tikzpicture}   
=
 \begin{tikzpicture}[baseline=-3.0pt]
  \draw[sk/boundary] (-1.8,\BOUNDY) -- (1.8,\BOUNDY);
  \skbound{B1}{(-1.5,\BOUNDY)}
  \skbound{B2}{( -0.5,\BOUNDY)}
  \skbound{B3}{( 0.5,\BOUNDY)}
    \skbound{B4}{( 1.5,\BOUNDY)}
  \skvertexp{V1}{(-1,-1)}
    \skvertexp{V2}{(1,-1)}
  \G{V1}{B1}
  \G{V1}{B2}
  \G{V2}{B3}
    \G{V2}{B4}
        \Gmix{V1}{V2}
  \node[above=2pt] at (B1) {$\k_1$};
  \node[above=2pt] at (B2) {$\k_2$};
  \node[above=2pt] at (B3) {$\k_3$};
  \node[above=2pt] at (B4) {$\k_4$};
\end{tikzpicture}  
.
\ee
On the other hand, the diagram in Eq.~(\ref{exchange-new-prop-2}) can be redrawn as
\be
 \begin{tikzpicture}[baseline=-3.0pt]
  \draw[sk/boundary] (-1.8,\BOUNDY) -- (1.8,\BOUNDY);
  \skbound{B1}{(-1.5,\BOUNDY)}
  \skbound{B2}{( -0.5,\BOUNDY)}
  \skbound{B3}{( 0.5,\BOUNDY)}
    \skbound{B4}{( 1.5,\BOUNDY)}
  \skvertexpdot{V1}{(-1,-1)}
    \skvertexd{V2}{(1,-1)}
  \Gdashed{V1}{B1}
  \Gdashed{V1}{B2}
  \Gdashed{V2}{B3}
    \Gdashed{V2}{B4}
        \Gdashed{V1}{V2}
  \node[above=2pt] at (B1) {$\k_1$};
  \node[above=2pt] at (B2) {$\k_2$};
  \node[above=2pt] at (B3) {$\k_3$};
  \node[above=2pt] at (B4) {$\k_4$};
     \node[above=2pt] at (-1.0,-2.0) {$\psi_{3}^{(1)}$};
   \node[above=2pt] at (1.0,-2.0) {$\psi_{3}^{(1)*}$};
\end{tikzpicture}   
=
 \begin{tikzpicture}[baseline=-3.0pt]
  \draw[sk/boundary] (-1.8,\BOUNDY) -- (1.8,\BOUNDY);
  \skbound{B1}{(-1.5,\BOUNDY)}
  \skbound{B2}{( -0.5,\BOUNDY)}
  \skbound{B3}{( 0.5,\BOUNDY)}
    \skbound{B4}{( 1.5,\BOUNDY)}
  \skvertexp{V1}{(-1,-1)}
    \skvertexm{V2}{(1,-1)}
  \G{V1}{B1}
  \G{V1}{B2}
  \G{V2}{B3}
    \G{V2}{B4}
        \G{V1}{V2}
  \node[above=2pt] at (B1) {$\k_1$};
  \node[above=2pt] at (B2) {$\k_2$};
  \node[above=2pt] at (B3) {$\k_3$};
  \node[above=2pt] at (B4) {$\k_4$};
\end{tikzpicture}  
.
\ee

Using Eqs.~(\ref{external-K-fourier}) and (\ref{external-K-fourier-white}), it is straightforward to verify that the analytic rules associated with these composite propagators are
\bea
t_1 \, \begin{tikzpicture}[baseline=-3.0pt]
  \skvertexp{V1}{(-1,0)}
    \skvertexp{V2}{(1,0)}
             \Gmix{V1}{V2}
\end{tikzpicture} 
\,\, t_2 
\quad
&\longrightarrow&  \quad \frac{  \phi_k (t_f) }{ \phi^*_k (t_f)} \phi_k^* (t_1)  \phi_k^* (t_2)  ,  \label{mix-1}
\\ 
t_1 \,  \begin{tikzpicture}[baseline=-3.0pt]
  \skvertexm{V1}{(-1,0)}
    \skvertexm{V2}{(1,0)}
             \Gmix{V1}{V2}
\end{tikzpicture} 
\,\, t_2 
\quad
&\longrightarrow&   \quad  \frac{  \phi^*_k (t_f) }{ \phi_k (t_f)} \phi_k (t_1)  \phi_k (t_2)  ,  \label{mix-2}
\\
t_1 \,  \begin{tikzpicture}[baseline=-3.0pt]
  \skvertexp{V1}{(-1,0)}
    \skvertexm{V2}{(1,0)}
             \G{V1}{V2}
\end{tikzpicture} 
\,\, t_2 
\quad
&\longrightarrow&  \quad \phi_k^* (t_1) \phi_k (t_2) , \label{mix-3}
\\
t_1 \,  \begin{tikzpicture}[baseline=-3.0pt]
  \skvertexm{V1}{(-1,0)}
    \skvertexp{V2}{(1,0)}
             \G{V1}{V2}
\end{tikzpicture} 
\,\, t_2 
\quad
&\longrightarrow&    \quad   \phi_k (t_1) \phi_k^* (t_2) , \label{mix-4}
 \\
\k \,\,\,
\begin{tikzpicture}[baseline=-3.0pt]
  \skbound{s1}{(0.0,  0.0)} 
   \skvertexp{v1}{(2,  0.0)} 
  \draw[sk/prop] (s1) -- (v1); 
\end{tikzpicture} 
\,\,\,
t'
\quad 
 &\longrightarrow &
\quad    \phi_k^* (t')   \phi_k (t_f)  ,     \label{mix-5}
\\
\k \,\,\,
\begin{tikzpicture}[baseline=-3.0pt]
  \skbound{s1}{(0.0,  0.0)} 
   \skvertexm{v1}{(2,  0.0)} 
  \draw[sk/prop] (s1) -- (v1); 
\end{tikzpicture} 
\,\,\,
t'
\quad 
& \longrightarrow &
\quad    \phi_k (t')   \phi_k^* (t_f)  .     \label{mix-6}
\eea
Together with the bulk-to-bulk propagators specified in Eqs.~(\ref{G-bb}) and (\ref{G-ww}), we therefore have a total of eight distinct propagators with which to assemble correlator diagrams built from the two types of bulk vertices.

\subsection{Color-grouping of diagrams}
\label{sec:color-grouping}

The challenge now is to systematically group diagrams in a unique way that reproduces the diagrammatic structure obtained in the Schwinger--Keldysh approach.

To proceed, let us return to the partitioning procedure introduced in Section~\ref{sec:partitioning}. Recall that any collection of diagrams contributing to an $n$-point correlator at order $V$ may be visualized as arising from a partition of a connected graph with $V$ bulk vertices and $n$ external legs. Since each partition can be mapped to the structure of a wavefunction coefficient, we may assign a color (black or white) to the vertices within each partition. The only requirement is that all vertices belonging to a given partition share the same color. Consequently, for a graph with $P$ partitions, there are $2^{P}$ possible colorings.

\begin{figure}[b!]
\centering
\includegraphics[width=\linewidth]{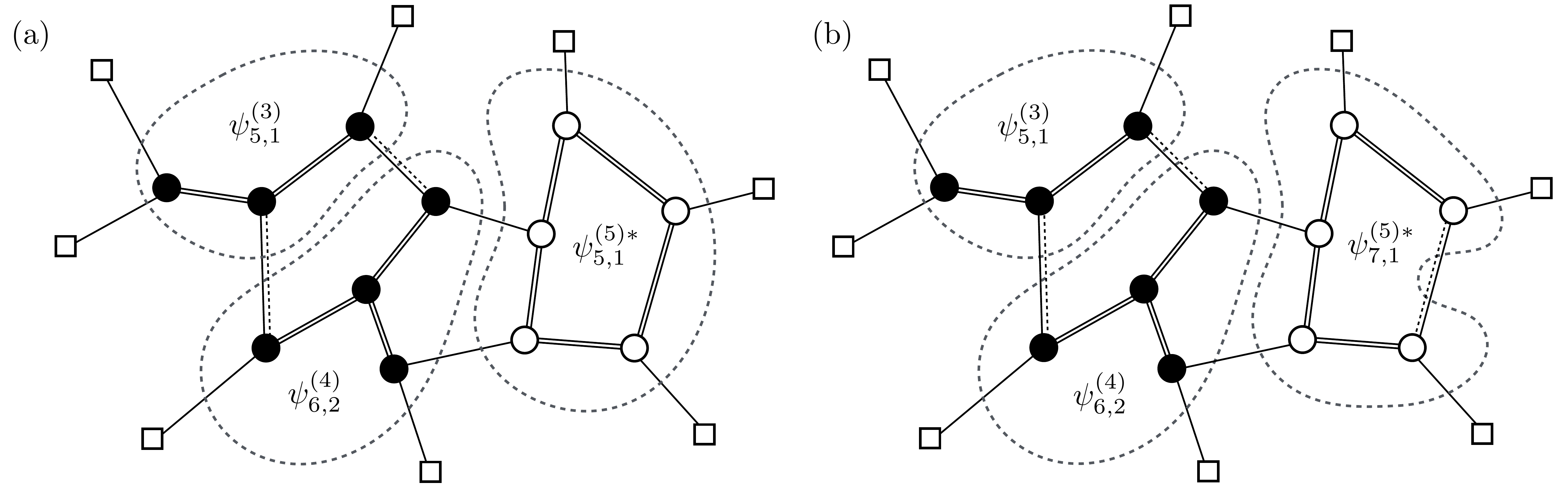}
\caption{\footnotesize (a) A possible coloring of the partition examined in Fig.~\ref{figure_01}. (b) The edge belonging to the partition $\psi_7^{(5)}$, which crosses the partition boundary twice, must be denoted with a double solid--dashed line.}
\label{figure_02}
\end{figure}

Once a coloring is chosen, vertices may be connected by edges representing the propagators introduced in Section~\ref{sec:new-prop}, according to the following rules:
\begin{itemize}

\item Edges fully enclosed within a partition must be drawn as double solid lines, representing bulk-to-bulk propagators, either (\ref{G-bb}) or (\ref{G-ww}).

\item Edges joining vertices of the same color but belonging to different partitions must be drawn as double solid--dashed lines, representing the composite propagators (\ref{mix-1}) and (\ref{mix-2}) connecting different wavefunction coefficients.

\item Edges joining vertices of different colors must be represented by single solid lines, corresponding to the composite propagators (\ref{mix-3}) and (\ref{mix-4}).

\item Finally, edges that cross the boundary of a given partition twice must also be represented by double solid--dashed lines, corresponding to propagators that form loops whose endpoints attach to the same wavefunction coefficient.

\end{itemize}
Figure~\ref{figure_02} illustrates these rules for one particular coloring of the two partitions previously examined in Fig.~\ref{figure_01}. 

\begin{figure}[t!]
\centering
\includegraphics[width=\linewidth]{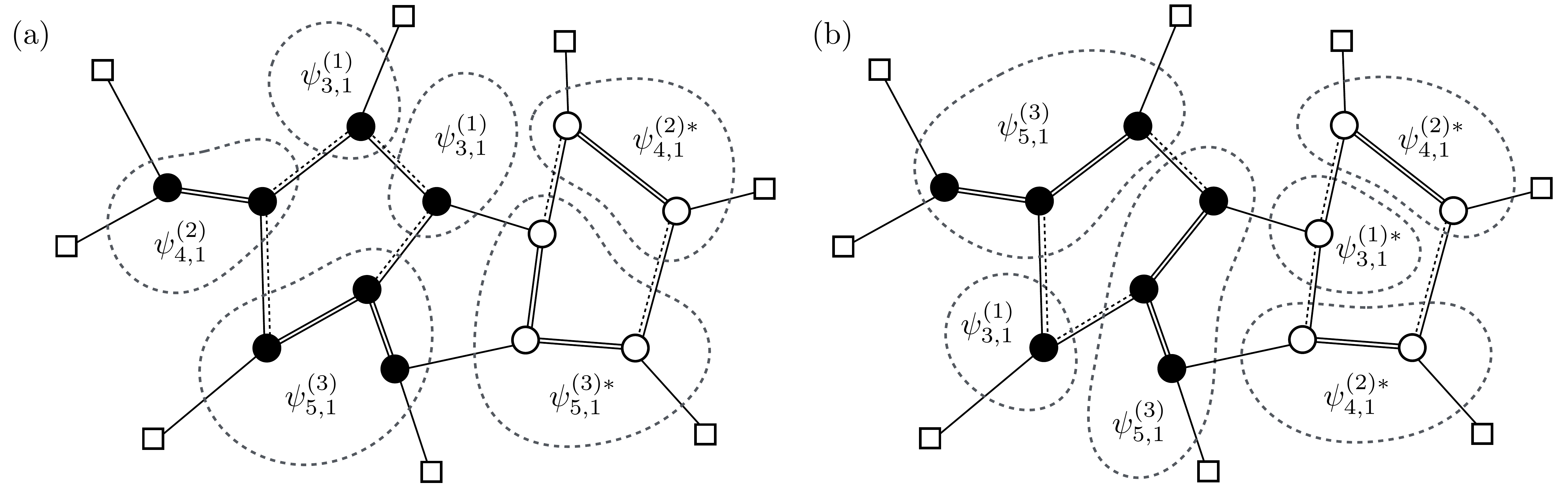}
\caption{\footnotesize Two examples of the same coloring pattern as in Fig.~\ref{figure_02}, but with different choices of partitions.}
\label{figure_03}
\end{figure}

Crucially, for a fixed coloring scheme, the same graph may admit different partitions, provided that no partition encloses vertices of different colors. For example, Fig.~\ref{figure_03} shows the same coloring as in Fig.~\ref{figure_02}, but with a different choice of partitions.
Each admissible partition reshuffles the way pairs of same-color vertices are connected, while never affecting links between vertices of opposite color.

\begin{figure}[h!]
\centering
\includegraphics[width=\linewidth]{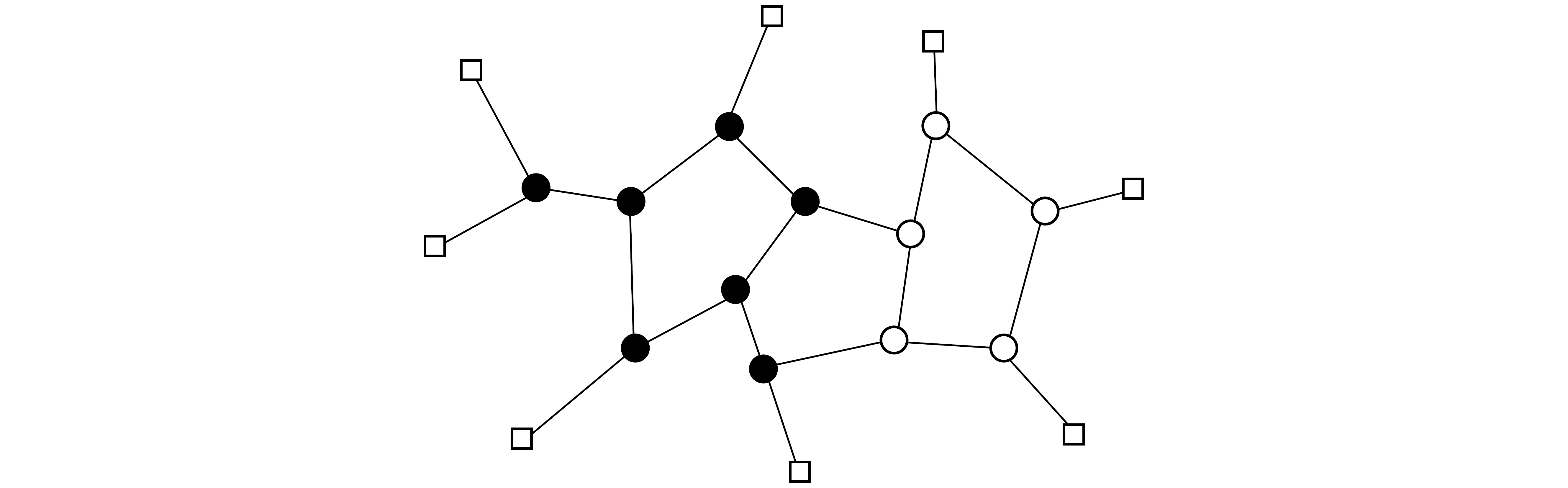}
\caption{\footnotesize A single graph representing the group of diagrams sharing the same coloring scheme, with vertices connected by effective single-line propagators.}
\label{figure_04}
\end{figure}

Now comes the central part of the analysis: For a fixed topology and coloring scheme, a graph represents a collection of boundary diagrams contributing to an $n$-point correlation function at order $V$, \emph{summed together}, with each diagram uniquely specified by a particular admissible partition of the graph. This collection contains every allowed way of drawing propagators consistent with the chosen coloring: a single solid line joining vertices of opposite color, and two distinct double-line structures joining vertices of the same color. Because all diagrams sharing the same topology and coloring scheme are summed together, and because the double-line propagators between same-color vertices appear in all possible combinations exactly once within this sum, the contribution from same-color connections factorizes.
As a result, the two double-line possibilities may be combined into a single propagator, defined as
\bea \label{G-bb-ww-new}
t_1 \,\, 
\begin{tikzpicture}[baseline=-3.0pt]
\coordinate (V1) at (-1, 0);
\coordinate (V2) at (1, 0);
     \G{V1}{V2}
     \skvertexp{v1}{(-1,  0.0)} 
     \skvertexp{v2}{(1,  0.0)} 
\end{tikzpicture}  
\,\, t_2
\quad
&=&
\quad
t_1 \,\, 
\begin{tikzpicture}[baseline=-3.0pt]
\coordinate (V1) at (-1, 0);
\coordinate (V2) at (1, 0);
     \Gdouble{V1}{V2}
     \skvertexp{v1}{(-1,  0.0)} 
     \skvertexp{v2}{(1,  0.0)} 
\end{tikzpicture}  
\,\, t_2
\quad
+
\quad
t_1 \,\, 
\begin{tikzpicture}[baseline=-3.0pt]
\coordinate (V1) at (-1, 0);
\coordinate (V2) at (1, 0);
     \Gmix{V1}{V2}
     \skvertexp{v1}{(-1,  0.0)} 
     \skvertexp{v2}{(1,  0.0)} 
\end{tikzpicture}  
\,\, t_2 , \\
t_1 \,\, 
\begin{tikzpicture}[baseline=-3.0pt]
\coordinate (V1) at (-1, 0);
\coordinate (V2) at (1, 0);
     \G{V1}{V2}
     \skvertexm{v1}{(-1,  0.0)} 
     \skvertexm{v2}{(1,  0.0)} 
\end{tikzpicture}  
\,\, t_2
\quad
&=&
\quad
t_1 \,\, 
\begin{tikzpicture}[baseline=-3.0pt]
\coordinate (V1) at (-1, 0);
\coordinate (V2) at (1, 0);
     \Gdouble{V1}{V2}
     \skvertexm{v1}{(-1,  0.0)} 
     \skvertexm{v2}{(1,  0.0)} 
\end{tikzpicture}  
\,\, t_2
\quad
+
\quad
t_1 \,\, 
\begin{tikzpicture}[baseline=-3.0pt]
\coordinate (V1) at (-1, 0);
\coordinate (V2) at (1, 0);
     \Gmix{V1}{V2}
     \skvertexm{v1}{(-1,  0.0)} 
     \skvertexm{v2}{(1,  0.0)} 
\end{tikzpicture}  
\,\, t_2 .
\eea
With these new single solid-line propagators, each fixed coloring scheme is represented by a single diagram. For instance, the diagrams associated with the partitions shown in Figs.~\ref{figure_02} and~\ref{figure_03} are all encoded in the single graph displayed in Fig.~\ref{figure_04}, which effectively sums over every admissible partition consistent with the constraint that no partition encloses vertices of different colors.

Finally, recalling Eq.~(\ref{b-t-b-k}) for $G(k,t_1,t_2)$, together with the rules (\ref{mix-1}) and (\ref{mix-2}), we find that the single solid lines connecting vertices of the same color obey
\bea 
t_1 \,\, 
\begin{tikzpicture}[baseline=-3.0pt]
\coordinate (V1) at (-1, 0);
\coordinate (V2) at (1, 0);
     \G{V1}{V2}
     \skvertexp{v1}{(-1,  0.0)} 
     \skvertexp{v2}{(1,  0.0)} 
\end{tikzpicture}  
\,\, t_2
\quad
& \longrightarrow &
\quad
 \phi_k (t_2) \phi_k^* (t_1) \theta (t_2 - t_1)
 + \phi_k (t_1) \phi_k^* (t_2)  \theta (t_1 - t_2) , 
 \\
t_1 \,\, 
\begin{tikzpicture}[baseline=-3.0pt]
\coordinate (V1) at (-1, 0);
\coordinate (V2) at (1, 0);
     \G{V1}{V2}
     \skvertexm{v1}{(-1,  0.0)} 
     \skvertexm{v2}{(1,  0.0)} 
\end{tikzpicture}  
\,\, t_2
\quad
& \longrightarrow & 
\quad
 \phi_k^* (t_2) \phi_k (t_1) \theta (t_2 - t_1)
 + \phi_k^* (t_1) \phi_k (t_2)  \theta (t_1 - t_2) .
\eea

\subsection{Schwinger--Keldysh diagrams}

The grouping of diagrams according to their topology and coloring scheme introduced in Section~\ref{sec:color-grouping} is unique. Moreover, we have seen that all diagrams belonging to a given group can be factorized into a single diagram obeying a new set of diagrammatic rules.

Let me now summarize these rules. There are two classes of three-legged bulk vertices, black and white, which obey the following assignments:
\bea
\label{bulk-vertex-black-SK}
\begin{tikzpicture}[baseline=-3.0pt]
\coordinate (v) at (0,0);
  \G{v}{-1.0, 0.0}
  \G{v}{0.4, 0.7}
  \G{v}{0.4,-0.7}
   \skvertexp{v}{(0,0)}
\end{tikzpicture} t
\quad
 &\longrightarrow  &
\quad 
- i (2 \pi)^{3} \delta^{(3)} (\k_1 + \k_2 + \k_3)
\int_{-\infty}^{t_f} dt\, \alpha (t) \Big[ \cdots \Big] , \\
\label{bulk-vertex-white-SK}
\begin{tikzpicture}[baseline=-3.0pt]
\coordinate (v) at (0,0);
  \G{v}{-1.0, 0.0}
  \G{v}{0.4, 0.7}
  \G{v}{0.4,-0.7}
   \skvertexm{v}{(0,0)}
\end{tikzpicture} t
\quad
 &\longrightarrow&  
\quad 
+ i (2 \pi)^{3} \delta^{(3)} (\k_1 + \k_2 + \k_3)
\int_{-\infty}^{t_f} dt \, \alpha (t) \Big[ \cdots \Big] .
\eea
These vertices are joined to each other, and to the boundary at time $t_f$, by bulk-to-bulk and bulk-to-boundary propagators, respectively, according to the following rules:
\bea 
t  \,\, 
\begin{tikzpicture}[baseline=-3.0pt]
\coordinate (V1) at (-1, 0);
\coordinate (V2) at (1, 0);
     \G{V1}{V2}
     \skvertexp{v1}{(-1,  0.0)} 
     \skvertexp{v2}{(1,  0.0)} 
\end{tikzpicture}  
\,\, t'
\quad
& \longrightarrow &
\quad
G_{++}(k,t,t') =
\phi_k(t') \phi_k^*(t)\, \theta(t' - t)
+ \phi_k(t) \phi_k^*(t')\, \theta(t - t') ,  \qquad\\
t \,\, 
\begin{tikzpicture}[baseline=-3.0pt]
\coordinate (V1) at (-1, 0);
\coordinate (V2) at (1, 0);
     \G{V1}{V2}
     \skvertexm{v1}{(-1,  0.0)} 
     \skvertexm{v2}{(1,  0.0)} 
\end{tikzpicture}  
\,\, t'
\quad
& \longrightarrow & 
\quad
G_{--}(k,t,t') =
\phi_k^*(t') \phi_k(t)\, \theta(t' - t)
+ \phi_k^*(t) \phi_k(t')\, \theta(t - t') , \\ 
t \,  
\begin{tikzpicture}[baseline=-3.0pt]
  \skvertexp{V1}{(-1,0)}
  \skvertexm{V2}{(1,0)}
  \G{V1}{V2}
\end{tikzpicture} 
\,\, t'
\quad
&\longrightarrow&  
\quad
G_{+-}(k,t,t') = \phi_k^*(t)\, \phi_k(t') , \\
t \,  
\begin{tikzpicture}[baseline=-3.0pt]
  \skvertexm{V1}{(-1,0)}
  \skvertexp{V2}{(1,0)}
  \G{V1}{V2}
\end{tikzpicture} 
\,\, t'
\quad
&\longrightarrow&    
\quad
G_{-+}(k,t,t') = \phi_k(t)\, \phi_k^*(t') , \\
\k \,\, 
\begin{tikzpicture}[baseline=-3.0pt]
  \skbound{s1}{(0.0,  0.0)} 
  \skvertexp{v1}{(2,  0.0)} 
  \draw[sk/prop] (s1) -- (v1); 
\end{tikzpicture} 
\,\, t
\quad 
& \longrightarrow &
\quad
G_{+}(k,t) = \phi_k^*(t)\, \phi_k(t_f) , \\
\k \,\, 
\begin{tikzpicture}[baseline=-3.0pt]
  \skbound{s1}{(0.0,  0.0)} 
  \skvertexm{v1}{(2,  0.0)} 
  \draw[sk/prop] (s1) -- (v1); 
\end{tikzpicture} 
\,\, t
\quad 
& \longrightarrow &
\quad
G_{-}(k,t) = \phi_k(t)\, \phi_k^*(t_f) .
\eea
These are nothing but the Schwinger--Keldysh rules for computing correlation functions. An $n$-point correlation function at order $V$ consists of the sum of all diagrams constructed using these rules, including all possible color assignments and topologies. This is the promised result.


\section{Examples}
\label{sec:Examples}

In this section I present a few explicit examples illustrating how Schwinger--Keldysh diagrams can be reorganized into Wavefunction of the Universe diagrams, and vice versa.

\subsection{Tree-level four-point function}
\label{sec:Example-1}

Let me begin by revisiting the first example discussed in the introduction, namely the tree-level four-point function shown in Eq.~(\ref{intro:example-1}). Here I will be more careful with the notation and with the underlying diagrammatic structure.

\begin{figure}[b]
\centering
\includegraphics[width=\linewidth]{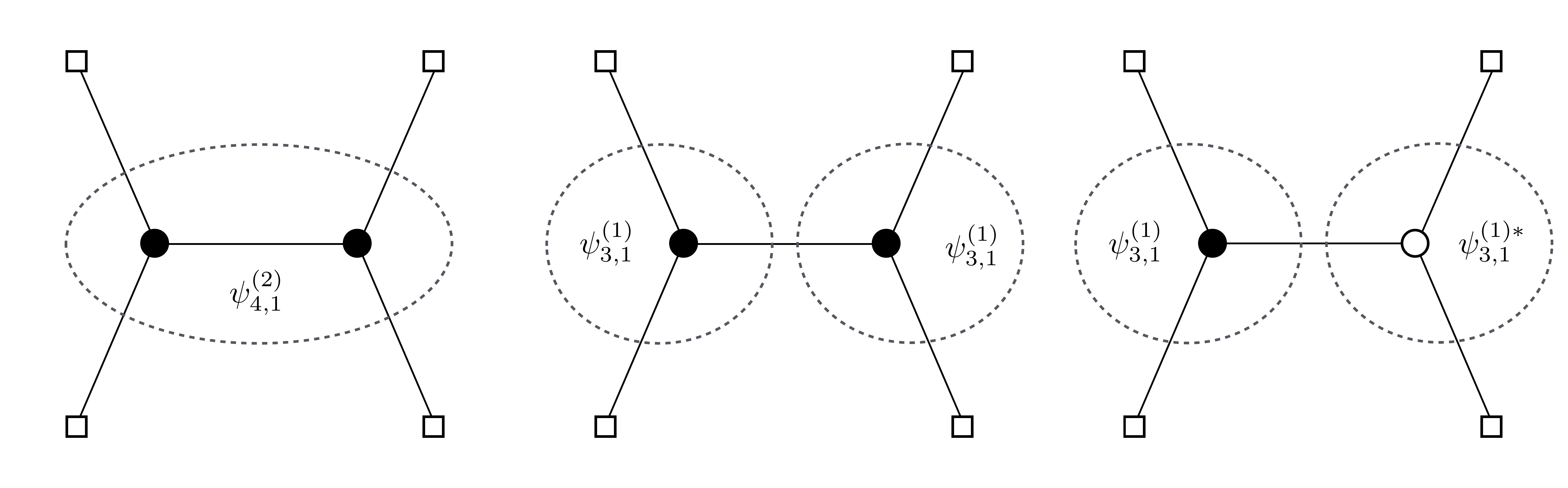}
\caption{\footnotesize  The figure shows the relevant colored graphs and their possible partitions representing diagrams contributing to the tree-level four-point correlation function.}
\label{figure_05}
\end{figure}

Since some of the diagrams contributing to this correlator are complex conjugates of others, it is sufficient to focus on the first and third diagrams on the right-hand side of Eq.~(\ref{intro:example-1}). To rewrite these diagrams in terms of wavefunction coefficients, we simply need to identify how many distinct ways each colored graph can be partitioned. Figure~\ref{figure_05} displays the admissible partitions for each diagram, together with the wavefunction coefficient associated with each partition.
The first diagram in Eq.~(\ref{intro:example-1}), which contains two black vertices, admits two distinct partitionings. The first partioning consists of a single group and corresponds to the four-point wavefunction coefficient $\psi_{4,1}^{(2)}$, constructed from two bulk vertices (see Appendix~\ref{Sec:App}). The second partitioning consists of two disconnected groups, each corresponding to a three-point wavefunction coefficient $\psi_{3,1}^{(1)}$. As a result, the first diagram in Eq.~(\ref{intro:example-1}) can be rewritten in terms of Wavefunction of the Universe diagrams as
\bea 
 \begin{tikzpicture}[baseline=-3.0pt]
  \draw[sk/boundary] (-1.8,\BOUNDY) -- (1.8,\BOUNDY);
  \skbound{B1}{(-1.5,\BOUNDY)}
  \skbound{B2}{( -0.5,\BOUNDY)}
  \skbound{B3}{( 0.5,\BOUNDY)}
    \skbound{B4}{( 1.5,\BOUNDY)}
  \skvertexp{V1}{(-1,-1)}
    \skvertexp{V2}{(1,-1)}
  \G{V1}{B1}
  \G{V1}{B2}
  \G{V2}{B3}
    \G{V2}{B4}
        \G{V1}{V2}
  \node[above=2pt] at (B1) {$\k_1$};
  \node[above=2pt] at (B2) {$\k_2$};
  \node[above=2pt] at (B3) {$\k_3$};
  \node[above=2pt] at (B4) {$\k_4$};
\end{tikzpicture}  
\,\,
= 
\,\,
 \begin{tikzpicture}[baseline=-3.0pt]
  \draw[sk/boundary] (-1.8,\BOUNDY) -- (1.8,\BOUNDY);
  \skbound{B1}{(-1.5,\BOUNDY)}
  \skbound{B2}{( -0.5,\BOUNDY)}
  \skbound{B3}{( 0.5,\BOUNDY)}
    \skbound{B4}{( 1.5,\BOUNDY)}
  \skvertexpdot{V1}{(0,-1)}
  \Gdashed{V1}{B1}
  \Gdashed{V1}{B2}
  \Gdashed{V1}{B3}
    \Gdashed{V1}{B4}
  \node[above=2pt] at (B1) {$\k_1$};
  \node[above=2pt] at (B2) {$\k_2$};
  \node[above=2pt] at (B3) {$\k_3$};
  \node[above=2pt] at (B4) {$\k_4$};
      \node[above=2pt] at (0.6,-1.5) {$\psi_{4,1}^{(2)}$};
\end{tikzpicture} 
+
 \begin{tikzpicture}[baseline=-3.0pt]
  \draw[sk/boundary] (-1.8,\BOUNDY) -- (1.8,\BOUNDY);
  \skbound{B1}{(-1.5,\BOUNDY)}
  \skbound{B2}{( -0.5,\BOUNDY)}
  \skbound{B3}{( 0.5,\BOUNDY)}
    \skbound{B4}{( 1.5,\BOUNDY)}
  \skvertexpdot{V1}{(-1,-1)}
    \skvertexpdot{V2}{(1,-1)}
  \Gdashed{V1}{B1}
  \Gdashed{V1}{B2}
  \Gdashed{V2}{B3}
    \Gdashed{V2}{B4}
        \Gdashed{V1}{V2}
  \node[above=2pt] at (B1) {$\k_1$};
  \node[above=2pt] at (B2) {$\k_2$};
  \node[above=2pt] at (B3) {$\k_3$};
  \node[above=2pt] at (B4) {$\k_4$};
     \node[above=2pt] at (-1.6,-1.5) {$\psi_{3,1}^{(1)}$};
   \node[above=2pt] at (1.6,-1.5) {$\psi_{3,1}^{(1)}$};
\end{tikzpicture}   
. 
\eea
Despite appearances, note that the first diagram on the right-hand side is not fully symmetric under permutations of the external momenta, since $\psi_{4,1}^{(2)}$ itself does not possess full permutation symmetry.

Turning now to the third Schwinger--Keldysh diagram in Eq.~(\ref{intro:example-1}), which contains vertices of opposite color, we find that it admits only a single partition. The two groups in this partition correspond to the three-point wavefunction coefficient $\psi_{3}^{(1)}$ and its complex conjugate $\psi_{3}^{(1)*}$. Consequently, this diagram can only be expressed as
\bea 
 \begin{tikzpicture}[baseline=-3.0pt]
  \draw[sk/boundary] (-1.8,\BOUNDY) -- (1.8,\BOUNDY);
  \skbound{B1}{(-1.5,\BOUNDY)}
  \skbound{B2}{( -0.5,\BOUNDY)}
  \skbound{B3}{( 0.5,\BOUNDY)}
    \skbound{B4}{( 1.5,\BOUNDY)}
  \skvertexp{V1}{(-1,-1)}
    \skvertexm{V2}{(1,-1)}
  \G{V1}{B1}
  \G{V1}{B2}
  \G{V2}{B3}
    \G{V2}{B4}
        \G{V1}{V2}
  \node[above=2pt] at (B1) {$\k_1$};
  \node[above=2pt] at (B2) {$\k_2$};
  \node[above=2pt] at (B3) {$\k_3$};
  \node[above=2pt] at (B4) {$\k_4$};
\end{tikzpicture}   
\,\,
=
 \begin{tikzpicture}[baseline=-3.0pt]
  \draw[sk/boundary] (-1.8,\BOUNDY) -- (1.8,\BOUNDY);
  \skbound{B1}{(-1.5,\BOUNDY)}
  \skbound{B2}{( -0.5,\BOUNDY)}
  \skbound{B3}{( 0.5,\BOUNDY)}
    \skbound{B4}{( 1.5,\BOUNDY)}
  \skvertexpdot{V1}{(-1,-1)}
    \skvertexd{V2}{(1,-1)}
  \Gdashed{V1}{B1}
  \Gdashed{V1}{B2}
  \Gdashed{V2}{B3}
    \Gdashed{V2}{B4}
        \Gdashed{V1}{V2}
  \node[above=2pt] at (B1) {$\k_1$};
  \node[above=2pt] at (B2) {$\k_2$};
  \node[above=2pt] at (B3) {$\k_3$};
  \node[above=2pt] at (B4) {$\k_4$};
     \node[above=2pt] at (-1.6,-1.5) {$\psi_{3,1}^{(1)}$};
   \node[above=2pt] at (1.7,-1.5) {$\psi_{3,1}^{(1)*}$};
\end{tikzpicture}   
.
\eea
Finally, upon summing all contributions, including the appropriate permutations of the external momenta, we recover Eq.~(\ref{four-point-example-gray-2}), which is the wavefunction diagrammatic representation of the tree-level four-point function. It is only after this sum is performed that the fully symmetric coefficient $\psi_{4}^{(2)}$ emerges as the sum of the topology-dependent contributions $\psi_{4,1}^{(2)}$.

\subsection{Three-point correlation function at one-loop}
\label{sec:Example-2}

\begin{figure}[b]
\centering
\includegraphics[width=\linewidth]{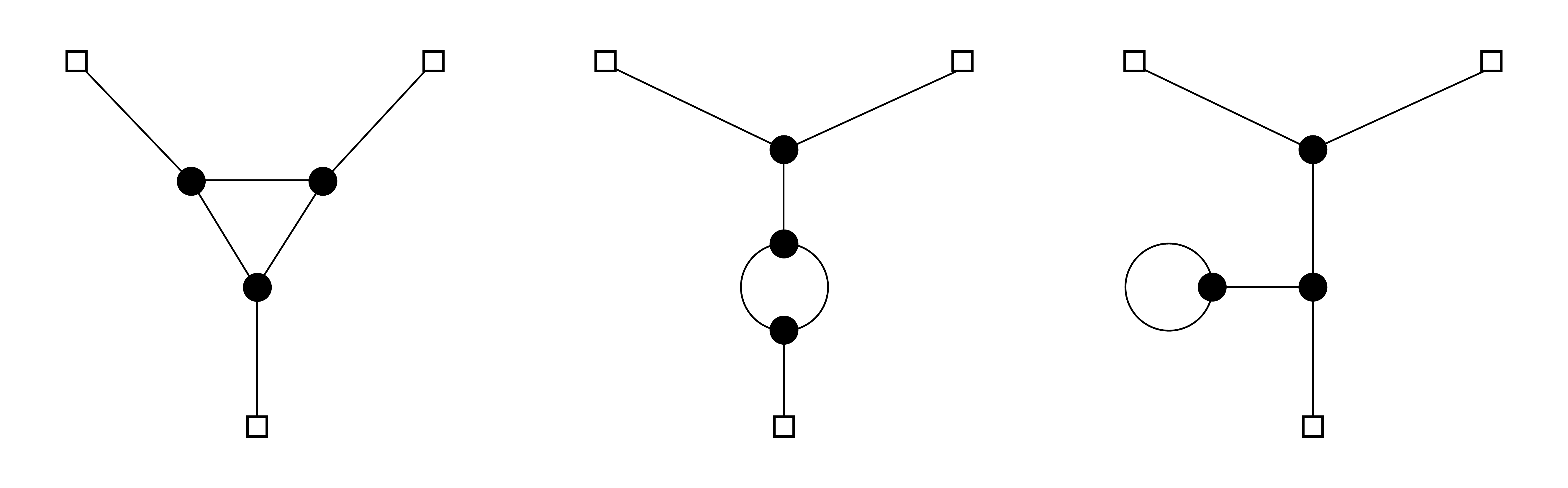}
\caption{\footnotesize The three topologies for diagrams contributing to the three-point function at one-loop.}
\label{figure_06}
\end{figure}

\begin{figure}[h]
\centering
\includegraphics[width=\linewidth]{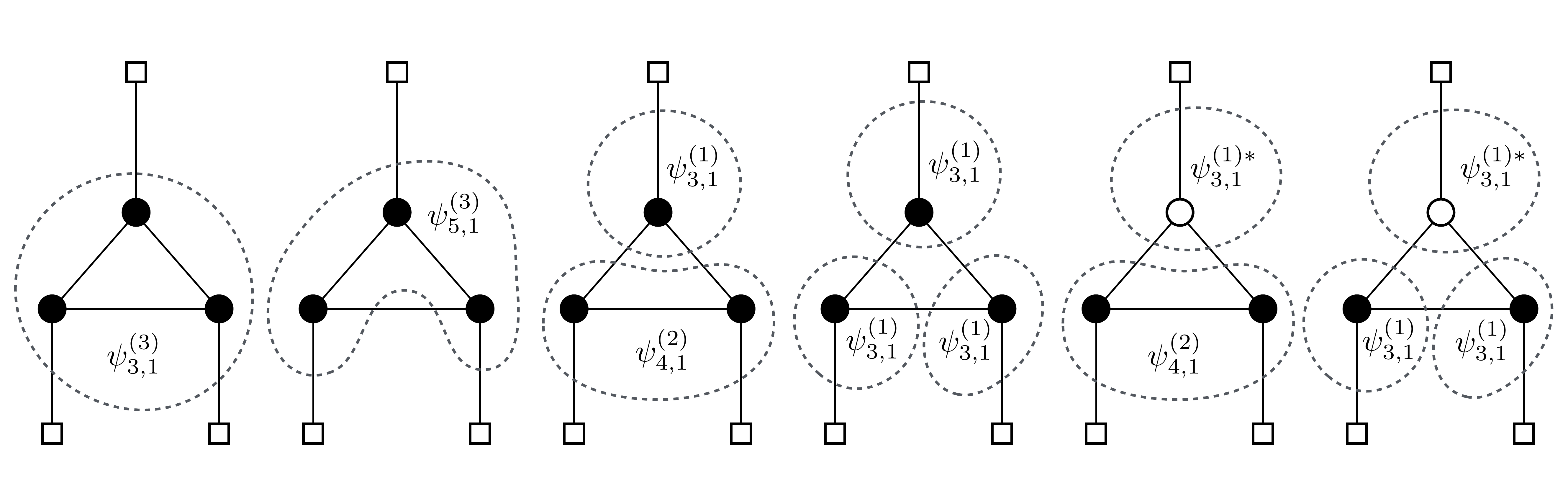}
\caption{\footnotesize The graphs and their possible partitions representing the one-loop three-point correlation function in the Schwinger-Keldysh formalism. The diagram with three black vertices allows four different partitions whereas the diagram with one white vertex and two black vertices leads to two possible partitions.}
\label{figure_07}
\end{figure}

In the Schwinger--Keldysh representation, the one-loop three-point function (equivalently, the contribution with $V=3$ bulk vertices) receives contributions from three distinct diagram topologies. It is therefore convenient to decompose it as
\be \label{three-point-three-toplogies}
\Big\langle \phi(\k_1)\,\phi(\k_2)\,\phi(\k_3) \Big\rangle^{(3)}_c
=
\sum_{t=1}^{3}
\Big\langle \phi(\k_1)\,\phi(\k_2)\,\phi(\k_3) \Big\rangle^{(3)}_{t} \,,
\ee
where $t$ labels the topology of the diagrams contributing to each term. The three possible topologies are displayed in Fig.~\ref{figure_06}.
Notice that the first topology in Fig.~\ref{figure_06} coincides with the second example discussed in the introduction. In what follows, I analyze in detail the map between Schwinger--Keldysh and Wavefunction of the Universe diagrams for this topology, and then simply summarize the corresponding results for the remaining two cases. As reviewed in the introduction, the Schwinger--Keldysh diagrams sharing this topology are
\bea
\label{examples:ex-2}
\Big\langle \phi (\k_1) \phi (\k_2) \phi (\k_3)  \Big\rangle_1 
&=&
\,\,
 \begin{tikzpicture}[baseline=-3.0pt]
  \draw[sk/boundary] (-1.3,\BOUNDY) -- (1.3,\BOUNDY);
  \skbound{B1}{(-1.0,\BOUNDY)}
  \skbound{B2}{( -0.0,\BOUNDY)}
  \skbound{B3}{( 1.0,\BOUNDY)}
  \skvertexp{V1}{(-0.7,-1)}
    \skvertexp{V2}{(0.7,-1)}
      \skvertexp{V3}{(0 ,-0.65)}
  \G{V1}{B1}
  \G{V3}{B2}
  \G{V2}{B3}
          \Garc{V1}{V2}{-45}{-135}  
           \Garc{V1}{V2}{45}{135}  
  \node[above=2pt] at (B1) {$\k_1$};
  \node[above=2pt] at (B2) {$\k_2$};
  \node[above=2pt] at (B3) {$\k_3$};
\end{tikzpicture}  
+ 
\begin{tikzpicture}[baseline=-3.0pt]
  \draw[sk/boundary] (-1.3,\BOUNDY) -- (1.3,\BOUNDY);
  \skbound{B1}{(-1.0,\BOUNDY)}
  \skbound{B2}{( -0.0,\BOUNDY)}
  \skbound{B3}{( 1.0,\BOUNDY)}
  \skvertexm{V1}{(-0.7,-1)}
    \skvertexm{V2}{(0.7,-1)}
  \G{V1}{B1}
  \G{V3}{B2}
  \G{V2}{B3}
          \Garc{V1}{V2}{-45}{-135}  
           \Garc{V1}{V2}{45}{135}  
             \skvertexm{V3}{(0 ,-0.65)}
  \node[above=2pt] at (B1) {$\k_1$};
  \node[above=2pt] at (B2) {$\k_2$};
  \node[above=2pt] at (B3) {$\k_3$};
\end{tikzpicture} 
\nn 
\\
&&
\,\,
 + 
 \begin{tikzpicture}[baseline=-3.0pt]
  \draw[sk/boundary] (-1.3,\BOUNDY) -- (1.3,\BOUNDY);
  \skbound{B1}{(-1.0,\BOUNDY)}
  \skbound{B2}{( -0.0,\BOUNDY)}
  \skbound{B3}{( 1.0,\BOUNDY)}
  \skvertexm{V1}{(-0.7,-1)}
    \skvertexp{V2}{(0.7,-1)}
  \G{V1}{B1}
  \G{V3}{B2}
  \G{V2}{B3}
          \Garc{V1}{V2}{-45}{-135}  
           \Garc{V1}{V2}{45}{135}  
                 \skvertexp{V3}{(0 ,-0.65)}
  \node[above=2pt] at (B1) {$\k_1$};
  \node[above=2pt] at (B2) {$\k_2$};
  \node[above=2pt] at (B3) {$\k_3$};
\end{tikzpicture}  
+ 
\begin{tikzpicture}[baseline=-3.0pt]
  \draw[sk/boundary] (-1.3,\BOUNDY) -- (1.3,\BOUNDY);
  \skbound{B1}{(-1.0,\BOUNDY)}
  \skbound{B2}{( -0.0,\BOUNDY)}
  \skbound{B3}{( 1.0,\BOUNDY)}
  \skvertexp{V1}{(-0.7,-1)}
    \skvertexm{V2}{(0.7,-1)}
  \G{V1}{B1}
  \G{V3}{B2}
  \G{V2}{B3}
          \Garc{V1}{V2}{-45}{-135}  
           \Garc{V1}{V2}{45}{135}  
             \skvertexm{V3}{(0 ,-0.65)}
  \node[above=2pt] at (B1) {$\k_1$};
  \node[above=2pt] at (B2) {$\k_2$};
  \node[above=2pt] at (B3) {$\k_3$};
\end{tikzpicture} 
  + {\rm perms } .
\eea
Note that, for our purposes, there are only two independent diagrams to analyze: the first one, with three black vertices, and the third one, which contains two black vertices and one white vertex. All remaining diagrams are obtained from these two either by complex conjugation and/or by permuting the external momenta. The next step is to enumerate all admissible partitions of the colored graphs representing these diagrams. Figure~\ref{figure_07} displays the possible partitionings, together with the wavefunction coefficient associated with each configuration.

In this example, the first partition gives rise to the coefficient $\psi_{3,1}^{(3)}$. In addition, the second partition allows an internal edge of the graph to cross the partition boundary twice. This produces a diagram in which a pair of legs of the same wavefunction coefficient is contracted into a loop, and the corresponding contribution is therefore encoded by the tree-level five-point coefficient $\psi_{5,1}^{(3)}$ (see Appendix~\ref{Sec:App}). As a result, the first Schwinger--Keldysh diagram on the right-hand side of Eq.~(\ref{intro:example-2}) can be rewritten as the following sum of Wavefunction of the Universe diagrams:
\bea 
 \begin{tikzpicture}[baseline=-3.0pt]
  \draw[sk/boundary] (-1.3,\BOUNDY) -- (1.3,\BOUNDY);
  \skbound{B1}{(-1.0,\BOUNDY)}
  \skbound{B2}{( -0.0,\BOUNDY)}
  \skbound{B3}{( 1.0,\BOUNDY)}
  \skvertexp{V1}{(-0.7,-1)}
    \skvertexp{V2}{(0.7,-1)}
      \skvertexp{V3}{(0 ,-0.65)}
  \G{V1}{B1}
  \G{V3}{B2}
  \G{V2}{B3}
          \Garc{V1}{V2}{-45}{-135}  
           \Garc{V1}{V2}{45}{135}  
  \node[above=2pt] at (B1) {$\k_1$};
  \node[above=2pt] at (B2) {$\k_2$};
  \node[above=2pt] at (B3) {$\k_3$};
\end{tikzpicture}  
\,\,
&=&
\,\,
 \begin{tikzpicture}[baseline=-3.0pt]
  \draw[sk/boundary] (-1.3,\BOUNDY) -- (1.3,\BOUNDY);
  \skbound{B1}{(-1.0,\BOUNDY)}
  \skbound{B2}{( 0.0,\BOUNDY)}
  \skbound{B3}{( 1.0,\BOUNDY)}
  \skvertexpdot{V1}{(0,-1)}
  \Gdashed{V1}{B1}
  \Gdashed{V1}{B2}
  \Gdashed{V1}{B3}
  \node[above=2pt] at (B1) {$\k_1$};
  \node[above=2pt] at (B2) {$\k_2$};
  \node[above=2pt] at (B3) {$\k_3$};
      \node[above=2pt] at (0.6,-1.5) {$\psi_{3,1}^{(3)}$};
\end{tikzpicture} 
+
 \begin{tikzpicture}[baseline=-3.0pt]
  \draw[sk/boundary] (-1.3,\BOUNDY) -- (1.3,\BOUNDY);
  \skbound{B1}{(-1.0,\BOUNDY)}
  \skbound{B2}{( -0.0,\BOUNDY)}
  \skbound{B3}{( 1.0,\BOUNDY)}
  \skvertexpdot{V1}{(0,-1)}
  \Gdashed{V1}{B1}
  \Gdashed{V1}{B2}
  \Gdashed{V1}{B3}
  \node[above=2pt] at (B1) {$\k_1$};
  \node[above=2pt] at (B2) {$\k_2$};
  \node[above=2pt] at (B3) {$\k_3$};
    \node[above=2pt] at (-0.8,-1.7) {$\psi_{5,1}^{(3)}$};
     \draw[sk/propdashed] (V1) to[out=-55,in=-125,looseness=20] (V1);
\end{tikzpicture} 
+
\begin{tikzpicture}[baseline=-3.0pt]
  \draw[sk/boundary] (-1.3,\BOUNDY) -- (1.3,\BOUNDY);
  \skbound{B1}{(-1.0,\BOUNDY)}
  \skbound{B2}{( -0.0,\BOUNDY)}
  \skbound{B3}{( 1.0,\BOUNDY)}
  \skvertexpdot{V1}{(-0.5,-1)}
    \skvertexpdot{V2}{(0.5,-1)}
  \Gdashed{V1}{B1}
  \Gdashed{V1}{B2}
  \Gdashed{V2}{B3}
    \Gdashedarc{V1}{V2}{-25}{-155}  
          \Gdashedarc{V1}{V2}{25}{155}  
  \node[above=2pt] at (B1) {$\k_1$};
  \node[above=2pt] at (B2) {$\k_2$};
  \node[above=2pt] at (B3) {$\k_3$};
   \node[above=2pt] at (-1.1,-1.5) {$\psi_{4,1}^{(2)}$};
   \node[above=2pt] at (1.1,-1.5) {$\psi_{3,1}^{(1)}$};
\end{tikzpicture}  
\nn \\
&&
\,\,
+
\,
 \begin{tikzpicture}[baseline=-3.0pt]
  \draw[sk/boundary] (-1.3,\BOUNDY) -- (1.3,\BOUNDY);
  \skbound{B1}{(-1.0,\BOUNDY)}
  \skbound{B2}{( -0.0,\BOUNDY)}
  \skbound{B3}{( 1.0,\BOUNDY)}
  \skvertexpdot{V1}{(-0.7,-1)}
    \skvertexpdot{V2}{(0.7,-1)}
  \Gdashed{V1}{B1}
  \Gdashed{V3}{B2}
          \Gdashedarc{V1}{V2}{-35}{-145}  
           \Gdashedarc{V1}{V2}{35}{145}  
                \skvertexpdot{V3}{(0 ,-0.7)}
                  \Gdashed{V2}{B3}
  \node[above=2pt] at (B1) {$\k_1$};
  \node[above=2pt] at (B2) {$\k_2$};
  \node[above=2pt] at (B3) {$\k_3$};
     \node[above=2pt] at (-1.3,-1.5) {$\psi_{3,1}^{(1)}$};
   \node[above=2pt] at (1.3,-1.5) {$\psi_{3,1}^{(1)}$};
     \node[above=2pt] at (0.45,-0.85) {$\psi_{3,1}^{(1)}$};
\end{tikzpicture}  
+  {\rm perms} . \quad
\eea
On the other hand, the Schwinger--Keldysh diagram with one white vertex and two black vertices can be rewritten in terms of wavefunction diagrams as
\be
\begin{tikzpicture}[baseline=-3.0pt]
  \draw[sk/boundary] (-1.3,\BOUNDY) -- (1.3,\BOUNDY);
  \skbound{B1}{(-1.0,\BOUNDY)}
  \skbound{B2}{( -0.0,\BOUNDY)}
  \skbound{B3}{( 1.0,\BOUNDY)}
  \skvertexm{V1}{(-0.7,-1)}
    \skvertexp{V2}{(0.7,-1)}
  \G{V1}{B1}
  \G{V3}{B2}
  \G{V2}{B3}
          \Garc{V1}{V2}{-45}{-135}  
           \Garc{V1}{V2}{45}{135}  
                 \skvertexp{V3}{(0 ,-0.65)}
  \node[above=2pt] at (B1) {$\k_1$};
  \node[above=2pt] at (B2) {$\k_2$};
  \node[above=2pt] at (B3) {$\k_3$};
\end{tikzpicture}  
+ 
{\rm perms}
\,\,
=
\,
\begin{tikzpicture}[baseline=-3.0pt]
  \draw[sk/boundary] (-1.3,\BOUNDY) -- (1.3,\BOUNDY);
  \skbound{B1}{(-1.0,\BOUNDY)}
  \skbound{B2}{( -0.0,\BOUNDY)}
  \skbound{B3}{( 1.0,\BOUNDY)}
 \skvertexd{V1}{(-0.5,-1)}
    \skvertexpdot{V2}{(0.5,-1)}
  \Gdashed{V1}{B1}
  \Gdashed{V2}{B2}
  \Gdashed{V2}{B3}
    \Gdashedarc{V1}{V2}{-25}{-155}  
          \Gdashedarc{V1}{V2}{25}{155}  
  \node[above=2pt] at (B1) {$\k_1$};
  \node[above=2pt] at (B2) {$\k_2$};
  \node[above=2pt] at (B3) {$\k_3$};
   \node[above=2pt] at (-1.2,-1.5) {$\psi_{3,1}^{(1) * }$};
   \node[above=2pt] at (1.1,-1.5) {$\psi_{4,1}^{(2)}$};
\end{tikzpicture}  
\,
+
 \begin{tikzpicture}[baseline=-3.0pt]
  \draw[sk/boundary] (-1.3,\BOUNDY) -- (1.3,\BOUNDY);
  \skbound{B1}{(-1.0,\BOUNDY)}
  \skbound{B2}{( -0.0,\BOUNDY)}
  \skbound{B3}{( 1.0,\BOUNDY)}
  \skvertexd{V1}{(-0.7,-1)}
   \skvertexpdot{V2}{(0.7,-1)}
  \Gdashed{V1}{B1}
  \Gdashed{V3}{B2}
          \Gdashedarc{V1}{V2}{-35}{-145}  
           \Gdashedarc{V1}{V2}{35}{145}  
               \skvertexpdot{V3}{(0 ,-0.7)}
                  \Gdashed{V2}{B3}
  \node[above=2pt] at (B1) {$\k_1$};
  \node[above=2pt] at (B2) {$\k_2$};
  \node[above=2pt] at (B3) {$\k_3$};
     \node[above=2pt] at (-1.4,-1.5) {$\psi_{3,1}^{(1)*}$};
   \node[above=2pt] at (1.3,-1.5) {$\psi_{3,1}^{(1)}$};
     \node[above=2pt] at (0.45,-0.85) {$\psi_{3,1}^{(1)}$};
\end{tikzpicture}
+ 
{\rm perms}. 
\ee
With these ingredients in place, we may now sum all Schwinger--Keldysh diagrams entering Eq.~(\ref{intro:example-2}) and reexpress them in terms of Wavefunction of the Universe diagrams. Since each contribution appears exactly once, each wavefunction coefficient combines with its complex conjugate into a real quantity. This yields
\bea 
\label{example-result-2}
\Big\langle \varphi(\k_1) \varphi(\k_2)  \varphi(\k_3) \Big\rangle_1^{(3)}
&=&
\,\,
 \begin{tikzpicture}[baseline=-3.0pt]
  \draw[sk/boundary] (-1.3,\BOUNDY) -- (1.3,\BOUNDY);
  \skbound{B1}{(-1.0,\BOUNDY)}
  \skbound{B2}{( 0.0,\BOUNDY)}
  \skbound{B3}{( 1.0,\BOUNDY)}
  \skvertexgd{V1}{(0,-1)}
  \Gdashed{V1}{B1}
  \Gdashed{V1}{B2}
  \Gdashed{V1}{B3}
  \node[above=2pt] at (B1) {$\k_1$};
  \node[above=2pt] at (B2) {$\k_2$};
  \node[above=2pt] at (B3) {$\k_3$};
      \node[above=2pt] at (0.6,-1.5) {$\psi_{3,1}^{(3)}$};
\end{tikzpicture} 
+
 \begin{tikzpicture}[baseline=-3.0pt]
  \draw[sk/boundary] (-1.3,\BOUNDY) -- (1.3,\BOUNDY);
  \skbound{B1}{(-1.0,\BOUNDY)}
  \skbound{B2}{( -0.0,\BOUNDY)}
  \skbound{B3}{( 1.0,\BOUNDY)}
  \skvertexgd{V1}{(0,-1)}
  \Gdashed{V1}{B1}
  \Gdashed{V1}{B2}
  \Gdashed{V1}{B3}
  \node[above=2pt] at (B1) {$\k_1$};
  \node[above=2pt] at (B2) {$\k_2$};
  \node[above=2pt] at (B3) {$\k_3$};
    \node[above=2pt] at (-0.8,-1.7) {$\psi_{5,1}^{(3)}$};
     \draw[sk/propdashed] (V1) to[out=-55,in=-125,looseness=20] (V1);
\end{tikzpicture} 
+
\begin{tikzpicture}[baseline=-3.0pt]
  \draw[sk/boundary] (-1.3,\BOUNDY) -- (1.3,\BOUNDY);
  \skbound{B1}{(-1.0,\BOUNDY)}
  \skbound{B2}{( -0.0,\BOUNDY)}
  \skbound{B3}{( 1.0,\BOUNDY)}
  \skvertexgd{V1}{(-0.5,-1)}
    \skvertexgd{V2}{(0.5,-1)}
  \Gdashed{V1}{B1}
  \Gdashed{V1}{B2}
  \Gdashed{V2}{B3}
    \Gdashedarc{V1}{V2}{-25}{-155}  
          \Gdashedarc{V1}{V2}{25}{155}  
  \node[above=2pt] at (B1) {$\k_1$};
  \node[above=2pt] at (B2) {$\k_2$};
  \node[above=2pt] at (B3) {$\k_3$};
   \node[above=2pt] at (-1.1,-1.5) {$\psi_{4,1}^{(2)}$};
   \node[above=2pt] at (1.1,-1.5) {$\psi_{3,1}^{(1)}$};
\end{tikzpicture}   
\nn \\[-10pt]
&&
\,\,
+
 \begin{tikzpicture}[baseline=-3.0pt]
  \draw[sk/boundary] (-1.3,\BOUNDY) -- (1.3,\BOUNDY);
  \skbound{B1}{(-1.0,\BOUNDY)}
  \skbound{B2}{( -0.0,\BOUNDY)}
  \skbound{B3}{( 1.0,\BOUNDY)}
  \skvertexgd{V1}{(-0.7,-1)}
    \skvertexgd{V2}{(0.7,-1)}
  \Gdashed{V1}{B1}
  \Gdashed{V3}{B2}
          \Gdashedarc{V1}{V2}{-35}{-145}  
           \Gdashedarc{V1}{V2}{35}{145}  
              \skvertexgd{V3}{(0 ,-0.7)}
                  \Gdashed{V2}{B3}
  \node[above=2pt] at (B1) {$\k_1$};
  \node[above=2pt] at (B2) {$\k_2$};
  \node[above=2pt] at (B3) {$\k_3$};
     \node[above=2pt] at (-1.3,-1.5) {$\psi_{3,1}^{(1)}$};
   \node[above=2pt] at (1.3,-1.5) {$\psi_{3,1}^{(1)}$};
     \node[above=2pt] at (0.45,-0.85) {$\psi_{3,1}^{(1)}$};
\end{tikzpicture}  
+ {\rm perms} . \quad
\eea
This is precisely the result quoted in the introduction.

There is, however, one subtlety in interpreting Eq.~(\ref{example-result-2}). Recall that both $\psi_{4,1}^{(2)}$ and $\psi_{5,1}^{(3)}$ are not fully symmetric under permutations of their arguments (see Appendix~\ref{Sec:App}). This means that, strictly speaking, I should specify which external momenta are assigned to which arguments of these coefficients---that is, which momenta correspond to the legs that connect to the rest of the diagram. For simplicity, this assignment is left implicit.

For completeness, and following the same steps outlined above, it is straightforward to show that the remaining two topologies displayed in Fig.~\ref{figure_06} give rise to the following contributions to the three-point function at third order:
\bea 
\label{example-result-3}
\Big\langle \varphi(\k_1) \varphi(\k_2)  \varphi(\k_3) \Big\rangle_2^{(3)}
&=&
\,\,
 \begin{tikzpicture}[baseline=-3.0pt]
  \draw[sk/boundary] (-1.3,\BOUNDY) -- (1.3,\BOUNDY);
  \skbound{B1}{(-1.0,\BOUNDY)}
  \skbound{B2}{( 0.0,\BOUNDY)}
  \skbound{B3}{( 1.0,\BOUNDY)}
  \skvertexgd{V1}{(0,-1)}
  \Gdashed{V1}{B1}
  \Gdashed{V1}{B2}
  \Gdashed{V1}{B3}
  \node[above=2pt] at (B1) {$\k_1$};
  \node[above=2pt] at (B2) {$\k_2$};
  \node[above=2pt] at (B3) {$\k_3$};
      \node[above=2pt] at (0.6,-1.5) {$\psi_{3,2}^{(3)}$};
\end{tikzpicture} 
+
 \begin{tikzpicture}[baseline=-3.0pt]
  \draw[sk/boundary] (-1.3,\BOUNDY) -- (1.3,\BOUNDY);
  \skbound{B1}{(-1.0,\BOUNDY)}
  \skbound{B2}{( 0.0,\BOUNDY)}
  \skbound{B3}{( 1.0,\BOUNDY)}
  \skvertexgd{V1}{(0,-1)}
  \Gdashed{V1}{B1}
  \Gdashed{V1}{B2}
  \Gdashed{V1}{B3}
  \node[above=2pt] at (B1) {$\k_1$};
  \node[above=2pt] at (B2) {$\k_2$};
  \node[above=2pt] at (B3) {$\k_3$};
      \node[above=2pt] at (0.6,-1.5) {$\psi_{5,1}^{(3)}$};
        \draw[sk/propdashed] (V1) to[out=-55,in=-125,looseness=20] (V1);
\end{tikzpicture} 
+
 \begin{tikzpicture}[baseline=-3.0pt]
  \draw[sk/boundary] (-1.3,\BOUNDY) -- (1.3,\BOUNDY);
  \skbound{B1}{(-1.0,\BOUNDY)}
  \skbound{B2}{( -0.0,\BOUNDY)}
  \skbound{B3}{( 1.0,\BOUNDY)}
  \skvertexgd{V1}{(-0.5,-1)}
    \skvertexgd{V2}{(0.5,-1)}
  \Gdashed{V1}{B1}
  \Gdashed{V1}{B2}
  \Gdashed{V2}{B3}
    \Gdashedarc{V1}{V2}{-25}{-155}  
          \Gdashedarc{V1}{V2}{25}{155}  
  \node[above=2pt] at (B1) {$\k_1$};
  \node[above=2pt] at (B2) {$\k_2$};
  \node[above=2pt] at (B3) {$\k_3$};
   \node[above=2pt] at (-1.1,-1.5) {$\psi_{4,1}^{(2)}$};
   \node[above=2pt] at (1.1,-1.5) {$\psi_{3,1}^{(1)}$};
\end{tikzpicture}  
\nn
\\[-10pt]
&&
\,\,
+
\begin{tikzpicture}[baseline=-3.0pt]
  \draw[sk/boundary] (-1.3,\BOUNDY) -- (1.3,\BOUNDY);
  \skbound{B1}{(-1.0,\BOUNDY)}
  \skbound{B2}{( -0.0,\BOUNDY)}
  \skbound{B3}{( 1.0,\BOUNDY)}
  \skvertexgd{V1}{(-0.5,-1)}
    \skvertexgd{V2}{(0.5,-1)}
  \Gdashed{V1}{B1}
  \Gdashed{V1}{B2}
  \Gdashed{V2}{B3}
  \Gdashed{V1}{V2}
  \node[above=2pt] at (B1) {$\k_1$};
  \node[above=2pt] at (B2) {$\k_2$};
  \node[above=2pt] at (B3) {$\k_3$};
   \node[above=2pt] at (-1.1,-1.5) {$\psi_{3,1}^{(1)}$};
   \node[above=2pt] at (1.1,-1.5) {$\psi_{2,1}^{(2)}$};
\end{tikzpicture}  
+
 \begin{tikzpicture}[baseline=-3.0pt]
  \draw[sk/boundary] (-1.8,\BOUNDY) -- (0.8,\BOUNDY);
  \skbound{B1}{(-1.5,\BOUNDY)}
  \skbound{B2}{( -0.5 ,\BOUNDY)}
  \skbound{B3}{( 0.5,\BOUNDY)}
  \skvertexgd{V1}{(-1.5,-1)}
    \skvertexgd{V2}{(-0.7,-1)}
      \skvertexgd{V3}{(0 ,-1)}
  \Gdashed{V1}{B1}
  \Gdashed{V3}{B2}
    \Gdashed{V2}{V3}
      \Gdashed{V3}{B3}
     \Gdashedarc{V1}{V2}{-45}{-135}  
          \Gdashedarc{V1}{V2}{45}{135}  
  \node[above=2pt] at (B1) {$\k_1$};
  \node[above=2pt] at (B2) {$\k_2$};
  \node[above=2pt] at (B3) {$\k_3$};
  \node[above=2pt] at (-1.7,-2.0) {$\psi_{3,1}^{(1)}$};
   \node[above=2pt] at (-0.7,-2.0) {$\psi_{3,1}^{(1)}$};
    \node[above=2pt] at (0.2,-2.0) {$\psi_{3,1}^{(1)}$};
\end{tikzpicture} 
+ \quad {\rm perms} . \quad
\eea
and
\bea 
\label{example-result-4}
\Big\langle \varphi(\k_1) \varphi(\k_2)  \varphi(\k_3) \Big\rangle_3^{(3)}
&=&
\,\,
 \begin{tikzpicture}[baseline=-3.0pt]
  \draw[sk/boundary] (-1.3,\BOUNDY) -- (1.3,\BOUNDY);
  \skbound{B1}{(-1.0,\BOUNDY)}
  \skbound{B2}{( 0.0,\BOUNDY)}
  \skbound{B3}{( 1.0,\BOUNDY)}
  \skvertexgd{V1}{(0,-1)}
  \Gdashed{V1}{B1}
  \Gdashed{V1}{B2}
  \Gdashed{V1}{B3}
  \node[above=2pt] at (B1) {$\k_1$};
  \node[above=2pt] at (B2) {$\k_2$};
  \node[above=2pt] at (B3) {$\k_3$};
      \node[above=2pt] at (0.6,-1.5) {$\psi_{3,3}^{(3)}$};
\end{tikzpicture} 
+
 \begin{tikzpicture}[baseline=-3.0pt]
  \draw[sk/boundary] (-1.3,\BOUNDY) -- (1.3,\BOUNDY);
  \skbound{B1}{(-1.0,\BOUNDY)}
  \skbound{B2}{( 0.0,\BOUNDY)}
  \skbound{B3}{( 1.0,\BOUNDY)}
  \skvertexgd{V1}{(0,-1)}
  \Gdashed{V1}{B1}
  \Gdashed{V1}{B2}
  \Gdashed{V1}{B3}
  \node[above=2pt] at (B1) {$\k_1$};
  \node[above=2pt] at (B2) {$\k_2$};
  \node[above=2pt] at (B3) {$\k_3$};
      \node[above=2pt] at (0.6,-1.5) {$\psi_{5,1}^{(3)}$};
        \draw[sk/propdashed] (V1) to[out=-55,in=-125,looseness=20] (V1);
\end{tikzpicture} 
+
 \begin{tikzpicture}[baseline=-3.0pt]
  \draw[sk/boundary] (-1.3,\BOUNDY) -- (1.3,\BOUNDY);
  \skbound{B1}{(-1.0,\BOUNDY)}
  \skbound{B2}{( -0.0,\BOUNDY)}
  \skbound{B3}{( 1.0,\BOUNDY)}
  \skvertexgd{V1}{(-0,-0.8)}
  \skvertexgd{V2}{(-0.0,-1.4)}
  \Gdashed{V1}{B1}
  \Gdashed{V1}{B2}
  \Gdashed{V1}{B3}
    \Gdashed{V1}{V2}
     \draw[sk/propdashed] (V2.25) to[out=45,in=-45,looseness=20] (V2.-25);
  \node[above=2pt] at (B1) {$\k_1$};
  \node[above=2pt] at (B2) {$\k_2$};
  \node[above=2pt] at (B3) {$\k_3$};
  \node[above=2pt] at (-0.6,-1.3) {$\psi_{4,1}^{(2)}$};
   \node[above=2pt] at (-0.5,-2.0) {$\psi_{3,1}^{(1)}$};
\end{tikzpicture}  
\nn
\\[-10pt]
&&
\,\,
+
 \begin{tikzpicture}[baseline=-3.0pt]
  \draw[sk/boundary] (-1.8,\BOUNDY) -- (0.8,\BOUNDY);
  \skbound{B1}{(-1.5,\BOUNDY)}
  \skbound{B2}{( -0.5,\BOUNDY)}
  \skbound{B3}{( 0.5,\BOUNDY)}
  \skvertexgd{V1}{(-1,-0.8)}
  \skvertexgd{V2}{(-1.0,-1.4)}
   \skvertexgd{V3}{(0 ,-0.8)}
  \Gdashed{V1}{B1}
  \Gdashed{V3}{B2}
    \Gdashed{V3}{B3}
    \Gdashed{V1}{V3}
    \Gdashed{V1}{V2}
     \draw[sk/propdashed] (V2.25) to[out=45,in=-45,looseness=20] (V2.-25);
  \node[above=2pt] at (B1) {$\k_1$};
  \node[above=2pt] at (B2) {$\k_2$};
  \node[above=2pt] at (B3) {$\k_3$};
  \node[above=2pt] at (-1.5,-1.2) {$\psi_{3,1}^{(1)}$};
   \node[above=2pt] at (-1.5,-2.0) {$\psi_{3,1}^{(1)}$};
     \node[above=2pt] at (0.5,-1.5) {$\psi_{3,1}^{(1)}$};
\end{tikzpicture} 
+
\begin{tikzpicture}[baseline=-3.0pt]
  \draw[sk/boundary] (-1.3,\BOUNDY) -- (1.3,\BOUNDY);
  \skbound{B1}{(-1.0,\BOUNDY)}
  \skbound{B2}{( -0.0,\BOUNDY)}
  \skbound{B3}{( 1.0,\BOUNDY)}
  \skvertexgd{V1}{(-0.5,-1)}
    \skvertexgd{V2}{(0.5,-1)}
  \Gdashed{V1}{B1}
  \Gdashed{V1}{B2}
  \Gdashed{V2}{B3}
  \Gdashed{V1}{V2}
  \node[above=2pt] at (B1) {$\k_1$};
  \node[above=2pt] at (B2) {$\k_2$};
  \node[above=2pt] at (B3) {$\k_3$};
   \node[above=2pt] at (-1.1,-1.5) {$\psi_{3,1}^{(1)}$};
   \node[above=2pt] at (1.1,-1.5) {$\psi_{2,2}^{(2)}$};
\end{tikzpicture}  
+  {\rm perms} . \quad
\eea
It is worth emphasizing that some diagrams appear to recur among the three contributions shown in Eqs.~(\ref{example-result-2}), (\ref{example-result-3}) and (\ref{example-result-4}). Specifically, the second diagram in~(\ref{example-result-2}), involving the coefficient $\psi_{5,1}^{(3)}$, reappears in (\ref{example-result-3}) and (\ref{example-result-4}). Likewise, the third diagram in~(\ref{example-result-2}), involving the coefficient $\psi_{4,1}^{(2)}$, reappears in (\ref{example-result-3}). However, these repetitions correspond to different assignments of external momenta to the legs of the wavefunction coefficients. Since these coefficients are not symmetric under permutations of their momenta, they give rise to distinct contributions, despite their similar diagrammatic appearance. As a result, each diagram appearing in Eqs.~(\ref{example-result-2})--(\ref{example-result-4}) represents a distinct function of the external momenta.

Finally, summing the three topology classes~(\ref{example-result-2}), (\ref{example-result-3}) and (\ref{example-result-4}) yields the full one-loop three-point function in the Wavefunction of the Universe representation, as already stated in Eq.~(\ref{example-result-three-point-full}). In this final expression, each coefficient represents the sum over all bulk topologies contributing to it, together with all admissible ways of attaching its legs to the rest of the diagram.

\subsection{Four-point correlation function at one loop}
\label{sec:Example-3}

\begin{figure}[h]
\centering
\includegraphics[width=\linewidth]{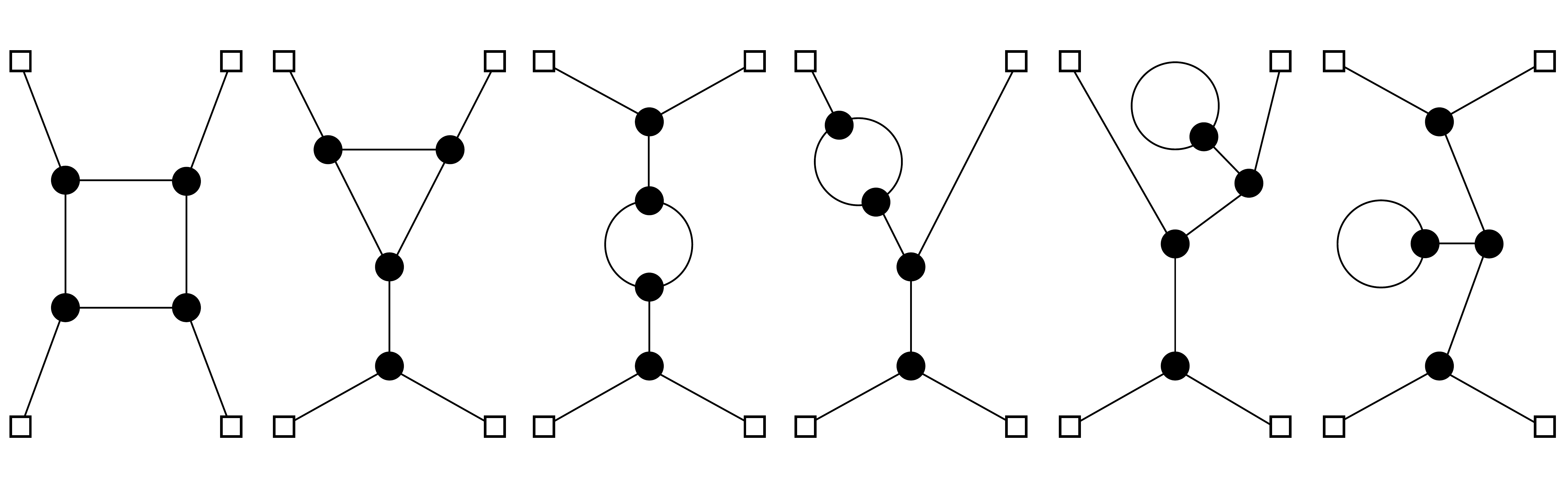}
\caption{\footnotesize The six topologies for diagrams contributing to the four-point function at one-loop.}
\label{figure_08}
\end{figure}

As a final example, let me consider the one-loop contributions to the four-point correlation function at fourth order in the number of vertices. Recall that the four-point function, expanded in terms of wavefunction coefficients, is given in Eq.~(\ref{four-point-example-gray-3}). Here our starting point is the same correlator written in terms of Schwinger--Keldysh diagrams. Since there are six distinct diagrammatic topologies at this order, it is convenient to decompose it as
\be \label{four-point-two-top}
\Big\langle \phi(\k_1)\cdots\phi(\k_4)\Big\rangle_c^{(4)}
=
\sum_{t=1}^{6}\Big\langle \phi(\k_1)\cdots\phi(\k_4)\Big\rangle_t^{(4)} \, .
\ee
The corresponding topologies are displayed in Fig.~\ref{figure_08}. To simplify the discussion, I will focus only on the first two, which already capture the most interesting features of the map.

The Schwinger--Keldysh diagrams associated with the first topology are:
\bea
\label{examples:ex-3}
\Big\langle \phi (\k_1) \cdots \phi (\k_4)  \Big\rangle_1^{(4)} 
&=& 
\,\,
 \begin{tikzpicture}[baseline=-3.0pt]
  \draw[sk/boundary] (-1.8,\BOUNDY) -- (1.8,\BOUNDY);
  \skbound{B1}{(-1.5,\BOUNDY)}
  \skbound{B2}{( -0.5,\BOUNDY)}
  \skbound{B3}{( 0.5,\BOUNDY)}
    \skbound{B4}{( 1.5,\BOUNDY)}
  \skvertexp{V1}{(-1,-1)}
    \skvertexp{V2}{(-0.35,-0.65)}
      \skvertexp{V3}{(0.35 ,-0.65)}
       \skvertexp{V4}{(1 ,-1)}
  \G{V1}{B1}
  \G{V2}{B2}
  \G{V3}{B3}
    \G{V4}{B4}
          \Garc{V1}{V4}{-35}{-145}  
           \Garc{V1}{V4}{35}{145}  
  \node[above=2pt] at (B1) {$\k_1$};
  \node[above=2pt] at (B2) {$\k_2$};
  \node[above=2pt] at (B3) {$\k_3$};
  \node[above=2pt] at (B4) {$\k_4$};
\end{tikzpicture}  
+ 
 \begin{tikzpicture}[baseline=-3.0pt]
  \draw[sk/boundary] (-1.8,\BOUNDY) -- (1.8,\BOUNDY);
  \skbound{B1}{(-1.5,\BOUNDY)}
  \skbound{B2}{( -0.5,\BOUNDY)}
  \skbound{B3}{( 0.5,\BOUNDY)}
    \skbound{B4}{( 1.5,\BOUNDY)}
  \skvertexm{V1}{(-1,-1)}
  \skvertexp{V4}{(1 ,-1)}
         \Garc{V1}{V4}{-35}{-145}  
           \Garc{V1}{V4}{35}{145}  
    \skvertexp{V2}{(-0.35,-0.65)}
      \skvertexp{V3}{(0.35 ,-0.65)}
  \G{V1}{B1}
  \G{V2}{B2}
  \G{V3}{B3}
    \G{V4}{B4}
  \node[above=2pt] at (B1) {$\k_1$};
  \node[above=2pt] at (B2) {$\k_2$};
  \node[above=2pt] at (B3) {$\k_3$};
  \node[above=2pt] at (B4) {$\k_4$};
\end{tikzpicture}  
\nn \\
&&
\!\!\!\!\!\!\!\!\!\!\!\!\!\!\!\!\!\!\!\!\!\!\!\!\!\!\!\!\!\!\!\!\!
+ 
 \begin{tikzpicture}[baseline=-3.0pt]
  \draw[sk/boundary] (-1.8,\BOUNDY) -- (1.8,\BOUNDY);
  \skbound{B1}{(-1.5,\BOUNDY)}
  \skbound{B2}{( -0.5,\BOUNDY)}
  \skbound{B3}{( 0.5,\BOUNDY)}
    \skbound{B4}{( 1.5,\BOUNDY)}
  \skvertexm{V1}{(-1,-1)}
  \skvertexp{V4}{(1 ,-1)}
         \Garc{V1}{V4}{-35}{-145}  
           \Garc{V1}{V4}{35}{145}  
    \skvertexm{V2}{(-0.35,-0.65)}
      \skvertexp{V3}{(0.35 ,-0.65)}
  \G{V1}{B1}
  \G{V2}{B2}
  \G{V3}{B3}
    \G{V4}{B4}
  \node[above=2pt] at (B1) {$\k_1$};
  \node[above=2pt] at (B2) {$\k_2$};
  \node[above=2pt] at (B3) {$\k_3$};
  \node[above=2pt] at (B4) {$\k_4$};
\end{tikzpicture}  
+ 
 \begin{tikzpicture}[baseline=-3.0pt]
  \draw[sk/boundary] (-1.8,\BOUNDY) -- (1.8,\BOUNDY);
  \skbound{B1}{(-1.5,\BOUNDY)}
  \skbound{B2}{( -0.5,\BOUNDY)}
  \skbound{B3}{( 0.5,\BOUNDY)}
    \skbound{B4}{( 1.5,\BOUNDY)}
  \skvertexm{V1}{(-1,-1)}
  \skvertexp{V4}{(1 ,-1)}
         \Garc{V1}{V4}{-35}{-145}  
           \Garc{V1}{V4}{35}{145}  
    \skvertexp{V2}{(-0.35,-0.65)}
      \skvertexm{V3}{(0.35 ,-0.65)}
  \G{V1}{B1}
  \G{V2}{B2}
  \G{V3}{B3}
    \G{V4}{B4}
  \node[above=2pt] at (B1) {$\k_1$};
  \node[above=2pt] at (B2) {$\k_2$};
  \node[above=2pt] at (B3) {$\k_3$};
  \node[above=2pt] at (B4) {$\k_4$};
\end{tikzpicture}  
+
 {\rm c.c.}  +   {\rm perms} , \quad
\eea
where ``c.c.'' denotes additional diagrams obtained by complex conjugation of those explicitly shown, and ``perms'' denotes additional diagrams obtained by permuting the external momenta. 

\begin{figure}[h!]
\centering
\includegraphics[width=\linewidth]{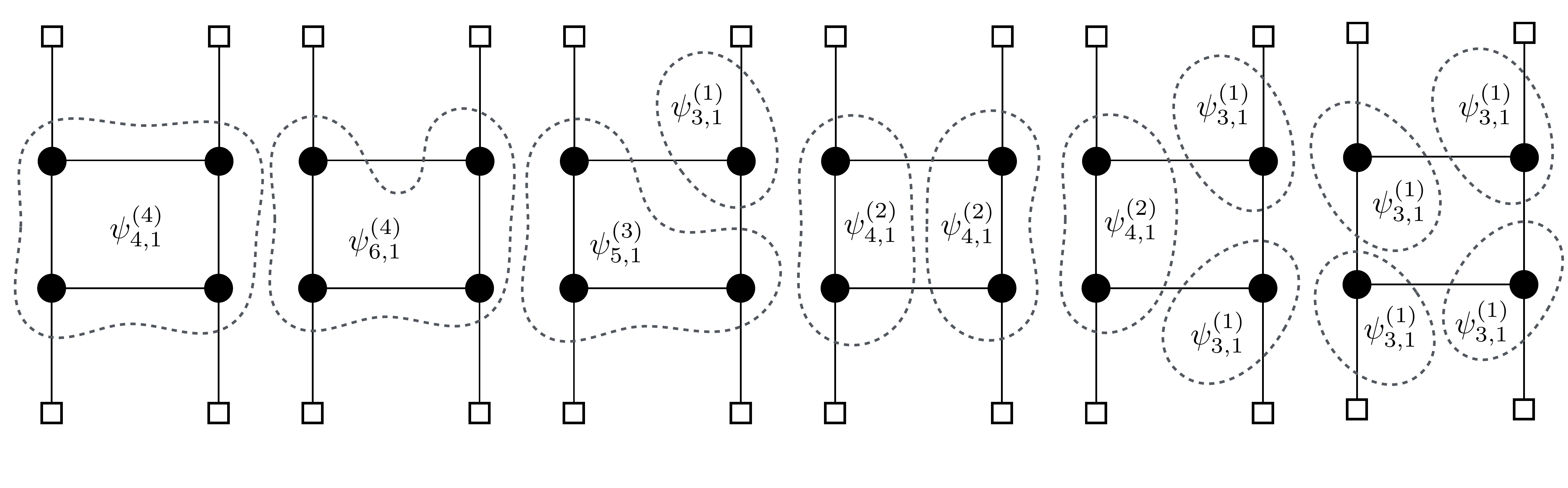}
\caption{\footnotesize The different partitions of the graph representing the first diagram in (\ref{examples:ex-3}), together with the wavefunction coefficient associated with each partition. The partitions of the other graphs in (\ref{examples:ex-3}) are obtained as subsets of these, with the appropriate coloring.}
\label{figure_09}
\end{figure}

As shown in Fig.~\ref{figure_09}, the graph associated with the first diagram in Eq.~(\ref{examples:ex-3}) admits six distinct partitions. 
The figure also indicates the wavefunction coefficient associated with each partition. Note that the partitions of colored graphs corresponding to the remaining diagrams in Eq.~(\ref{examples:ex-3}) are subsets of those already present in Fig.~\ref{figure_09}, subject to the constraint that every partition contains vertices of a single color. For this reason, it is not necessary to draw them explicitly. 

With the partitions in Fig.~\ref{figure_09} in hand, and the partitions of other graphs mixing colors, each Schwinger--Keldysh diagram in Eq.~(\ref{examples:ex-3}) can be rewritten in terms of wavefunction coefficients (and their complex conjugates). Summing all such contributions then yields
\bea \label{first-topology-four-point-function}
\Big\langle \phi (\k_1) \cdots \phi (\k_4)  \Big\rangle_1^{(4)} 
&=& 
\,\,
\begin{tikzpicture}[baseline=-3.0pt]
  \draw[sk/boundary] (-1.8,\BOUNDY) -- (1.8,\BOUNDY);
  \skbound{B1}{(-1.5,\BOUNDY)}
  \skbound{B2}{( -0.5,\BOUNDY)}
  \skbound{B3}{( 0.5,\BOUNDY)}
    \skbound{B4}{( 1.5,\BOUNDY)}
  \skvertexgd{V1}{(0,-1)}
  \Gdashed{V1}{B1}
  \Gdashed{V1}{B2}
  \Gdashed{V1}{B3}
    \Gdashed{V1}{B4}
  \node[above=2pt] at (B1) {$\k_1$};
  \node[above=2pt] at (B2) {$\k_2$};
  \node[above=2pt] at (B3) {$\k_3$};
  \node[above=2pt] at (B4) {$\k_4$};
      \node[above=2pt] at (0.6,-1.5) {$\psi_{4,1}^{(4)}$};
\end{tikzpicture} 
+ 
\begin{tikzpicture}[baseline=-3.0pt]
  \draw[sk/boundary] (-1.8,\BOUNDY) -- (1.8,\BOUNDY);
  \skbound{B1}{(-1.5,\BOUNDY)}
  \skbound{B2}{( -0.5,\BOUNDY)}
  \skbound{B3}{( 0.5,\BOUNDY)}
    \skbound{B4}{( 1.5,\BOUNDY)}
  \skvertexgd{V1}{(0,-1)}
  \Gdashed{V1}{B1}
  \Gdashed{V1}{B2}
  \Gdashed{V1}{B3}
    \Gdashed{V1}{B4}
  \node[above=2pt] at (B1) {$\k_1$};
  \node[above=2pt] at (B2) {$\k_2$};
  \node[above=2pt] at (B3) {$\k_3$};
  \node[above=2pt] at (B4) {$\k_4$};
      \node[above=2pt] at (0.6,-1.5) {$\psi_{6,1}^{(4)}$};
           \draw[sk/propdashed] (V1) to[out=-55,in=-125,looseness=20] (V1);
\end{tikzpicture} 
\nn \\[-10pt]
&&
\,\,
+ 
  \begin{tikzpicture}[baseline=-3.0pt]
  \draw[sk/boundary] (-1.8,\BOUNDY) -- (1.8,\BOUNDY);
  \skbound{B1}{(-1.5,\BOUNDY)}
  \skbound{B2}{( -0.5,\BOUNDY)}
  \skbound{B3}{( 0.5,\BOUNDY)}
    \skbound{B4}{( 1.5,\BOUNDY)}
 \skvertexgd{V1}{(-0.5,-1)}
    \skvertexgd{V2}{(1,-1)}
  \Gdashed{V1}{B1}
  \Gdashed{V1}{B2}
  \Gdashed{V1}{B3}
    \Gdashed{V2}{B4}
    \Gdashedarc{V1}{V2}{-25}{-155}  
          \Gdashedarc{V1}{V2}{25}{155}  
  \node[above=2pt] at (B1) {$\k_1$};
  \node[above=2pt] at (B2) {$\k_2$};
  \node[above=2pt] at (B3) {$\k_3$};
  \node[above=2pt] at (B4) {$\k_4$};
   \node[above=2pt] at (-1.1,-1.5) {$\psi_{5,1}^{(3)}$};
   \node[above=2pt] at (1.6,-1.5) {$\psi_{3,1}^{(1)}$};
\end{tikzpicture}   
+ 
  \begin{tikzpicture}[baseline=-3.0pt]
  \draw[sk/boundary] (-1.8,\BOUNDY) -- (1.8,\BOUNDY);
  \skbound{B1}{(-1.5,\BOUNDY)}
  \skbound{B2}{( -0.5,\BOUNDY)}
  \skbound{B3}{( 0.5,\BOUNDY)}
    \skbound{B4}{( 1.5,\BOUNDY)}
 \skvertexgd{V1}{(-1,-1)}
   \skvertexgd{V2}{(1,-1)}
  \Gdashed{V1}{B1}
  \Gdashed{V1}{B2}
  \Gdashed{V2}{B3}
    \Gdashed{V2}{B4}
    \Gdashedarc{V1}{V2}{-25}{-155}  
          \Gdashedarc{V1}{V2}{25}{155}  
  \node[above=2pt] at (B1) {$\k_1$};
  \node[above=2pt] at (B2) {$\k_2$};
  \node[above=2pt] at (B3) {$\k_3$};
  \node[above=2pt] at (B4) {$\k_4$};
   \node[above=2pt] at (-1.6,-1.5) {$\psi_{4,1}^{(2)}$};
   \node[above=2pt] at (1.6,-1.5) {$\psi_{4,1}^{(2)}$};
\end{tikzpicture}   
\nn 
\\
&&
\,\,
+
 \begin{tikzpicture}[baseline=-3.0pt]
  \draw[sk/boundary] (-1.8,\BOUNDY) -- (1.8,\BOUNDY);
  \skbound{B1}{(-1.5,\BOUNDY)}
  \skbound{B2}{( -0.5,\BOUNDY)}
  \skbound{B3}{( 0.5,\BOUNDY)}
    \skbound{B4}{( 1.5,\BOUNDY)}
  \skvertexgd{V1}{(-1,-1)}
    \skvertexgd{V4}{(1 ,-1)}
  \Gdashed{V1}{B1}
    \Gdashed{V4}{B4}
          \Gdashedarc{V1}{V4}{-35}{-145}  
           \Gdashedarc{V1}{V4}{35}{145}  
            \skvertexgd{V2}{(-0.35,-0.65)}
     \skvertexgd{V3}{(0.35 ,-0.65)}
        \Gdashed{V2}{B2}
  \Gdashed{V3}{B3}
  \node[above=2pt] at (B1) {$\k_1$};
  \node[above=2pt] at (B2) {$\k_2$};
  \node[above=2pt] at (B3) {$\k_3$};
  \node[above=2pt] at (B4) {$\k_4$};
  \node[above=2pt] at (-1.6,-1.5) {$\psi_{3,1}^{(1)}$};
   \node[above=2pt] at (1.6,-1.5) {$\psi_{3,1}^{(1)}$};
     \node[above=2pt] at (0.82,-0.9) {$\psi_{3,1}^{(1)}$};
     \node[above=2pt] at (-0.8,-0.9) {$\psi_{3,1}^{(1)}$};
\end{tikzpicture}  
+
  \begin{tikzpicture}[baseline=-3.0pt]
  \draw[sk/boundary] (-1.8,\BOUNDY) -- (1.8,\BOUNDY);
  \skbound{B1}{(-1.5,\BOUNDY)}
  \skbound{B2}{( -0.5,\BOUNDY)}
  \skbound{B3}{( 0.5,\BOUNDY)}
    \skbound{B4}{( 1.5,\BOUNDY)}
 \skvertexgd{V1}{(-1,-1)}
  \skvertexgd{V3}{(1,-1)}
  \Gdashed{V1}{B1}
    \Gdashed{V3}{B4}
    \Gdashedarc{V1}{V3}{-25}{-155}  
          \Gdashedarc{V1}{V3}{25}{155}  
            \skvertexgd{V2}{(0,-0.73)}
               \Gdashed{V2}{B2}
  \Gdashed{V2}{B3}
  \node[above=2pt] at (B1) {$\k_1$};
  \node[above=2pt] at (B2) {$\k_2$};
  \node[above=2pt] at (B3) {$\k_3$};
  \node[above=2pt] at (B4) {$\k_4$};
   \node[above=2pt] at (-1.6,-1.5) {$\psi_{3,1}^{(1)}$};
   \node[above=2pt] at (1.6,-1.5) {$\psi_{3,1}^{(1)}$};
    \node[above=2pt] at (0.7,-0.9) {$\psi_{4,1}^{(2)}$};
\end{tikzpicture}   
+ {\rm perms}, \qquad
\eea
where, as usual, ``perms'' denotes the additional diagrams obtained by permuting the external momenta whenever this produces a distinct contribution. As before, some care is required when interpreting the diagrams in (\ref{first-topology-four-point-function}), since several of them involve wavefunction coefficients that are not symmetric under permutations of their inflowing momenta. For simplicity, I have not indicated explicitly which propagator is attached to which leg of each such vertex; this assignment is nevertheless fixed, and can be unambiguously inferred from the partitioning shown in Fig.~\ref{figure_09}.

\begin{figure}[b!]
\centering
\includegraphics[width=\linewidth]{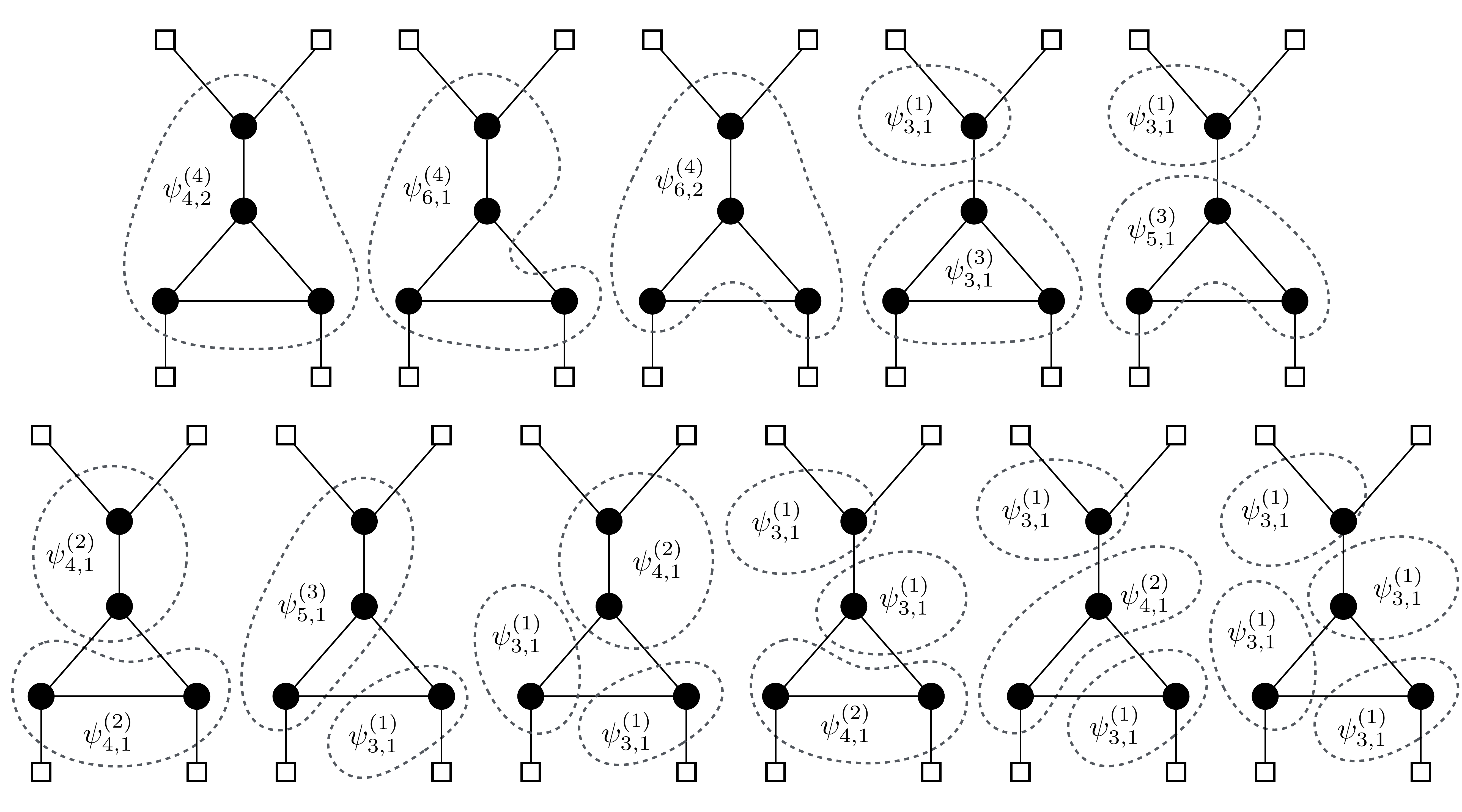}
\caption{\footnotesize The figure shows the different ways in which the graph representing the first diagram in (\ref{examples:four-point-4-1}) can be partitioned, and the corresponding wavefunction coefficient labeling each partition. Recall that the partitions of other graphs are subsets of those already shown here, but respecting the appropriate coloring.}
\label{figure_10}
\end{figure}

Moving on, the second contribution in Eq.~(\ref{four-point-two-top}) is represented in the Schwinger--Keldysh formalism by
\bea
\label{examples:four-point-4-1}
\Big\langle \phi (\k_1) \cdots \phi (\k_4)  \Big\rangle_2^{(4)} 
&=& 
\,\,
 \begin{tikzpicture}[baseline=-3.0pt]
  \draw[sk/boundary] (-1.8,\BOUNDY) -- (1.8,\BOUNDY);
  \skbound{B1}{(-1.5,\BOUNDY)}
  \skbound{B2}{( -0.5,\BOUNDY)}
  \skbound{B3}{( 0.5,\BOUNDY)}
    \skbound{B4}{( 1.5,\BOUNDY)}
    \skvertexp{V1}{(-1,-1)}
     \skvertexp{V2}{(0.2,-1)}
     \skvertexp{V4}{(1 ,-1)}
  \G{V1}{B1}
  \G{V4}{B3}
    \G{V4}{B4}
    \G{V2}{V4}
     \Garc{V1}{V2}{-35}{-145}  
          \Garc{V1}{V2}{35}{145}  
          \skvertexp{V3}{(-0.4 ,-0.75)}
  \G{V3}{B2}
  \node[above=2pt] at (B1) {$\k_1$};
  \node[above=2pt] at (B2) {$\k_2$};
  \node[above=2pt] at (B3) {$\k_3$};
  \node[above=2pt] at (B4) {$\k_4$};
\end{tikzpicture}
+ 
 \begin{tikzpicture}[baseline=-3.0pt]
  \draw[sk/boundary] (-1.8,\BOUNDY) -- (1.8,\BOUNDY);
  \skbound{B1}{(-1.5,\BOUNDY)}
  \skbound{B2}{( -0.5,\BOUNDY)}
  \skbound{B3}{( 0.5,\BOUNDY)}
    \skbound{B4}{( 1.5,\BOUNDY)}
    \skvertexm{V1}{(-1,-1)}
     \skvertexp{V2}{(0.2,-1)}
     \skvertexp{V4}{(1 ,-1)}
  \G{V1}{B1}
  \G{V4}{B3}
    \G{V4}{B4}
    \G{V2}{V4}
     \Garc{V1}{V2}{-35}{-145}  
          \Garc{V1}{V2}{35}{145}  
          \skvertexp{V3}{(-0.4 ,-0.75)}
  \G{V3}{B2}
  \node[above=2pt] at (B1) {$\k_1$};
  \node[above=2pt] at (B2) {$\k_2$};
  \node[above=2pt] at (B3) {$\k_3$};
  \node[above=2pt] at (B4) {$\k_4$};
\end{tikzpicture}
+
 \begin{tikzpicture}[baseline=-3.0pt]
  \draw[sk/boundary] (-1.8,\BOUNDY) -- (1.8,\BOUNDY);
  \skbound{B1}{(-1.5,\BOUNDY)}
  \skbound{B2}{( -0.5,\BOUNDY)}
  \skbound{B3}{( 0.5,\BOUNDY)}
    \skbound{B4}{( 1.5,\BOUNDY)}
    \skvertexp{V1}{(-1,-1)}
     \skvertexm{V2}{(0.2,-1)}
     \skvertexp{V4}{(1 ,-1)}
  \G{V1}{B1}
  \G{V4}{B3}
    \G{V4}{B4}
    \G{V2}{V4}
     \Garc{V1}{V2}{-35}{-145}  
          \Garc{V1}{V2}{35}{145}  
          \skvertexp{V3}{(-0.4 ,-0.75)}
  \G{V3}{B2}
  \node[above=2pt] at (B1) {$\k_1$};
  \node[above=2pt] at (B2) {$\k_2$};
  \node[above=2pt] at (B3) {$\k_3$};
  \node[above=2pt] at (B4) {$\k_4$};
\end{tikzpicture}
\nn \\
&& \!\!\!\!\!\!\!\!\!\!\!\!\!\!\!
+
 \begin{tikzpicture}[baseline=-3.0pt]
  \draw[sk/boundary] (-1.8,\BOUNDY) -- (1.8,\BOUNDY);
  \skbound{B1}{(-1.5,\BOUNDY)}
  \skbound{B2}{( -0.5,\BOUNDY)}
  \skbound{B3}{( 0.5,\BOUNDY)}
    \skbound{B4}{( 1.5,\BOUNDY)}
    \skvertexp{V1}{(-1,-1)}
     \skvertexp{V2}{(0.2,-1)}
     \skvertexm{V4}{(1 ,-1)}
  \G{V1}{B1}
  \G{V4}{B3}
    \G{V4}{B4}
    \G{V2}{V4}
     \Garc{V1}{V2}{-35}{-145}  
          \Garc{V1}{V2}{35}{145}  
          \skvertexp{V3}{(-0.4 ,-0.75)}
  \G{V3}{B2}
  \node[above=2pt] at (B1) {$\k_1$};
  \node[above=2pt] at (B2) {$\k_2$};
  \node[above=2pt] at (B3) {$\k_3$};
  \node[above=2pt] at (B4) {$\k_4$};
\end{tikzpicture}
+
 \begin{tikzpicture}[baseline=-3.0pt]
  \draw[sk/boundary] (-1.8,\BOUNDY) -- (1.8,\BOUNDY);
  \skbound{B1}{(-1.5,\BOUNDY)}
  \skbound{B2}{( -0.5,\BOUNDY)}
  \skbound{B3}{( 0.5,\BOUNDY)}
    \skbound{B4}{( 1.5,\BOUNDY)}
    \skvertexm{V1}{(-1,-1)}
     \skvertexp{V2}{(0.2,-1)}
     \skvertexp{V4}{(1 ,-1)}
  \G{V1}{B1}
  \G{V4}{B3}
    \G{V4}{B4}
    \G{V2}{V4}
     \Garc{V1}{V2}{-35}{-145}  
          \Garc{V1}{V2}{35}{145}  
          \skvertexm{V3}{(-0.4 ,-0.75)}
  \G{V3}{B2}
  \node[above=2pt] at (B1) {$\k_1$};
  \node[above=2pt] at (B2) {$\k_2$};
  \node[above=2pt] at (B3) {$\k_3$};
  \node[above=2pt] at (B4) {$\k_4$};
\end{tikzpicture}
+ 
 \begin{tikzpicture}[baseline=-3.0pt]
  \draw[sk/boundary] (-1.8,\BOUNDY) -- (1.8,\BOUNDY);
  \skbound{B1}{(-1.5,\BOUNDY)}
  \skbound{B2}{( -0.5,\BOUNDY)}
  \skbound{B3}{( 0.5,\BOUNDY)}
    \skbound{B4}{( 1.5,\BOUNDY)}
    \skvertexm{V1}{(-1,-1)}
     \skvertexm{V2}{(0.2,-1)}
     \skvertexp{V4}{(1 ,-1)}
  \G{V1}{B1}
  \G{V4}{B3}
    \G{V4}{B4}
    \G{V2}{V4}
     \Garc{V1}{V2}{-35}{-145}  
          \Garc{V1}{V2}{35}{145}  
          \skvertexp{V3}{(-0.4 ,-0.75)}
  \G{V3}{B2}
  \node[above=2pt] at (B1) {$\k_1$};
  \node[above=2pt] at (B2) {$\k_2$};
  \node[above=2pt] at (B3) {$\k_3$};
  \node[above=2pt] at (B4) {$\k_4$};
\end{tikzpicture}
\nn \\
&& \!\!\!\!\!\!\!\!\!\!\!\!\!\!\!
+
 \begin{tikzpicture}[baseline=-3.0pt]
  \draw[sk/boundary] (-1.8,\BOUNDY) -- (1.8,\BOUNDY);
  \skbound{B1}{(-1.5,\BOUNDY)}
  \skbound{B2}{( -0.5,\BOUNDY)}
  \skbound{B3}{( 0.5,\BOUNDY)}
    \skbound{B4}{( 1.5,\BOUNDY)}
    \skvertexm{V1}{(-1,-1)}
     \skvertexp{V2}{(0.2,-1)}
     \skvertexm{V4}{(1 ,-1)}
  \G{V1}{B1}
  \G{V4}{B3}
    \G{V4}{B4}
    \G{V2}{V4}
     \Garc{V1}{V2}{-35}{-145}  
          \Garc{V1}{V2}{35}{145}  
          \skvertexp{V3}{(-0.4 ,-0.75)}
  \G{V3}{B2}
  \node[above=2pt] at (B1) {$\k_1$};
  \node[above=2pt] at (B2) {$\k_2$};
  \node[above=2pt] at (B3) {$\k_3$};
  \node[above=2pt] at (B4) {$\k_4$};
\end{tikzpicture}
+ 
 {\rm c.c.} + {\rm perms} . 
\eea
Figure~\ref{figure_10} shows the partitions of the graph associated with the first diagram in Eq.~(\ref{examples:four-point-4-1}). 
As before, the partitions required for the remaining diagrams in (\ref{examples:four-point-4-1}) are subsets of those displayed in Fig.~\ref{figure_10}, subject to the appropriate coloring, so we omit them. Reading off the coefficients directly from Fig.~\ref{figure_10}, one arrives at:
\bea
\label{examples:four-point-4-2}
\Big\langle \phi (\k_1) \cdots \phi (\k_4)  \Big\rangle_2^{(4)} 
&=& 
\begin{tikzpicture}[baseline=-3.0pt]
  \draw[sk/boundary] (-1.8,\BOUNDY) -- (1.8,\BOUNDY);
  \skbound{B1}{(-1.5,\BOUNDY)}
  \skbound{B2}{( -0.5,\BOUNDY)}
  \skbound{B3}{( 0.5,\BOUNDY)}
    \skbound{B4}{( 1.5,\BOUNDY)}
  \skvertexgd{V1}{(0,-1)}
  \Gdashed{V1}{B1}
  \Gdashed{V1}{B2}
  \Gdashed{V1}{B3}
    \Gdashed{V1}{B4}
  \node[above=2pt] at (B1) {$\k_1$};
  \node[above=2pt] at (B2) {$\k_2$};
  \node[above=2pt] at (B3) {$\k_3$};
  \node[above=2pt] at (B4) {$\k_4$};
      \node[above=2pt] at (0.6,-1.5) {$\psi_{4,2}^{(4)}$};
\end{tikzpicture} 
+ 
\begin{tikzpicture}[baseline=-3.0pt]
  \draw[sk/boundary] (-1.8,\BOUNDY) -- (1.8,\BOUNDY);
  \skbound{B1}{(-1.5,\BOUNDY)}
  \skbound{B2}{( -0.5,\BOUNDY)}
  \skbound{B3}{( 0.5,\BOUNDY)}
    \skbound{B4}{( 1.5,\BOUNDY)}
  \skvertexgd{V1}{(0,-1)}
  \Gdashed{V1}{B1}
  \Gdashed{V1}{B2}
  \Gdashed{V1}{B3}
    \Gdashed{V1}{B4}
  \node[above=2pt] at (B1) {$\k_1$};
  \node[above=2pt] at (B2) {$\k_2$};
  \node[above=2pt] at (B3) {$\k_3$};
  \node[above=2pt] at (B4) {$\k_4$};
      \node[above=2pt] at (0.6,-1.5) {$\psi_{6,1}^{(4)}$};
           \draw[sk/propdashed] (V1) to[out=-55,in=-125,looseness=20] (V1);
\end{tikzpicture} 
+
\begin{tikzpicture}[baseline=-3.0pt]
  \draw[sk/boundary] (-1.8,\BOUNDY) -- (1.8,\BOUNDY);
  \skbound{B1}{(-1.5,\BOUNDY)}
  \skbound{B2}{( -0.5,\BOUNDY)}
  \skbound{B3}{( 0.5,\BOUNDY)}
    \skbound{B4}{( 1.5,\BOUNDY)}
  \skvertexgd{V1}{(0,-1)}
  \Gdashed{V1}{B1}
  \Gdashed{V1}{B2}
  \Gdashed{V1}{B3}
    \Gdashed{V1}{B4}
  \node[above=2pt] at (B1) {$\k_1$};
  \node[above=2pt] at (B2) {$\k_2$};
  \node[above=2pt] at (B3) {$\k_3$};
  \node[above=2pt] at (B4) {$\k_4$};
      \node[above=2pt] at (0.6,-1.5) {$\psi_{6,2}^{(4)}$};
           \draw[sk/propdashed] (V1) to[out=-55,in=-125,looseness=20] (V1);
\end{tikzpicture} 
\nn \\[-10pt]
&& 
\!\!\!\!\!\!\!\!\!\!\!\!\!\!\!\!\!\!\!\!\!\!\!\!
+
 \begin{tikzpicture}[baseline=-3.0pt]
  \draw[sk/boundary] (-1.8,\BOUNDY) -- (1.8,\BOUNDY);
  \skbound{B1}{(-1.5,\BOUNDY)}
  \skbound{B2}{( -0.5,\BOUNDY)}
  \skbound{B3}{( 0.5,\BOUNDY)}
    \skbound{B4}{( 1.5,\BOUNDY)}
  \skvertexgd{V1}{(-1,-1)}
    \skvertexgd{V2}{(1,-1)}
  \Gdashed{V1}{B1}
  \Gdashed{V1}{B2}
  \Gdashed{V2}{B3}
    \Gdashed{V2}{B4}
        \Gdashed{V1}{V2}
  \node[above=2pt] at (B1) {$\k_1$};
  \node[above=2pt] at (B2) {$\k_2$};
  \node[above=2pt] at (B3) {$\k_3$};
  \node[above=2pt] at (B4) {$\k_4$};
   \node[above=2pt] at (-1.0,-2.0) {$\psi_{3,1}^{(1)}$};
   \node[above=2pt] at (1.0,-2.0) {$\psi_{3,1}^{(3)}$};
\end{tikzpicture}   
+
 \begin{tikzpicture}[baseline=-3.0pt]
  \draw[sk/boundary] (-1.8,\BOUNDY) -- (1.8,\BOUNDY);
  \skbound{B1}{(-1.5,\BOUNDY)}
  \skbound{B2}{( -0.5,\BOUNDY)}
  \skbound{B3}{( 0.5,\BOUNDY)}
    \skbound{B4}{( 1.5,\BOUNDY)}
  \skvertexgd{V1}{(-1,-1)}
    \skvertexgd{V2}{(1,-1)}
  \Gdashed{V1}{B1}
  \Gdashed{V1}{B2}
  \Gdashed{V2}{B3}
    \Gdashed{V2}{B4}
        \Gdashed{V1}{V2}
  \node[above=2pt] at (B1) {$\k_1$};
  \node[above=2pt] at (B2) {$\k_2$};
  \node[above=2pt] at (B3) {$\k_3$};
  \node[above=2pt] at (B4) {$\k_4$};
   \node[above=2pt] at (-1.0,-2.0) {$\psi_{3,1}^{(1)}$};
   \node[above=2pt] at (1.7,-1.5) {$\psi_{5,1}^{(3)}$};
        \draw[sk/propdashed] (V2) to[out=-55,in=-125,looseness=20] (V2); 
\end{tikzpicture}   
+
  \begin{tikzpicture}[baseline=-3.0pt]
  \draw[sk/boundary] (-1.8,\BOUNDY) -- (1.8,\BOUNDY);
  \skbound{B1}{(-1.5,\BOUNDY)}
  \skbound{B2}{( -0.5,\BOUNDY)}
  \skbound{B3}{( 0.5,\BOUNDY)}
    \skbound{B4}{( 1.5,\BOUNDY)}
 \skvertexgd{V1}{(-1,-1)}
   \skvertexgd{V2}{(1,-1)}
  \Gdashed{V1}{B1}
  \Gdashed{V1}{B2}
  \Gdashed{V2}{B3}
    \Gdashed{V2}{B4}
    \Gdashedarc{V1}{V2}{-25}{-155}  
          \Gdashedarc{V1}{V2}{25}{155}  
  \node[above=2pt] at (B1) {$\k_1$};
  \node[above=2pt] at (B2) {$\k_2$};
  \node[above=2pt] at (B3) {$\k_3$};
  \node[above=2pt] at (B4) {$\k_4$};
   \node[above=2pt] at (-1.2,-2) {$\psi_{4,1}^{(2)}$};
   \node[above=2pt] at (1.2,-2) {$\psi_{4,1}^{(2)}$};
\end{tikzpicture}    
\nn
\\
&& 
\!\!\!\!\!\!\!\!\!\!\!\!\!\!\!\!\!\!\!\!\!\!\!\!
+
  \begin{tikzpicture}[baseline=-3.0pt]
  \draw[sk/boundary] (-1.8,\BOUNDY) -- (1.8,\BOUNDY);
  \skbound{B1}{(-1.5,\BOUNDY)}
  \skbound{B2}{( -0.5,\BOUNDY)}
  \skbound{B3}{( 0.5,\BOUNDY)}
    \skbound{B4}{( 1.5,\BOUNDY)}
 \skvertexgd{V1}{(-0.5,-1)}
    \skvertexgd{V2}{(1,-1)}
  \Gdashed{V1}{B1}
  \Gdashed{V1}{B2}
  \Gdashed{V1}{B3}
    \Gdashed{V2}{B4}
    \Gdashedarc{V1}{V2}{-25}{-155}  
          \Gdashedarc{V1}{V2}{25}{155}  
  \node[above=2pt] at (B1) {$\k_1$};
  \node[above=2pt] at (B2) {$\k_2$};
  \node[above=2pt] at (B3) {$\k_3$};
  \node[above=2pt] at (B4) {$\k_4$};
   \node[above=2pt] at (-0.9,-2) {$\psi_{5,1}^{(3)}$};
   \node[above=2pt] at (1.2,-2) {$\psi_{3,1}^{(1)}$};
\end{tikzpicture}   
+ 
  \begin{tikzpicture}[baseline=-3.0pt]
  \draw[sk/boundary] (-1.8,\BOUNDY) -- (1.8,\BOUNDY);
  \skbound{B1}{(-1.5,\BOUNDY)}
  \skbound{B2}{( -0.5,\BOUNDY)}
  \skbound{B3}{( 0.5,\BOUNDY)}
    \skbound{B4}{( 1.5,\BOUNDY)}
 \skvertexgd{V1}{(-1,-1)}
  \skvertexgd{V3}{(1,-1)}
  \Gdashed{V1}{B1}
    \Gdashed{V3}{B4}
    \Gdashedarc{V1}{V3}{-25}{-155}  
          \Gdashedarc{V1}{V3}{25}{155}  
            \skvertexgd{V2}{(0,-0.73)}
               \Gdashed{V2}{B2}
  \Gdashed{V2}{B3}
  \node[above=2pt] at (B1) {$\k_1$};
  \node[above=2pt] at (B2) {$\k_2$};
  \node[above=2pt] at (B3) {$\k_3$};
  \node[above=2pt] at (B4) {$\k_4$};
   \node[above=2pt] at (-1.2,-2) {$\psi_{3,1}^{(1)}$};
   \node[above=2pt] at (1.2,-2) {$\psi_{3,1}^{(1)}$};
    \node[above=2pt] at (0.7,-0.9) {$\psi_{4,1}^{(2)}$};
\end{tikzpicture}   
+
 \begin{tikzpicture}[baseline=-3.0pt]
  \draw[sk/boundary] (-1.8,\BOUNDY) -- (1.8,\BOUNDY);
  \skbound{B1}{(-1.5,\BOUNDY)}
  \skbound{B2}{( -0.5,\BOUNDY)}
  \skbound{B3}{( 0.5,\BOUNDY)}
    \skbound{B4}{( 1.5,\BOUNDY)}
  \skvertexgd{V1}{(-1,-1)}
    \skvertexgd{V2}{(0,-1)}
     \skvertexgd{V3}{(1,-1)}
  \Gdashed{V1}{B1}
  \Gdashed{V1}{B2}
  \Gdashed{V3}{B3}
    \Gdashed{V3}{B4}
        \Gdashed{V2}{V3}
         \Gdashedarc{V1}{V2}{-25}{-155}  
          \Gdashedarc{V1}{V2}{25}{155}  
  \node[above=2pt] at (B1) {$\k_1$};
  \node[above=2pt] at (B2) {$\k_2$};
  \node[above=2pt] at (B3) {$\k_3$};
  \node[above=2pt] at (B4) {$\k_4$};
   \node[above=2pt] at (-1.0,-2.0) {$\psi_{4,1}^{(2)}$};
    \node[above=2pt] at (0.0,-2.0) {$\psi_{3,1}^{(1)}$};
   \node[above=2pt] at (1.0,-2.0) {$\psi_{3,1}^{(1)}$};
\end{tikzpicture}  
\nn 
\\
&&
\!\!\!\!\!\!\!\!\!\!\!\!\!\!\!\!\!\!\!\!\!\!\!\!
+
 \begin{tikzpicture}[baseline=-3.0pt]
  \draw[sk/boundary] (-1.8,\BOUNDY) -- (1.8,\BOUNDY);
  \skbound{B1}{(-1.5,\BOUNDY)}
  \skbound{B2}{( -0.5,\BOUNDY)}
  \skbound{B3}{( 0.5,\BOUNDY)}
    \skbound{B4}{( 1.5,\BOUNDY)}
  \skvertexgd{V1}{(-1,-1)}
    \skvertexgd{V2}{(0,-1)}
     \skvertexgd{V3}{(1,-1)}
  \Gdashed{V1}{B1}
  \Gdashed{V2}{B2}
  \Gdashed{V3}{B3}
    \Gdashed{V3}{B4}
        \Gdashed{V2}{V3}
         \Gdashedarc{V1}{V2}{-25}{-155}  
          \Gdashedarc{V1}{V2}{25}{155}  
  \node[above=2pt] at (B1) {$\k_1$};
  \node[above=2pt] at (B2) {$\k_2$};
  \node[above=2pt] at (B3) {$\k_3$};
  \node[above=2pt] at (B4) {$\k_4$};
   \node[above=2pt] at (-1.2,-2.0) {$\psi_{3,1}^{(1)}$};
    \node[above=2pt] at (0.0,-2.0) {$\psi_{4,1}^{(2)}$};
   \node[above=2pt] at (1.0,-2.0) {$\psi_{3,1}^{(1)}$};
\end{tikzpicture}  
+
 \begin{tikzpicture}[baseline=-3.0pt]
  \draw[sk/boundary] (-1.8,\BOUNDY) -- (1.8,\BOUNDY);
  \skbound{B1}{(-1.5,\BOUNDY)}
  \skbound{B2}{( -0.5,\BOUNDY)}
  \skbound{B3}{( 0.5,\BOUNDY)}
    \skbound{B4}{( 1.5,\BOUNDY)}
   \skvertexgd{V1}{(-1,-1)}
    \skvertexgd{V2}{(0.2,-1)}
      \skvertexgd{V4}{(1 ,-1)}
  \Gdashed{V1}{B1}
  \Gdashed{V4}{B3}
    \Gdashed{V4}{B4}
    \Gdashed{V2}{V4}
     \Gdashedarc{V1}{V2}{-35}{-145}  
          \Gdashedarc{V1}{V2}{35}{145}  
          \skvertexgd{V3}{(-0.4 ,-0.75)}
  \Gdashed{V3}{B2}
  \node[above=2pt] at (B1) {$\k_1$};
  \node[above=2pt] at (B2) {$\k_2$};
  \node[above=2pt] at (B3) {$\k_3$};
  \node[above=2pt] at (B4) {$\k_4$};
   \node[above=2pt] at (-1.2,-2.0) {$\psi_{3,1}^{(1)}$};
   \node[above=2pt] at (0.25,-2.0) {$\psi_{3,1}^{(1)}$};
    \node[above=2pt] at (0.0,-0.85) {$\psi_{3,1}^{(1)}$};
     \node[above=2pt] at (1.2,-2.0) {$\psi_{3,1}^{(1)}$};
\end{tikzpicture} 
+
{\rm perms}. \qquad
\eea
Just as in the one-loop three-point example of Section~\ref{sec:Example-2}, some diagrams appear to recur in Eqs.~(\ref{examples:ex-3}) and (\ref{examples:four-point-4-2}). This repetition is only apparent: whenever a diagram involves a wavefunction coefficient that is not symmetric under permutations of its inflowing momenta, the same structure can represent distinct contributions, depending on how external legs are assigned to the arguments of that coefficient (equivalently, to different channels).

To close this example, there remain four additional terms in the decomposition (\ref{four-point-two-top}) that I will not work out explicitly, as they can be computed in exactly the same way. After summing all six topological contributions, one recovers the full result in Eq.~(\ref{four-point-example-gray-3}), expressed in terms of fully symmetric wavefunction coefficients.


\setcounter{equation}{0}
\section{Conclusions}
\label{sec:Conclusions}

The Wavefunction of the Universe and the Schwinger--Keldysh in--in formalism have become standard tools for analyzing the structure of primordial $n$-point correlation functions. On the one hand, the Wavefunction of the Universe approach is particularly efficient for deriving general, non-trivial relations obeyed by $n$-point functions, by exploiting how unitarity, locality, and the symmetries of the system constrain wavefunction coefficients~\cite{Arkani-Hamed:2015bza, Arkani-Hamed:2018kmz, Pajer:2020wnj, Baumann:2021fxj, Baumann:2022jpr, Xianyu:2022jwk, Wang:2022eop}. On the other hand, the Schwinger--Keldysh in--in formalism provides a powerful framework for addressing the regularization of loop divergences arising from well-motivated bulk theories~\cite{Chen:2016nrs, Wang:2021qez, Qin:2023bjk, Qin:2023nhv, Xianyu:2023ytd, Huenupi:2024ksc, Qin:2024gtr, Ballesteros:2024cef, Palma:2025oux, Zhang:2025nzd, Ballesteros:2025nhz}. A map between the two approaches should allow for a more complete understanding of how these complementary techniques overlap.

Although the relation between these two approaches has been discussed previously~\cite{Cespedes:2020xqq, Bzowski:2023nef, Stefanyszyn:2024msm, Bucciotti:2024lvb, Arkani-Hamed:2025mce}, I am not aware of any prior work that presents a systematic, order-by-order procedure mapping diagrams computed in the Schwinger--Keldysh formalism to the more fundamental wavefunction coefficients. In this article, I have provided such a procedure, showing explicitly how diagrams arising in the Wavefunction of the Universe framework can be reorganized into Schwinger--Keldysh diagrams, and vice versa.

The method developed here is based on the use of graphs that encode the topology of the diagrams contributing to a given $n$-point correlation function. A key ingredient of the analysis is the correspondence between different partitions of a graph and different wavefunction coefficients. While the explicit examples presented in this work focused on a bulk theory consisting of a single scalar field with cubic interactions, the underlying logic is completely general and can be straightforwardly extended to more elaborate theories involving multiple fields, higher spins, or more intricate interaction structures.

More generally, the procedure begins by drawing a graph representing a given interaction topology, characterized by $V$ bulk vertices and $n$ external legs. One then assigns a color (black or white) to each vertex, thereby specifying a particular Schwinger--Keldysh diagram contributing to the connected correlation function $\big\langle \varphi(\k_1)\cdots\varphi(\k_n) \big\rangle_c^{(V)}$ at order $V$ in the number of interaction vertices. Such a Schwinger--Keldysh diagram corresponds not to a single object in the Wavefunction of the Universe approach, but rather to a sum of wavefunction diagrams. These are obtained by considering all admissible partitions of the graph such that each partition encloses only vertices of the same color. Each admissible partition is then uniquely associated with a specific combination of wavefunction coefficients.

I expect that this map will be useful for clarifying the relationship between the various cutting rules that have been proposed in the literature, both at the level of correlation functions and at the level of wavefunction coefficients ~\cite{Melville:2021lst, Goodhew:2021oqg, Tong:2021wai, Albayrak:2023hie, AguiSalcedo:2023nds, Stefanyszyn:2023qov, Donath:2024utn, Ghosh:2024aqd, Ema:2024hkj, DuasoPueyo:2024usw, Jain:2025maa, Lee:2025kgs, Das:2025qsh, Colipi-Marchant:2025oin}. In this context, it is tempting to speculate that the correspondence between graph partitions and wavefunction coefficients may ultimately be understood as a manifestation of unitarity at the level of cosmological correlation functions. Moreover, this map may also provide a systematic framework for relating and comparing the different procedures used to regulate loop integrals within each formalism.

\subsection*{Acknowledgements}

I am particularly indebted to Spyros Sypsas for numerous discussions that played a crucial role in shaping the ideas behind this work. I would also like to thank Sebasti\'an C\'espedes, Francisco Colip\'i­, Javier Huenupi, Ellie Hughes, Gabriel Mar\'i­n, Nicol\'as Parra, Francisco Rojas, and Danilo Tapia for useful discussions on various aspects related to the subject of this article. This work was supported by Fondecyt Regular project No. 1251511.


\begin{appendix}

\renewcommand{\theequation}{\Alph{section}.\arabic{equation}}

\setcounter{equation}{0}
\section{Some wavefunction coefficient diagrams}
\label{Sec:App}

In Section~\ref{sec:topology-wf}, I introduced the notation $\psi_{n,t}^{(V)}(\k_1,\ldots,\k_n)$, where $t$ labels the topology of a specific diagram contributing to the computation of $\psi_n^{(V)}(\k_1,\ldots,\k_n)$ through Eq.~(\ref{wfc-topology}). In this appendix, I list several additional examples of topology-sensitive wavefunction coefficients that appear throughout the main text. 

At second order, there is only one diagram contributing to the four-point wavefunction coefficient $\psi_4^{(2)}$, which is given by
\bea
 \psi_{4,1}^{(2)} (\k_1 , \ldots , \k_4 )  &=&  \,\,
 \begin{tikzpicture}[baseline=-3.0pt]
  \draw[sk/boundary] (-1.8,\BOUNDY) -- (1.8,\BOUNDY);
  \skbound{B1}{(-1.5,\BOUNDY)}
  \skbound{B2}{( -0.5,\BOUNDY)}
  \skbound{B3}{( 0.5,\BOUNDY)}
    \skbound{B4}{( 1.5,\BOUNDY)}
  \skvertexp{V1}{(-1,-1)}
    \skvertexp{V2}{(1,-1)}
  \Gdouble{V1}{B1}
  \Gdouble{V1}{B2}
  \Gdouble{V2}{B3}
    \Gdouble{V2}{B4}
        \Gdouble{V1}{V2}
  \node[above=2pt] at (B1) {$\k_1$};
  \node[above=2pt] at (B2) {$\k_2$};
  \node[above=2pt] at (B3) {$\k_3$};
  \node[above=2pt] at (B4) {$\k_4$};
    \skvertexp{V1}{(-1,-1)}
    \skvertexp{V2}{(1,-1)}
    \skbound{B1}{(-1.5,\BOUNDY)}
  \skbound{B2}{( -0.5,\BOUNDY)}
  \skbound{B3}{( 0.5,\BOUNDY)}
    \skbound{B4}{( 1.5,\BOUNDY)}
\end{tikzpicture} 
.
\eea
In contrast, there are six distinct diagrams contributing to the fourth-order coefficient $\psi_4^{(4)}$. These are given by
\bea
\psi_{4,1}^{(4)} (\k_1, \ldots, \k_4)  &=&  \,\,
 \begin{tikzpicture}[baseline=-3.0pt]
  \draw[sk/boundary] (-1.8,\BOUNDY) -- (1.8,\BOUNDY);
  \skbound{B1}{(-1.5,\BOUNDY)}
  \skbound{B2}{( -0.5,\BOUNDY)}
  \skbound{B3}{( 0.5,\BOUNDY)}
    \skbound{B4}{( 1.5,\BOUNDY)}
  \skvertexp{V1}{(-1,-0.8)}
    \skvertexp{V2}{(-0.35,-0.55)}
      \skvertexp{V3}{(0.35 ,-0.55)}
       \skvertexp{V4}{(1 ,-0.8)}
  \Gdouble{V1}{B1}
  \Gdouble{V2}{B2}
  \Gdouble{V3}{B3}
    \Gdouble{V4}{B4}
          \Gdoublearc{V1}{V4}{-25}{-155}  
           \Gdoublearc{V1}{V4}{25}{155}  
            \skbound{B1}{(-1.5,\BOUNDY)}
  \skbound{B2}{( -0.5,\BOUNDY)}
  \skbound{B3}{( 0.5,\BOUNDY)}
    \skbound{B4}{( 1.5,\BOUNDY)}
  \skvertexp{V1}{(-1,-0.8)}
    \skvertexp{V2}{(-0.35,-0.55)}
      \skvertexp{V3}{(0.35 ,-0.55)}
       \skvertexp{V4}{(1 ,-0.8)}
  \node[above=2pt] at (B1) {$\k_1$};
  \node[above=2pt] at (B2) {$\k_2$};
  \node[above=2pt] at (B3) {$\k_3$};
  \node[above=2pt] at (B4) {$\k_4$};
\end{tikzpicture}  
,
\\
\psi_{4,2}^{(4)}  (\k_1, \ldots, \k_4) 
&=&   \,\,
 \begin{tikzpicture}[baseline=-3.0pt]
  \draw[sk/boundary] (-1.8,\BOUNDY) -- (1.8,\BOUNDY);
  \skbound{B1}{(-1.5,\BOUNDY)}
  \skbound{B2}{( -0.5,\BOUNDY)}
  \skbound{B3}{( 0.5,\BOUNDY)}
    \skbound{B4}{( 1.5,\BOUNDY)}
    \skvertexp{V1}{(-1,-1)}
     \skvertexp{V2}{(0.2,-1)}
     \skvertexp{V4}{(1 ,-1)}
  \Gdouble{V1}{B1}
  \Gdouble{V4}{B3}
    \Gdouble{V4}{B4}
    \Gdouble{V2}{V4}
     \Gdoublearc{V1}{V2}{-35}{-145}  
          \Gdoublearc{V1}{V2}{35}{145}  
          \skvertexp{V3}{(-0.4 ,-0.75)}
  \Gdouble{V3}{B2}
      \skvertexp{V1}{(-1,-1)}
     \skvertexp{V2}{(0.2,-1)}
     \skvertexp{V4}{(1 ,-1)}
     \skvertexp{V3}{(-0.4 ,-0.75)}
    \skbound{B1}{(-1.5,\BOUNDY)}
  \skbound{B2}{( -0.5,\BOUNDY)}
  \skbound{B3}{( 0.5,\BOUNDY)}
    \skbound{B4}{( 1.5,\BOUNDY)}
  \node[above=2pt] at (B1) {$\k_1$};
  \node[above=2pt] at (B2) {$\k_2$};
  \node[above=2pt] at (B3) {$\k_3$};
  \node[above=2pt] at (B4) {$\k_4$};
\end{tikzpicture}
,
\\
\psi_{4,3}^{(4)}  (\k_1, \ldots, \k_4) 
&=&    \,\,
 \begin{tikzpicture}[baseline=-3.0pt]
  \draw[sk/boundary] (-1.8,\BOUNDY) -- (2.8,\BOUNDY);
  \skbound{B1}{(-1.5,\BOUNDY)}
  \skbound{B2}{( -0.5,\BOUNDY)}
  \skbound{B3}{( 1.5,\BOUNDY)}
    \skbound{B4}{( 2.5,\BOUNDY)}
  \skvertexp{V1}{(-1,-1)}
    \skvertexp{V2}{(0,-1)}
      \skvertexp{V3}{(1 ,-1)}
       \skvertexp{V4}{(2 ,-1)}
  \Gdouble{V1}{B1}
  \Gdouble{V1}{B2}
  \Gdouble{V4}{B3}
    \Gdouble{V4}{B4}
    \Gdouble{V1}{V2}
    \Gdouble{V3}{V4}
     \Gdoublearc{V2}{V3}{-45}{-135}  
          \Gdoublearc{V2}{V3}{45}{135}  
      \skbound{B1}{(-1.5,\BOUNDY)}
  \skbound{B2}{( -0.5,\BOUNDY)}
  \skbound{B3}{( 1.5,\BOUNDY)}
    \skbound{B4}{( 2.5,\BOUNDY)}
  \skvertexp{V1}{(-1,-1)}
    \skvertexp{V2}{(0,-1)}
      \skvertexp{V3}{(1 ,-1)}
       \skvertexp{V4}{(2 ,-1)}
  \node[above=2pt] at (B1) {$\k_1$};
  \node[above=2pt] at (B2) {$\k_2$};
  \node[above=2pt] at (B3) {$\k_3$};
  \node[above=2pt] at (B4) {$\k_4$};
\end{tikzpicture}  
,
\\
\psi_{4,4}^{(4)}  (\k_1, \ldots, \k_4) 
&=&   \,\,
 \begin{tikzpicture}[baseline=-3.0pt]
  \draw[sk/boundary] (-1.8,\BOUNDY) -- (2.8,\BOUNDY);
  \skbound{B1}{(-1.5,\BOUNDY)}
  \skbound{B2}{( 0.5 ,\BOUNDY)}
  \skbound{B3}{( 1.5,\BOUNDY)}
    \skbound{B4}{( 2.5,\BOUNDY)}
  \skvertexp{V1}{(-1,-1)}
    \skvertexp{V2}{(0,-1)}
      \skvertexp{V3}{(1 ,-1)}
       \skvertexp{V4}{(2 ,-1)}
  \Gdouble{V1}{B1}
  \Gdouble{V3}{B2}
  \Gdouble{V4}{B3}
    \Gdouble{V4}{B4}
    \Gdouble{V2}{V3}
    \Gdouble{V3}{V4}
     \Gdoublearc{V1}{V2}{-45}{-135}  
          \Gdoublearc{V1}{V2}{45}{135}  
      \skbound{B1}{(-1.5,\BOUNDY)}
  \skbound{B2}{(0.5,\BOUNDY)}
  \skbound{B3}{( 1.5,\BOUNDY)}
    \skbound{B4}{( 2.5,\BOUNDY)}
  \skvertexp{V1}{(-1,-1)}
    \skvertexp{V2}{(0,-1)}
      \skvertexp{V3}{(1 ,-1)}
       \skvertexp{V4}{(2 ,-1)}
  \node[above=2pt] at (B1) {$\k_1$};
  \node[above=2pt] at (B2) {$\k_2$};
  \node[above=2pt] at (B3) {$\k_3$};
  \node[above=2pt] at (B4) {$\k_4$};
\end{tikzpicture} 
,
\\
\psi_{4,5}^{(4)}  (\k_1, \ldots, \k_4) 
&=&    \,\,
 \begin{tikzpicture}[baseline=-3.0pt]
  \draw[sk/boundary] (-1.8,\BOUNDY) -- (1.8,\BOUNDY);
  \skbound{B1}{(-1.5,\BOUNDY)}
  \skbound{B2}{( -0.5,\BOUNDY)}
  \skbound{B3}{( 0.5,\BOUNDY)}
    \skbound{B4}{( 1.5,\BOUNDY)}
  \skvertexp{V1}{(-1,-0.7)}
    \skvertexp{V2}{(0.0,-1.3)}
      \skvertexp{V3}{(0 ,-0.7)}
       \skvertexp{V4}{(1 ,-0.7)}
  \Gdouble{V1}{B1}
  \Gdouble{V1}{B2}
  \Gdouble{V4}{B3}
    \Gdouble{V4}{B4}
    \Gdouble{V1}{V3}
    \Gdouble{V3}{V2}
    \Gdouble{V3}{V4}
     \draw[sk/propdouble] (V2.25) to[out=45,in=-45,looseness=20] (V2.-25);
      \skbound{B1}{(-1.5,\BOUNDY)}
  \skbound{B2}{( -0.5,\BOUNDY)}
  \skbound{B3}{( 0.5,\BOUNDY)}
    \skbound{B4}{( 1.5,\BOUNDY)}
  \skvertexp{V1}{(-1,-0.7)}
     \skvertexp{V2}{(0.0,-1.3)}
      \skvertexp{V3}{(0 ,-0.7)}
       \skvertexp{V4}{(1 ,-0.7)}
  \node[above=2pt] at (B1) {$\k_1$};
  \node[above=2pt] at (B2) {$\k_2$};
  \node[above=2pt] at (B3) {$\k_3$};
  \node[above=2pt] at (B4) {$\k_4$};
\end{tikzpicture}  ,
\eea
and
\be
\psi_{4,6}^{(4)}  (\k_1, \ldots, \k_4) 
=   \,\,
 \begin{tikzpicture}[baseline=-3.0pt]
  \draw[sk/boundary] (-1.8,\BOUNDY) -- (1.8,\BOUNDY);
  \skbound{B1}{(-1.5,\BOUNDY)}
  \skbound{B2}{( -0.5,\BOUNDY)}
  \skbound{B3}{( 0.5,\BOUNDY)}
    \skbound{B4}{( 1.5,\BOUNDY)}
  \skvertexp{V1}{(-1,-0.7)}
    \skvertexp{V2}{(-1.0,-1.3)}
      \skvertexp{V3}{(0 ,-0.7)}
       \skvertexp{V4}{(1 ,-0.7)}
  \Gdouble{V1}{B1}
  \Gdouble{V3}{B2}
  \Gdouble{V4}{B3}
    \Gdouble{V4}{B4}
    \Gdouble{V1}{V3}
    \Gdouble{V1}{V2}
    \Gdouble{V3}{V4}
     \draw[sk/propdouble] (V2.25) to[out=45,in=-45,looseness=20] (V2.-25);
      \skbound{B1}{(-1.5,\BOUNDY)}
  \skbound{B2}{( -0.5,\BOUNDY)}
  \skbound{B3}{( 0.5,\BOUNDY)}
    \skbound{B4}{( 1.5,\BOUNDY)}
  \skvertexp{V1}{(-1,-0.7)}
     \skvertexp{V2}{(-1.0,-1.3)}
      \skvertexp{V3}{(0 ,-0.7)}
       \skvertexp{V4}{(1 ,-0.7)}
  \node[above=2pt] at (B1) {$\k_1$};
  \node[above=2pt] at (B2) {$\k_2$};
  \node[above=2pt] at (B3) {$\k_3$};
  \node[above=2pt] at (B4) {$\k_4$};
\end{tikzpicture}  .
\ee
To continue, there is only one diagram contributing to $\psi_5^{(3)}$, which is given by
\be
\psi_{5,1}^{(3)}  (\k_1, \ldots, \k_5) 
=    \,\,
 \begin{tikzpicture}[baseline=-3.0pt]
  \draw[sk/boundary] (-2.8,\BOUNDY) -- (1.8,\BOUNDY);
  \skbound{B1}{(-2.5,\BOUNDY)}
  \skbound{B2}{( -1.5,\BOUNDY)}
  \skbound{B3}{( -0.5,\BOUNDY)}
    \skbound{B4}{( 0.5,\BOUNDY)}
       \skbound{B5}{( 1.5,\BOUNDY)}
  \skvertexp{V1}{(-1.5,-1)}
    \skvertexp{V2}{(-0.5,-1)}
      \skvertexp{V3}{(0.5 ,-1)}
  \Gdouble{V1}{B1}
  \Gdouble{V1}{B2}
  \Gdouble{V2}{B3}
    \Gdouble{V3}{B4}
     \Gdouble{V3}{B5}
      \Gdouble{V1}{V3}
        \skbound{B1}{(-2.5,\BOUNDY)}
  \skbound{B2}{( -1.5,\BOUNDY)}
  \skbound{B3}{( -0.5,\BOUNDY)}
    \skbound{B4}{( 0.5,\BOUNDY)}
       \skbound{B5}{( 1.5,\BOUNDY)}
  \skvertexp{V1}{(-1.5,-1)}
    \skvertexp{V2}{(-0.5,-1)}
      \skvertexp{V3}{(0.5 ,-1)}
  \node[above=2pt] at (B1) {$\k_1$};
  \node[above=2pt] at (B2) {$\k_2$};
  \node[above=2pt] at (B3) {$\k_3$};
  \node[above=2pt] at (B4) {$\k_4$};
  \node[above=2pt] at (B5) {$\k_5$};
\end{tikzpicture}  .
\ee
Finally, there are two diagrams contributing to $\psi_6^{(4)}$, which are given by
\bea
\psi_{6,1}^{(4)}  (\k_1, \ldots, \k_6) 
&=&   \,\,
 \begin{tikzpicture}[baseline=-3.0pt]
  \draw[sk/boundary] (-2.8,\BOUNDY) -- (2.8,\BOUNDY);
  \skbound{B1}{(-2.5,\BOUNDY)}
  \skbound{B2}{( -1.5,\BOUNDY)}
  \skbound{B3}{( -0.5,\BOUNDY)}
    \skbound{B4}{( 0.5,\BOUNDY)}
       \skbound{B5}{( 1.5,\BOUNDY)}
          \skbound{B6}{(2.5,\BOUNDY)}
  \skvertexp{V1}{(-1.5,-1)}
    \skvertexp{V2}{(-0.5,-1)}
      \skvertexp{V3}{(0.5 ,-1)}
       \skvertexp{V4}{(1.5 ,-1)}
  \Gdouble{V1}{B1}
  \Gdouble{V1}{B2}
  \Gdouble{V2}{B3}
    \Gdouble{V3}{B4}
     \Gdouble{V4}{B5}
      \Gdouble{V4}{B6}
      \Gdouble{V1}{V4}
        \skbound{B1}{(-2.5,\BOUNDY)}
  \skbound{B2}{( -1.5,\BOUNDY)}
  \skbound{B3}{( -0.5,\BOUNDY)}
    \skbound{B4}{( 0.5,\BOUNDY)}
       \skbound{B5}{( 1.5,\BOUNDY)}
          \skbound{B6}{(2.5,\BOUNDY)}
  \skvertexp{V1}{(-1.5,-1)}
    \skvertexp{V2}{(-0.5,-1)}
      \skvertexp{V3}{(0.5 ,-1)}
       \skvertexp{V4}{(1.5 ,-1)}
  \node[above=2pt] at (B1) {$\k_1$};
  \node[above=2pt] at (B2) {$\k_2$};
  \node[above=2pt] at (B3) {$\k_3$};
  \node[above=2pt] at (B4) {$\k_4$};
  \node[above=2pt] at (B5) {$\k_5$};
  \node[above=2pt] at (B6) {$\k_6$};
\end{tikzpicture}  
,
\\
\psi_{6,2}^{(4)} (\k_1, \ldots, \k_6)  
&=&   \,\,
 \begin{tikzpicture}[baseline=-3.0pt]
  \draw[sk/boundary] (-2.8,\BOUNDY) -- (2.8,\BOUNDY);
  \skbound{B1}{(-2.5,\BOUNDY)}
  \skbound{B2}{( -1.5,\BOUNDY)}
  \skbound{B3}{( -0.5,\BOUNDY)}
    \skbound{B4}{( 0.5,\BOUNDY)}
       \skbound{B5}{( 1.5,\BOUNDY)}
          \skbound{B6}{(2.5,\BOUNDY)}
  \skvertexp{V1}{(-1.0,-0.6)}
    \skvertexp{V2}{(-0.0,-0.6)}
      \skvertexp{V3}{(1.0 ,-0.6)}
       \skvertexp{V4}{(0.0 ,-1.2)}
  \Gdouble{V1}{B1}
  \Gdouble{V1}{B2}
  \Gdouble{V2}{B3}
    \Gdouble{V2}{B4}
     \Gdouble{V3}{B5}
      \Gdouble{V3}{B6}
      \Gdouble{V4}{V1}
      \Gdouble{V4}{V2}
      \Gdouble{V4}{V3}
        \skbound{B1}{(-2.5,\BOUNDY)}
  \skbound{B2}{( -1.5,\BOUNDY)}
  \skbound{B3}{( -0.5,\BOUNDY)}
    \skbound{B4}{( 0.5,\BOUNDY)}
       \skbound{B5}{( 1.5,\BOUNDY)}
          \skbound{B6}{(2.5,\BOUNDY)}
  \skvertexp{V1}{(-1.0,-0.6)}
    \skvertexp{V2}{(-0.0,-0.6)}
      \skvertexp{V3}{(1.0 ,-0.6)}
       \skvertexp{V4}{(0.0 ,-1.2)}
  \node[above=2pt] at (B1) {$\k_1$};
  \node[above=2pt] at (B2) {$\k_2$};
  \node[above=2pt] at (B3) {$\k_3$};
  \node[above=2pt] at (B4) {$\k_4$};
  \node[above=2pt] at (B5) {$\k_5$};
  \node[above=2pt] at (B6) {$\k_6$};
\end{tikzpicture}  
.
\eea

\end{appendix}

\end{document}